\documentclass[aps,prd,reprint,superscriptaddress]{revtex4-1}

\usepackage{graphicx}  
\usepackage{subfigure}
\usepackage{latexsym}
\usepackage{slashed}
\usepackage{dcolumn}   
\usepackage{bm}        
\usepackage{amsmath}
\usepackage{amssymb}   
\usepackage{longtable}
\usepackage[section]{placeins}
\usepackage{float}
\def\askImath{\dot{\imath}}
\def\askPmu{(p - p')^\mu}
\def\askGmu{g^{\mu\nu}}
\def\askBbar{\overline{B}}
\newcommand{\ewnur}{EW$\nu_R$ }
\newcommand{\be}{\begin{equation}}
\newcommand{\ee}{\end{equation}}
\newcommand{\bea}{\begin{eqnarray}}
\newcommand{\eea}{\end{eqnarray}}
\newcommand{\bma}{\begin{matrix}}
\newcommand{\ema}{\end{matrix}}

\newcommand{\bml}{\begin{mathletters}}
\newcommand{\eml}{\end{mathletters}}
\newcommand{\bes}{\begin{subequations}}
\newcommand{\ees}{\end{subequations}}

\newcommand{\bi}{\begin{itemize}}
\newcommand{\ei}{\end{itemize}}
\newcommand{\gev}{~{\rm GeV}}

\newcommand{\uva}{\affiliation{Department of Physics, University of Virginia, Charlottesville, VA 22904-4714, USA}}
\newcommand{\hue}{\affiliation{Center for Theoretical and Computational Physics, Hue University College of Education, Hue, Vietnam}}

\begin{document}
\title{Electroweak precision constraints on the electroweak-scale right-handed neutrino model}
\author{Vinh Hoang}
\email{vvh9ux@virginia.edu}\uva
\author{Pham Q. Hung}
\email{pqh@virginia.edu}\uva\hue
\author{Ajinkya Shrish Kamat}
\email{ask4db@virginia.edu}\uva

\date{\today}

\begin{abstract}
A model of electroweak-scale right-handed neutrino (\ewnur) model was constructed five years ago in which the right-handed neutrinos are members of mirror
fermion weak doublets and where the Majorana masses of the right-handed neutrinos
are found to be {\em naturally} of the order of the electroweak scale.  These features facilitate their searches at the LHC through signals such
 as like-sign dilepton events. This model contains, in addition to the mirror quarks and leptons, extra scalars transforming as weak triplets. In this paper, we study the constraints imposed on these additional particles by the electroweak precision parameters S, T, and U. These constraints are crucial in determining the viability of the electroweak $\nu_R$ model and the allowed parameter space needed for a detailed phenomenology of the model. 

\end{abstract}

\maketitle

\section{Introduction}

Two of the most pressing problems in particle physics are, without any doubt, the nature of the spontaneous breaking of the electroweak symmetry and the nature of neutrino masses and mixings. It goes without saying that the discovery of a Higgs-like particle with a mass of 126 GeV at the LHC goes a long way in the attempt to answer to the first question although much remains to be determined if the 126 GeV object is truly a $0^{++}$ particle predicted by the Standard Model (SM) or it is something else beyond the Standard Model. As to the second question concerning neutrino masses, the general consensus is that the discovery of neutrino oscillations is best explained by neutrinos having a mass- albeit a very tiny one. One might say that this is the first sign of Physics Beyond the Standard Model since neutrinos are massless in the SM.
There has also been important advances in measuring mixing angles in the PMNS  matrix of the neutrino sector. In particular, the angle $\theta_{13}$ was found by the Daya Bay experiment \cite{daya} to be quite large, a number which was subsequently confirmed by the RENO experiment \cite{reno}. 

In spite of these successes, we still do not know whether the neutrinos are of the Dirac type or of the Majorana type. In either case, the simplest approach is to add  right-handed neutrinos which are definitely particles beyond the SM spectrum. What is the nature of these right-handed neutrinos? The standard assumption is one in which they are {\em SM singlets}: the so-called {\em sterile neutrinos}. For this singlet assumption to become a physical reality, one should be able to test it in order to either prove or disprove it. Presently, there is {\em no evidence} for these sterile neutrinos. Furthermore, since nothing is known about the possible existence and associated properties of right-handed neutrinos, it is prudent to entertain other logical possibilities. Why should right-handed neutrinos be sterile? Would the assumption of  {\em SM-non singlet} right-handed neutrinos be also reasonable? Could one test it? This latter assumption is one that was proposed by one of us in formulating the EW$\nu_R$ model \cite{pqnur} to which we will come back below.

The most elegant mechanism for generating tiny neutrino masses is the quintessential seesaw mechanism in which a large lepton-number-violating Majorana mass, $M_R$, typically of the order of some Grand Unified Theory mass scale, was given to the SM-singlet right-handed neutrino and a Dirac mass, $m_D \ll M_R$, was assumed to come form the electroweak sector, giving rise to a mass $\sim m_D^2/M_R \ll m_D$ which could be of the order of O($<eV$) \cite{seesaw}. How does one test this version of seesaw mechanism? One could either look for the right-handed neutrinos and/or search for lepton-number-violating processes. It is however practically impossible to directly ``detect'' the SM-singlet right-handed neutrino unless extreme fine-tuning is carried out to make the right-handed neutrinos much lighter than the GUT scale\cite{smirnov}. The most common way to test the seesaw mechanism is to look for signals where a lepton-number violating process such as the neutrino-less double beta decay is present. However, such a process is extremely hard to detect and so far one has not had much luck with it.

In a generic seesaw scenario, one has two scales: $M_R \sim M_{GUT} \sim 10^{16} \, GeV$ and $m_D \propto \Lambda_{EW} \sim 246 \, GeV$. Out of those two scales, only $\Lambda_{EW} $ is observable while $M_{GUT}$ is a hypothetical scale that may or may not exist. Without fine-tuning, the fate of the SM-singlet (sterile) right-handed neutrinos is linked to that of this hypothetical scale. 

The question that was asked in \cite{pqnur} was as follows: Is it possible to {\em naturally} make the Majorana mass of the right-handed neutrinos of the order of the electroweak scale? The answer is yes. One {\em only} needs to extend the SM in the fermion and scalar sectors. The gauge group is still $SU(3)_C \times SU(2) \times U(1)_Y$, where the usual subscript $L$ for $SU(2)$ is absent for reasons to be explained below. If $M_R \sim \Lambda_{EW}$ and because of $m_D^2/M_R$, one would need $m_D \sim O(keV)$ in order to have neutrino masses of the order of eV or less.  This then requires the introduction of a hypothetical scale, $m_S$, which, in contrast to $M_{GUT}$, is of O(keV). This scale may be related to the physics of dark matter \cite{pqnur2}. This is the model of electroweak-scale right-handed neutrinos presented in \cite{pqnur}. As one will see in the brief review of this model, this necessitates the introduction of mirror fermion doublets of the SM gauge group, of which the right-handed neutrinos are members. The right-handed neutrinos in the model of \cite{pqnur} acquire a Majorana mass naturally of $O(\Lambda_{EW} )$. Furthermore, they belong to weak doublets and couple to W's and Z and have electroweak production cross sections at colliders such as the LHC. Some of the signals are described briefly in \cite{pqnur}. In addition, the \ewnur model contains one Higgs doublet and two Higgs triplets, one of which contains a doubly-charged scalar. Some of the phenomenology of this sector of the model was explored in \cite{pqaranda}. We shall come back to the implication of this scalar sector on the 126 GeV object in a separate paper.

The model of \cite{pqnur} contains ``mirror '' quarks and leptons which are accessible at the LHC. The phenomenology of these fermions will be presented in \cite{pqaranda2}. Since, for every SM left-handed doublet, one has a right-handed doublet (and similarly for the SM right-handed singlets), the number of chiral doublets has increased by a factor of {\em two}. This raises the obvious question of potential problems with electroweak precision data through the S, T and U parameters. In particular, even if one artificially makes the top and bottom members of these mirror doublets degenerate, one is faced with a large contribution to the S parameter. (In fact, this was a big problem with Technicolor models \cite{technicolor}.) These large contributions from the extra chiral doublets would have to be offset by contributions from other sectors with the opposite sign in such a way that the sum falls within the experimental constraints. It was mentioned in \cite{pqnur} that such extra contributions can be found in the scalar sector, in particular the Higgs triplet sector where its contribution can be negative enough to offset the positive contribution from the mirror fermions to S. It is the purpose of the present manuscript to examine in detail the contributions of the mirror fermions and the extended Higgs sectors to the electroweak precision parameters. As we shall see below, the \ewnur model fits nicely with the electroweak precision constraints which, in turn, put limits on the mass splittings within the multiplets of mirror fermions and the Higgs multiplets and so on.

Finally, one should notice that there are aspects of the SM which are intrinsically {\em non-perturbative} such as the electroweak phase transition. The most common framework to study non-perturbative phenomena is through lattice regularization. It is known that one cannot put a chiral gauge theory such as the SM on the lattice without violating gauge invariance. However, a gauge-invariant formulation of the SM on the lattice is possible if one introduces mirror fermions \cite{montvay}. Is it possible that the mirror fermions of the EW$\nu_R$ model play such a role?

We end the Introduction by quoting part of a sentence in the famous paper about parity violation by Lee and Yang \cite{leeyang}: ``If such asymmetry is indeed found, the question could still be raised whether there could not exist corresponding elementary particles exhibiting opposite asymmetry such that in the broader sense there will still be over-all right-left symmetry..'' \cite{pqnur} is, in some sense, a response to this famous quote.

The plan of the manuscript will be as follows. First, we summarize the essential elements of the \ewnur model of \cite{pqnur}. Second, we present calculations of the electroweak precision parameters in the \ewnur model. Third, we discuss the implications coming from the constraints on the electroweak precision parameters on the various mass splittings and parameters of the mirror sector as well as of the extended Higgs sector. We conclude with some remarks concerning the 126 GeV boson.

\section{The \ewnur model}
\label{sec:ewnur}

\cite{pqnur} asked the following two questions: 1) Could one obtain the right-handed neutrino Majorana mass
strictly within the SM gauge group $SU(3)_c \otimes SU(2)_L \otimes U(1)_Y$
by just extending its particle content?; 2) If it is possible to do
so, what would be the constraints on the Dirac mass scale? The answer to the first question lies in the construction of the \ewnur model \cite{pqnur}.

In a generic seesaw scenario, $\nu_R$s are SM singlets and, as a result, a right-handed neutrino mass term of the form $M_R \, \nu_{R}^{T} \sigma_{2} \nu_R$ is also a singlet of the SM. As a result, $M_R$ can take on any value and is usually assumed to be of the order of some GUT scale if the SM is embedded in a GUT group such as $SO(10)$. To constrain $M_R$, one has to endow the right-handed neutrinos with some quantum numbers. For example, if $\nu_R$ belongs to a {\bf 16} of $SO(10)$ it is natural for $M_R$ to be of the order of the $SO(10)$ breaking scale. Another example is the left-right symmetric extension of the SM \cite{goran} where $\nu_R$ belongs to a doublet of $SU(2)_R$. The aforementioned Majorana mass term would still be a singlet under $SU(2)_L$ but it is no longer so under $SU(2)_R$. It is then natural that $M_R \sim M_{\tilde{R}} \gg M_{\tilde{L}}$, where $M_{\tilde{R},\tilde{L}}$ are the breaking scales of $SU(2)_R$ and $SU(2)_L$ respectively. In all of these scenarios, the value of the Dirac mass $m_D$ in $m_D^2/M_R$ usually comes from the breaking of the SM $SU(2)_L$ and is naturally proportional to the electroweak breaking scale. The smallness of neutrino masses gives rise, without fine-tuning,  to an ``energy gap'' $m_D \sim O(\Lambda_{EW}) \longrightarrow M_R \sim O(M_{GUT}) \, \text{or} \, O(M_{\tilde{R}})$. Without fine-tuning, the large value of $M_R \sim O(M_{GUT})$ would make it practically impossible to detect the SM-singlet right-handed neutrinos at machines such as the LHC and to directly test the seesaw mechanism. However, in the L-R model, the production of $\nu_R$ can proceed first through the production of $W_R$ as first shown in \cite{goran2}. The feasibility of such a process was discussed in \cite{Ferrari:2000sp}. (Other mechanisms proposed to make the SM-singlet right-handed neutrinos accessible at the LHC through SM $W$ are discussed in \cite{than} although it might be very difficult to do so due to the size of the Dirac Yukawa coupling.)

It is clear as presented in \cite{pqnur} that one of the natural and minimal ways (in terms of the gauge group) to test the seesaw mechanism and to detect the right-handed neutrinos at colliders such as the LHC is to make the right-handed neutrinos {\em non singlets} under the SM $SU(2)_L$ for two reasons. The first reason has to do with the mass scale $M_R$. If $\nu_R$'s are non-singlets under $SU(2)_L$ then $M_R$ necessarily comes from the breaking of $SU(2)_L$ and therefore would naturally be of the order of the electroweak scale. Energetically-speaking, it could be directly detected at the LHC \cite{pqnur}. The second reason has to do with the possible detection of $\nu_R$'s themselves. Being $SU(2)_L$ non-singlets, they can couple to the SM electroweak gauge bosons and the production cross sections would be naturally of the order of the electroweak cross sections \cite{pqnur}. 

The simplest way to make $\nu_R$'s $SU(2)_L$ non-singlets is to group them into $SU(2)_L$ right-handed doublets with the right-handed charged partners which are new charged leptons with opposite chirality to the SM charged leptons. Anomaly freedom would necessitate the introduction of $SU(2)_L$ doublets of right-handed quarks.  These new right-handed quarks and leptons are called {\em mirror fermions} in \cite{pqnur}. The right-handed quarks and charged leptons are accompanied by their {\em left-handed} partners which are $SU(2)_L$ singlets, a complete mirror image of the SM fermions, so mass terms can be formed by coupling to the Higgs doublet. The $SU(2)_L \times U(1)_Y$ fermion content of the \ewnur model of \cite{pqnur} is given, for each family,  as follows.

\bi

\item $SU(2)_L$ lepton doublets:

\be
\label{ldoublet}
SM: l_{L} = \left( \begin{array}{c}
\nu_L \\
e_{L}
\end{array} \right) \,; \,
Mirror: l^{M}_{R} = \left( \begin{array}{c}
\nu_R \\
e^{M}_{R}
\end{array} \right) \
\ee
for the SM left-handed lepton doublet and
for the right-handed mirror lepton doublet respectively.  

\item $SU(2)_L$ lepton singlets:
\be
\label{lsinglet}
SM: e_R \, ; \, Mirror: e^{M}_L \,,
\ee
for the right-handed SM lepton singlet and left-handed mirror lepton singlet respectively. 
\ei

Similarly, for the quarks, we have

\bi

\item $SU(2)_L$ quark doublets:

\be
\label{qdoublet}
SM: q_{L} = \left( \begin{array}{c}
u_{L} \\
d_{L}
\end{array} \right) \,; \,
Mirror: q^{M}_{R} = \left( \begin{array}{c}
u^{M}_{R} \\
d^{M}_{R}
\end{array} \right) \
\ee
for the SM left-handed quark doublet and
for the right-handed mirror quark doublet respectively. 

\item $SU(2)_L$ quark singlets:
\be
\label{qsinglet}
SM: u_R \,, \, d_R \, ; \, Mirror: u^{M}_L \,,\, d^{M}_L
\ee
for the right-handed SM quark singlets and left-handed mirror quark singlets respectively. 
\ei
Apart from chiralities, the $SU(2)_L \times U(1)_Y$ quantum numbers of the mirror fermions are identical to those of the SM fermions. A remark is in order at this point. What we refer to as mirror fermions are the particles listed above and they are not to be confused with particles in the literature which have similar names but which are entirely of a different kind. As the above listing shows, the mirror quarks and leptons are particles which are {\em different} from the SM ones. It is for this reason that a superscript $M$ was used in \cite{pqnur} and here in order to avoid possible confusion.
These chiral mirror fermions will necessarily contribute to the precision electroweak parameters and potentially could create disagreements unless contributions from other sectors are taken into account. This will be the main focus of the next sections.

As with the SM leptons, the interaction of mirror leptons with the $SU(2)_L \times U(1)_Y$ gauge bosons are found in the terms 
\be
\label{kinetic}
\bar{l}^{M}_{R} \slashed{D} l^{M}_{R}\, ; \, \bar{e}^{M}_{L} \slashed{D} e^{M}_{L} \, ,
\ee
where the covariant derivatives $\slashed{D}$ are the same as the ones used for the SM leptons and are listed explicitly in the Appendix \ref{sec:appfeynf}. The gauge interactions of the mirror quarks can similarly be found. 

We next review the salient point of the \ewnur model of \cite{pqnur}: The electroweak seesaw mechanism. For the sake of clarity, we repeat here the arguments given in \cite{pqnur}.
As discussed in \cite{pqnur}, a Majorana mass term of the type $M_R\, \nu_{R}^{T}\,  \sigma_{2} \, \nu_R$ necessarily breaks the electroweak gauge group. The reason is as follows. The bilinear $l_R^{M,T} \,\sigma_2\,l_R^{M}$ contains $\nu_{R}^{T}\,  \sigma_{2} \, \nu_R$ and transforms under $SU(2)_L \times U(1)_Y$
as $(1+3,Y/2=-1)$. For obvious reasons, the Higgs field which couples to this bilinear {\em cannot} be an $SU(2)_L$ singlet with the quantum number $(1, Y/2=+1)$ since this singlet charged scalar cannot develop a VEV. This leaves the {\em triplet} Higgs field $\tilde{\chi} =(3, Y/2=+1)$ as a suitable scalar which can couple to the aforementioned bilinear and whose neutral component can develop a VEV:
\be
\label{delta}
\tilde{\chi} = \frac{1}{\sqrt{2}}\,\vec{\tau}.\vec{\chi}=
\left( \begin{array}{cc}
\frac{1}{\sqrt{2}}\,\chi^{+} & \chi^{++} \\
\chi^{0} & -\frac{1}{\sqrt{2}}\,\chi^{+}
\end{array} \right) \,.
\ee
The Yukawa coupling of the bilinear to this Higgs field was given in \cite{pqnur} and is written down again here
\bea
\label{majorana}
{\cal L}_M &=& g_M \,(l^{M,T}_{R}\, \sigma_2) \,(\imath \, \tau_2 \,\tilde{\chi})\, l^{M}_{R}  \nonumber \\
&=& g_M( \nu_{R}^{T}\,  \sigma_{2} \, \nu_R \chi^{0} - \frac{1}{\sqrt{2}} \nu_{R}^{T}\,  \sigma_{2} e^{M}_{R} \chi^{+}- \frac{1}{\sqrt{2}} e_{R}^{M,T}\,  \sigma_{2} \nu_{R} \chi^{+} \nonumber \\
&&+e_{R}^{M,T}\,  \sigma_{2} e^{M}_{R} \chi^{++})\,,
\eea
From Eq.(\ref{majorana}), one notices the Yukawa term $g_M \, \nu_{R}^{T}\,  \sigma_{2} \, \nu_R \chi^{0}$ which upon having
\be
\label{delta0}
\langle \chi^{0} \rangle = v_M \,,
\ee
gives rise to the right-handed Majorana mass
\be
\label{majmass}
M_R = g_M\,v_M \,.
\ee
As it has been stressed in \cite{pqnur}, $M_R$ is naturally of the order of the electroweak scale since $v_M \sim O(\Lambda_{EW})$ and is constrained to be larger than $M_Z/2 \sim 46 \, GeV$ because of the constraint coming from the width of the Z boson (no more than three light neutrinos). A triplet Higgs field with such a large vacuum expectation value will destroy the ``custodial symmetry'' value $\rho=1$ at tree level. A nice remedy for this problem was given in \cite{pqnur} and will be reviewed below.

A Dirac mass term of the type $\bar{\nu}_L \nu_R$ comes from $\bar{l}_L l^{M}_{R}$  which is $1+ 3$ under $SU(2)_L$. It was argued in \cite{pqnur} why a singlet scalar field is the appropriate choice and why a triplet is phenomenologically ruled out \cite{triplet}.
As in \cite{pqnur}, the interaction with the singlet scalar is given as
\bea
\label{singlet}
{\cal L}_S &=& g_{Sl} \, \bar{l}_{L}\, \phi_S \, l^{M}_{R} + H.c. 
\nonumber \\
&=& g_{Sl}\,(\bar{\nu}_L \nu_R + \bar{e}_L \, e^{M}_{R})\,\phi_S
+ H.c. 
\eea 
With
\be
\label{vevs}
\langle \phi_S \rangle = v_S \,,
\ee
the neutrino Dirac mass is given by
\be
\label{neumass}
m_{\nu}^D = g_{Sl} \, v_S \,,
\ee
If $g_{Sl} \sim O(1)$, this implies that $v_S \sim O(10^5 \, eV)$. It has been discussed in \cite{pqnur}  that this value for $v_S$ is six orders of magnitude smaller then the electroweak scale $\Lambda_{EW}$ and this hierarchy requires a cross coupling between the singlet and the triplet scalars to be of order $\sim 10^{-12}$. To "evade" this fine tuning, it was proposed in \cite{pqnur} that the "classical" singlet scalar field takes a value $\phi_S(t_0) \sim (10^5 \, eV)$ at the present time and its value changes with time whose rate is dictated by a "slow-rolling" effective potential. It goes without saying that much remains to be worked out for this scenario. Alternatively, one can assume that $v_S \sim O(\Lambda_{EW})$ and set $g_{Sl} \sim 10^{-7}$ as suggested in \cite{pqnur2} to obtain a Dirac mass of the desired order. (This is actually not so unnatural as the example of the electron mass being $\sim  10^{-7} \Lambda_{EW}$ illustrates.) 

Eq.(\ref{majorana}) gives a Majorana mass to the right-handed neutrinos but one could easily have from gauge invariance a term such as $g_{L} \,(l^{T}_{L}\, \sigma_2) \,(\imath \, \tau_2 \,\tilde{\chi})\, l_{L}$ which would yield a large Majorana mass for the left-handed neutrinos unless fine-tuning is carried out. 

As discussed in \cite{pqnur}, in order to guarantee that left-handed neutrinos have vanishing Majorana masses at tree level, a ``mirror global symmetry'' $U(1)_M$ was imposed:
\be
\label{U1Ml}
(l^{M}_{R}, e^{M}_{L}) \rightarrow e^{i\,\theta_M} (l^{M}_{R}, e^{M}_{L}) \,,
\tilde{\chi} \rightarrow e^{-2\,i\,\theta_M}\,\tilde{\chi} \,,
\phi_S \rightarrow e^{-i\,\theta_M}\,\phi_S \,,
\ee
for the mirror leptons and triplet and singlet scalars and
\be
\label{U1Mq}
(q^{M}_{R}, u^{M}_{L}, d^{M}_{L}) \rightarrow e^{i\,\theta_M} (q^{M}_{R}, u^{M}_{L}, d^{M}_{L}) \,,
\ee
for the mirror quarks. In \cite{pqnur}, it was mentioned that the left-handed neutrinos can acquire a Majorana mass at one-loop of the type $M_L = \lambda \, \frac{1}{16\, \pi^2}\,\frac{m_{\nu}^{D\, 2}}{M_R}\,\ln \frac{M_R}{M_{\phi_S}}$, where $\lambda$ is the $\phi_S$ quartic coupling. This is smaller than the light neutrino mass by at least two orders of magnitude and can be neglected. 

Beside preventing the left-handed neutrinos from acquiring a large tree-level Majorana mass, this $U(1)_M$ symmetry also prevents terms such as $\bar{q}_L q^{M}_R$, $\bar{u}_R u^{M}_L$ and $\bar{d}_R d^{M}_L$. Therefore, as stressed in \cite{pqnur}, any bilinear mixing between SM fermions and mirror fermions will have to couple with the singlet scalar $\phi_S$ in order to be $U(1)_M$-invariant at tree level, namely $\bar{q}_L \, \phi_S \, q^{M}_R$, $\bar{u}_R \, \phi_S \, u^{M}_L$ and $\bar{d}_R \, \phi_S \, d^{M}_L$. Because of these mixings between the two sectors, the mass eigenstates are not pure left-handed SM quarks or right-handed mirror quarks. This was discussed in \cite{pqnur}. However, the deviation from the ``pure'' states, i.e. for example $\tilde{u}_L=u_{L} + O(v_{S}/\Lambda_{EW}) u^{M,c}_R$,  is proportional to $v_S / \Lambda_{EW} \sim 10^{-6}$ and, for most practical purposes, one can neglect this mixing.
 
To finish up with the review of the \ewnur model, we review the triplet scalar sector of \cite{pqnur}.  Let us recall that the $\rho$-parameter for arbitrary Higgs multiplets is given by $\rho= (\sum_{i}[T(T+1)-T_{3}^2]_i v_{i}^2 c_{T,Y})/(2 \, \sum_{i} T_{3i}^2 v_{i}^2)$, where $c_{T,Y} = 1$ for complex multiplet and $c_{T,Y} = 1/2$ for real multiplet \cite{higgshunter}. If one just has the triplet $\tilde{\chi}$ and nothing else, one would obtain $\rho=1/2$ in contradiction with the fact that experimentally $\rho \approx 1$. Pure Higgs doublets would give naturally $\rho=1$. A mixture with one triplet and one doublet would give $\rho \approx 1$ if the VEV of the triplet, $v_M$, is much less than that of the doublet, $v_2$, i.e. $v_M \ll v_2$. But this is not what we want since we would like to have $v_M$ and $v_2$ of $O(\Lambda_{EW})$. To preserve the custodial symmetry with a Higgs triplet, another triplet Higgs scalar $\xi=(3,Y/2=0)$ is needed in addition to the aforementioned $\tilde{\chi}(3,Y/2=1)$ and the usual doublet $\phi=(2, Y/2=-1/2)$. The potential for these three scalar multiplets and its minimization is given in the Appendix. This potential possesses a global $SU(2)_L \times SU(2)_R$ symmetry under which the two triplets are combined into the following $(3,3)$ representation \cite{ChanoGold,GeorgMach,pqaranda,godbole}:
\be
\label{chi}
\chi = \left( \begin{array}{ccc}
\chi^{0} &\xi^{+}& \chi^{++} \\
\chi^{-} &\xi^{0}&\chi^{+} \\
\chi^{--}&\xi^{-}& \chi^{0*}
\end{array} \right) \,.
\ee
Similarly, $\phi$ and $\tilde{\phi} = \imath \tau_2 \phi^{*}$ can be grouped into a $(2,2)$ representation:
\be
\label{Phi}
\Phi=\left( \begin{array}{cc}
\phi^{0*} & \phi^{+} \\
\phi^{-} & \phi^{0}
\end{array} \right) \,.
\ee
Proper vacuum alignment so that $SU(2)_L \times U(1)_Y \rightarrow U(1)_{em}$ gives
\be
\label{chivev}
\langle \chi \rangle = \left( \begin{array}{ccc}
v_M &0&0 \\
0&v_M&0 \\
0&0&v_M
\end{array} \right) \,,
\ee
and
\be
\label{phivev}
\langle \Phi \rangle = \left( \begin{array}{cc}
v_2/\sqrt{2} &0 \\
0&v_2/\sqrt{2} 
\end{array} \right) \,.
\ee
This breaks the global $SU(2)_L \times SU(2)_R$ down to the custodial $SU(2)_D$.
One obtains $M_W = g\,v/2$ and $M_Z = M_W/\cos \theta_W$, with $v= \sqrt{v_2^2+ 8\,v_M^2} \approx 246 \, GeV$ and, at tree level, $\rho= M_W/M_Z \, \cos \theta_W=1$ as desired. 

After spontaneous breaking of $SU(2)_L \times U(1)_Y$, beside the three Nambu-Goldstone bosons which are absorbed by W and Z, there are {\em ten} physical scalars which are grouped into {\bf 5} + {\bf 3} + {\bf 1} of the custodial $SU(2)_D$. (In fact there are two singlets of $SU(2)_D$.) These states are discussed in the next section and in the Appendix \ref{sec:appfeyns}.

\begin{figure}[H]
\centering
    \includegraphics[scale=0.19]{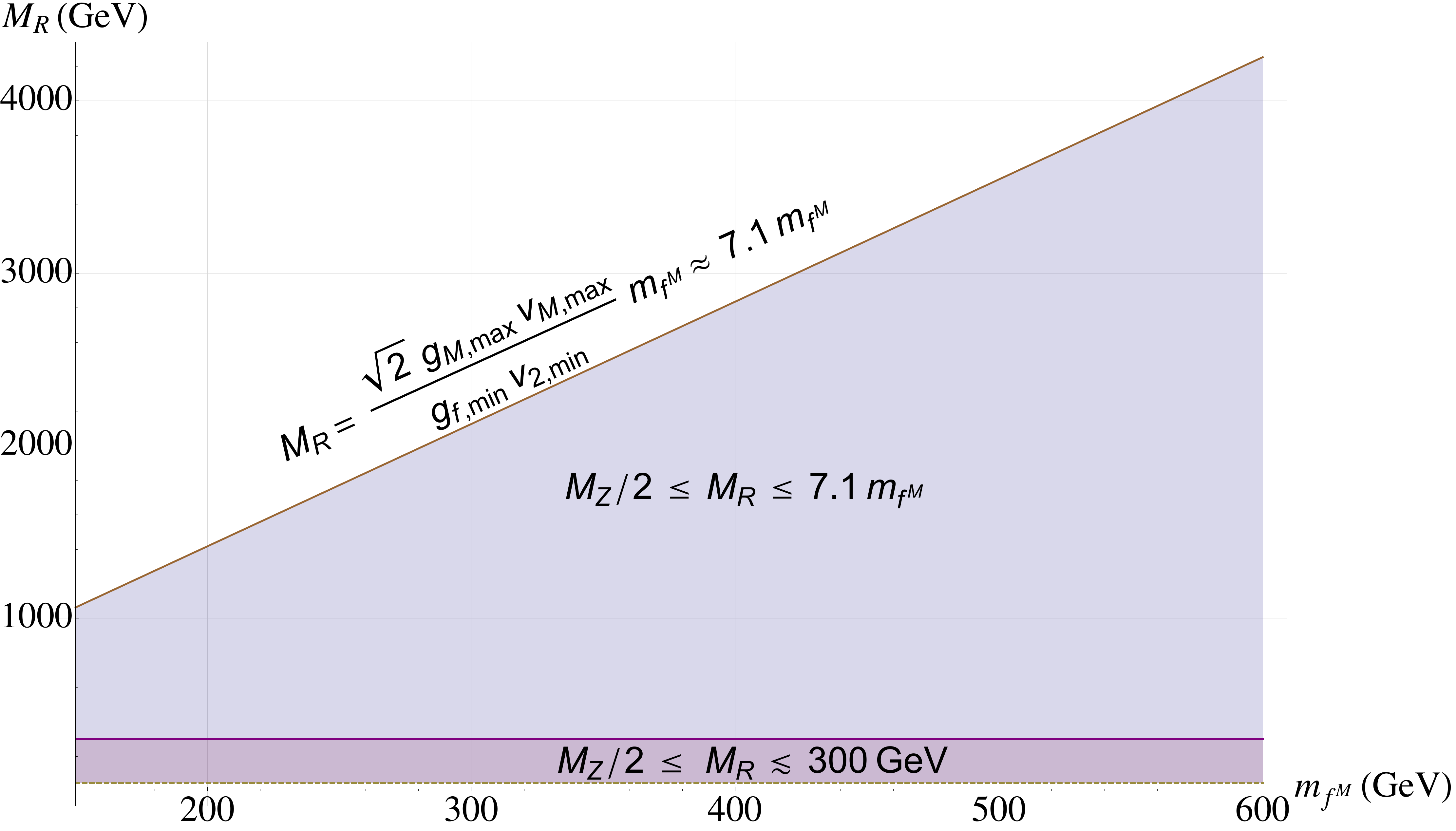} 
 \caption{\label{fig:MR_mf}{\small Mass of $\nu_R$ versus mass of charged mirror fermion $f^M$ with constraints due to perturbativity of the Yukawa couplings. Thus, the final constraints are $M_Z/2 \leq M_R \leq 300\, \gev$ and $m_{f^M} \leq 610\, \gev$ (small purple area).}}
\end{figure}

Last but not least in this mini review is the question of charged fermion masses, in particular the top quark and mirror fermion masses and the perturbativity of the Yukawa couplings they arise from. This is a topic on its own but a few words are in order here. It goes without saying that this interesting topic deserves a detailed investigation but such endeavor is beyond the scope of this manuscript. Since $v= \sqrt{v_2^2+ 8\,v_M^2} \approx 246 \, GeV$, it is evident that $v_2 < 246\, \gev$ and $v_M < 87 \, \gev$. This has implications regarding fermion masses, since charged fermion masses are proportional to $v_2$ while the $\nu_R$ masses are proportional to $v_M$. The requirement that the Yukawa couplings giving rise to these masses, namely $g_f$'s and $g_M$, are perturbative (i.e. $\alpha_{f,M} \equiv g_{f,M}^2/(4\,\pi) \leq 1$) imposes constraints on the allowed ranges of $v_2$ and $v_M$ respectively, and also on the allowed ranges of masses of the mirror fermions. Since $M_R = g_M\,v_M$ and since {\em naively} a charged fermion mass is given by (ignoring mixings in the mass matrix for now) $m_f = g_f \, v_2 / \sqrt{2}$, for a given mass of a charged mirror fermion ($m_{f^M}$) the upper limit on masses of $\nu_R$'s can be given by
\begin{equation}\label{eq:MR_mf}
	M_R\; \leq\; \dfrac{\sqrt{2}\,g_{M,\text{max}}\, v_{M,\text{max}}}{g_{f^M,\text{min}}\, v_{2,\text{min}}}\, m_{f^M}
\end{equation}
Let us estimate each quantity in the fraction on the right hand side of this equation. As mentioned before, $g_{M,\text{max}} = \sqrt{4 \pi}$. Because the top quark mass is known, ({\em naively} expressing it as $m_{\text{top}} = g_{\text{top}} \, v_2 / \sqrt{2} \approx 170 \gev$), perturbativity of $g_{\text{top}}$ gives $v_2 \geq 68\, \gev$. This constrains $v_M$ further such that $v_M \leq 84 \gev$. Since $M_R > M_Z /2$, it follows that $v_M \geq 13 \gev$ to ensure that $g_M \leq \sqrt{4\; \pi}$. This limit on $v_M$ implies that $v_2 \leq 243 \gev$. Hence, considering the charged mirror fermion masses to be heavier than $150\, \gev$ it is straight forward to see that $g_f \geq 0.87$ for $v_2 \sim 243 \gev$. Thus, Eq.~(\ref{eq:MR_mf}) becomes $M_R\, \leq\, 7.1\, m_{f^M}$. On the other hand $g_M \leq \sqrt{4\pi}$ and $v_M \leq 84 \gev$ also imply that $M_R \leq 300\, \gev$. Both these constraints are plotted in FIG.~\ref{fig:MR_mf}. In addition to any other constraints, the aforementioned constraints are also to be incorporated while while studying the phenomenology of the \ewnur model.
Considering $m_f = g_f \, v_2 / \sqrt{2} \leq g_f \, 148 \gev$, one expects a Yukawa coupling $g_{\text{top}} \sim 1.2$ for the top quark. This coupling can actually be even smaller if the SM quark mass matrix is of the "democratic type" i.e. having all matrix elements being equal to 1 \cite{democratic}. (A more "realistic" version differs slightly from this one.) The largest mass eigenvalue in such a model is $\sim 3 \, g_f \, 148 \gev$ giving $g_t \sim 0.4$. For very heavy mirror quarks, the Yukawa couplings might be larger, but, because the requirement for perturbativity is $ \alpha_{f} \equiv g_{f}^2/4\, \pi \leq 1$, a value of $g_f \sim 2\, \text{or}\, 3$ might not be problematic. There is also an interesting twist in the situation when the Yukawa couplings become large: A possibility that the electroweak symmetry can be broken dynamically by condensates of heavy fermions through the exchange of a fundamental scalar as it has been done for a heavy fourth generation \cite{pqchi}.

\section{Oblique Parameters}
\label{sec:obl}
As we have mentioned in the introduction, the contributions to the S parameter coming from the extra mirror fermions will be positive and will exceed the constraints imposed by electroweak precision data. These contributions will have to be cancelled by those coming from another sector such as the triplet Higgs present in the \ewnur model. This has been suggested in \cite{pqnur}. In this section, we carry out a detailed calculation of the electroweak precision parameters, the so-called oblique parameters, within the framework of the \ewnur model.

In Appendix \ref{sec:appfeyns}, we summarize the discussion of the minimization of the scalar potential given by Eq.(\ref{eq:pot}). In what follows, we list the expressions for the physical states and for the Nambu-Goldstone bosons in terms of the original scalar fields.

\par Physical observables like the oblique parameters are to be expressed using the masses of physical scalars. To express these physical states we use the subsidiary fields \cite{pqaranda}:
	\begin{eqnarray}\label{eq:subsid}
		\phi^0 &\equiv& \frac{1}{\sqrt{2}} \Big(v_2 + \phi^{0r} + \askImath \phi^{0\askImath}\Big),\nonumber\\[2mm]
		\chi^0 &\equiv& v_M + \frac{1}{\sqrt{2}} \Big(\chi^{0r} + \askImath \chi^{0\askImath}\Big);\\[2mm]
		\psi^\pm &\equiv& \frac{1}{\sqrt{2}} \Big(\chi^\pm + \xi^\pm \Big),\hspace{1em}\zeta^\pm \equiv \frac{1}{\sqrt{2}} \Big(\chi^\pm - \xi^\pm\Big)
	\end{eqnarray}
for the complex neutral and charged fields respectively. Here the quantities with superscripts `r' and `i' denote the `real' and the `imaginary' components, respectively. Note that the real components, $\phi^{0r}$ and $\chi^{0r}$, have {\em zero} vacuum expectation values. With these fields the Nambu-Goldstone bosons are given by
	\begin{eqnarray}\label{eq:goldstone}
		G_3^\pm &=& c_H \phi^\pm + s_H \psi^\pm,\nonumber\\[2mm]
		G_3^0 &=& \askImath \Big(-c_H \phi^{0\askImath} + s_H \chi^{0\askImath}\Big).
	\end{eqnarray}
The scalar potential in Eq. (\ref{eq:pot}) preserves the custodial $SU(2)_D$. Hence, the physical scalars can be grouped, as stated in the previous section, based on their transformation properties under $SU(2)_D$ as follows:
	\begin{eqnarray}
		\text{five-plet (quintet)} &\rightarrow& H_5^{\pm\pm},\; H_5^\pm,\; H_5^0;\nonumber\\[2mm]
		\text{triplet} &\rightarrow& H_3^\pm,\; H_3^0;\nonumber\\[2mm]
		\text{two singlets} &\rightarrow& H_1^0,\; H_1^{0\prime}\,,
	\end{eqnarray}
where
	\begin{eqnarray}\label{eq:higgs}
		H_5^{++} &=& \chi^{++},\; H_5^+ = \zeta^+,\; H_3^+ = c_H \psi^+ - s_H \phi^+,\nonumber\\[2mm]
		H_5^0 &=& \frac{1}{\sqrt{6}}\Big(2\xi^0 - \sqrt{2}\chi^{0r}\Big),\; H_3^0 = \askImath \Big(c_H \chi^{0\askImath} + s_H \phi^{0\askImath}\Big),\nonumber\\[2mm]
		H_1^0 &=& \phi^{0r},\; H_1^{0\prime} = \frac{1}{\sqrt{3}} \Big(\sqrt{2}\chi^{0r} + \xi^0\Big)\,,
	\end{eqnarray}
with $H_5^{--} = (H_5^{++})^\ast$, $H_5^- = -(H_5^+)^\ast$, $H_3^- = -(H_3^+)^\ast$, and $H_3^0 = -(H_3^0)^\ast$. The oblique parameters, the Feynman rules and the loop diagrams will be expressed in terms of these physical scalar five-plet, triplet, two scalars and their masses, $m_{H_5^{\pm\pm,\pm,0}}$, $m_{H_3^{\pm,0}}$, $m_{H_1}$, $m_{H_1^\prime}$ respectively. We will also use
	\begin{equation}
		s_H = \sin\theta_H = \frac{2 \sqrt{2}\; v_M}{v},\hspace{2em} c_H = \cos\theta_H = \frac{v_2}{v}
	\end{equation}
\par The effects of vacuum polarization diagrams (oblique corrections) on the electroweak-interaction observables can be described by three finite parameters $S$, $T$ and $U$, known as the \textit{Oblique Parameters}. Using these parameters one could probe the effects of new Physics on the electroweak interactions at the one-loop level, if the new Physics scale is much larger as compared to $M_Z$ \cite{PeskTak, pdg, Polon}. Hence, these parameters can be defined using perturbative expansion as \cite{PeskTak}:
  \begin{eqnarray}\label{eq:stu_pesk}
  	\alpha S &\equiv& 4 e^2 [ \Pi_{33}^\prime (0) - \Pi_{3Q}^\prime (0) ], \nonumber\\[2mm]
  	\alpha T &\equiv& \dfrac{e^2}{s_W^2 c_W^2 M_Z^2} [ \Pi_{11}(0) - \Pi_{33}(0) ], \nonumber\\[2mm]
  	\alpha U &\equiv& 4 e^2 [ \Pi_{11}^\prime (0) - \Pi_{33}^\prime (0) ]\,,
  \end{eqnarray}
where $s_W= \sin \theta_W,\;c_W= \cos \theta_W$ are the functions of the weak-mixing angle $\theta_W$. $\Pi_{11}$ and $\Pi_{33}$ are the vacuum polarizations of the isospin currents and $\Pi_{3Q}$ the vacuum polarization of one isospin and one electromagnetic current. The $\Pi^{\prime}$ functions are defined as $\Pi^{\prime}(0) = \left(\Pi(q^2) - \Pi(0)\right)/q^2$ in general, and we will be using $q^2 = M_Z^2$.
\par These parameters can be expressed in terms of the self-energies of the $W$, $Z$ and $\gamma$ bosons and the $Z\gamma$ mixing \cite{PeskTak}. For the purpose of the \ewnur model the constraints on the new Physics in \ewnur model from $S$, $T$, $U$ can be obtained by subtracting the the Standard Model (SM) contributions to $S$, $T$, $U$ from the corresponding total contributions due to the \ewnur model. Hence, the new Physics contributions to the $S$, $T$, $U$ due to \ewnur model are denoted by $\widetilde{S}$, $\widetilde{T}$, $\widetilde{U}$ respectively (following notation used in \cite{bhattroy}) and they can be expressed as
\begin{widetext}
  \begin{flalign}\label{eq:s}
  	\dfrac{\widehat{\alpha}}{4 \widehat{s}_W^2 \widehat{c}_W^2} \widetilde{S} &= \dfrac{1}{M_Z^2} \Bigg[ \overline{\Pi}_{ZZ}(M_Z^2) - \left( \dfrac{\widehat{c}_W^2 - \widehat{s}_W^2}{\widehat{c}_W^2 \widehat{s}_W^2} \right) \overline{\Pi}_{Z\gamma}(M_Z^2) - \overline{\Pi}_{\gamma\gamma}(M_Z^2) \Bigg]^{\text{\ewnur}}&\nonumber \\[4mm]
	&- \dfrac{1}{M_Z^2} \Bigg[ \overline{\Pi}_{ZZ}(M_Z^2) - \left( \dfrac{\widehat{c}_W^2 - \widehat{s}_W^2}{\widehat{c}_W^2 \widehat{s}_W^2} \right) \overline{\Pi}_{Z\gamma}(M_Z^2) - \overline{\Pi}_{\gamma\gamma}(M_Z^2) \Bigg]^{SM}& \\[4mm]
	\label{eq:t}
   	\widehat{\alpha} \widetilde{T} &= \dfrac{1}{M_W^2}\Bigg[\Pi_{11} (0) - \Pi_{33} (0)\Bigg]^{\text{\ewnur}} - \dfrac{1}{M_W^2}\Bigg[\Pi_{11} (0) - \Pi_{33} (0)\Bigg]^{SM}&\\[4mm]
  	\dfrac{\widehat{\alpha}}{4 \widehat{s}_W^2} \widetilde{U} &= \Bigg[\dfrac{\overline{\Pi}_{WW} (M_W^2)}{M_W^2} - \widehat{c}_W^2 \dfrac{\overline{\Pi}_{ZZ}(M_Z^2)}{M_Z^2} - 2 \widehat{s}_W \widehat{c}_W \dfrac{\overline{\Pi}_{Z\gamma}(M_Z^2)}{M_Z^2} - \widehat{s}_W^2 \dfrac{\overline{\Pi}_{\gamma\gamma}(M_Z^2)}{M_Z^2}\Bigg]^{\text{\ewnur}}&\nonumber \\[4mm]
	\label{eq:u}
	&- \Bigg[\dfrac{\overline{\Pi}_{WW} (M_W^2)}{M_W^2} - \widehat{c}_W^2 \dfrac{\overline{\Pi}_{ZZ}(M_Z^2)}{M_Z^2} - 2 \widehat{s}_W \widehat{c}_W \dfrac{\overline{\Pi}_{Z\gamma}(M_Z^2)}{M_Z^2} - \widehat{s}_W^2 \dfrac{\overline{\Pi}_{\gamma\gamma}(M_Z^2)}{M_Z^2}\Bigg]^{SM}\, ,&
  \end{flalign}
\end{widetext}
where all quantities with a hat on top ( $\widehat{~}$ ) i.e. $\widehat{s}_W$, $\widehat{c}_W$, $\widehat{\alpha}\equiv\; \widehat{g}^2 \widehat{s}_W^2/(4 \pi)$ are defined in the $\overline{MS}$ scheme evaluated at $M_Z$ \cite{pdg}. Hereafter, in this article the hats on top of these and other quantities are omitted, but implied. The notation $\overline{\Pi}(q^2) = \Pi(q^2) - \Pi(0)$ \cite{PeskTak} and the superscript '\ewnur' denotes the contribution due to \ewnur model. We can see that $S$ is associated with the difference between the $Z$ self-energy at $q^2=M_Z^2$ and $q^2=0$. $T$ is proportional to the difference between $W$ and $Z$ self-energies at $q^2=0$. The new physics contribution to $U$ in the \ewnur model is small as compared to that to $S$ and $T$. Also, this contribution is constrained only by the mass, $M_W$, and the width, $\Gamma_W$, of the $W$ boson. Thus, we can project the $STU$ parameter space on the 2-D $ST$ parameter space in the $U = 0$ plane \cite{gfitter}. Hence, in this paper our emphasis will be on the constraints on the $S$ and $T$ parameters only. The steps in the derivations of the new Physics contributions to $S$ and $T$ are provided in the Appendix \ref{sec:appst}. The new Physics contributions to $S$, $T$ from the scalar sector in \ewnur model (denoted by $\widetilde{S}_{scalar}$, $\widetilde{T}_{scalar}$ respectively) and the contributions from the mirror fermion sector in \ewnur model (denoted by $\widetilde{S}_{fermion}$, $\widetilde{T}_{fermion}$ respectively) are calculated separately and then added to obtain the total new Physics contributions in \ewnur model, contributions $\widetilde{S}$, $\widetilde{T}$. Note that the scalar sector contributions and mirror fermion sector contributions in \ewnur model are separately finite. Thus,
	\begin{flalign}
		\label{eq:sss}\hspace{2em}\widetilde{S} &= \widetilde{S}_{scalar} + \widetilde{S}_{fermion}& \\
		\label{eq:ttt}\hspace{2em}\widetilde{T} &= \widetilde{T}_{scalar} + \widetilde{T}_{fermion}\,.&
	\end{flalign}
\par The new Physics contributions to $S$ and $T$ due to the scalar sector of the \ewnur model is given in Eq. (\ref{eq:expss}) and Eq. (\ref{eq:expts}) respectively. The corresponding new Physics contributions to $S$ and $T$ due to the lepton sector in \ewnur model are given in Eq. (\ref{eq:expsl}) and Eq.(\ref{eq:exptl}) respectively. Similarly, the new Physics contributions to $S$ and $T$ due to the quarks in \ewnur model are given in Eq. (\ref{eq:expsq}) and Eq. (\ref{eq:exptq}) respectively.

It should be noted that in this paper we assume that the mixings between different mirror-quark and mirror-lepton generations are negligible. Thus, the mass matrices for these fermion sectors are already diagonal. To compare the new Physics contributions from the \ewnur model with the experimental constraints (refer to the plots in section \ref{numerical}) we have considered wide ranges of the mirror fermions masses. Hence, even if small non-zero mixings between different mirror fermion generations are included, it will only move individual points in the available parameter space, but will not significantly affect the total available parameter space and will not influence the conclusions of this paper.
\section{Numerical results}
\label{numerical}

In this section we will study numerically the results presented in Section \ref{sec:obl} and is organized as follows.
First, we present unconstrained scatter plots for the S and T parameters coming from the mirror fermion sector and from the scalar sector. These scatter plots are given in the $\tilde{T}$-$\tilde{S}$ plane for the scalar and mirror fermion sectors separately. The main desire is to observe possible regions where the two sectors can cancel each other. Second, we generate the scatter plots for $\tilde{T}$ and $\tilde{S}$ for the scalar sector as a function of the mass splittings among the scalars. In particular, we will notice below there is a ``significant'' region in the parameter space where $\tilde{S}$ can be quite negative if the mass splitting between the doubly-charged scalar with the other ones is large. Third, we combine two sectors and plot the scatter points of the EW$\nu_R$ model in the $\tilde{T}$-$\tilde{S}$ plane endowed with the 1$\sigma$ and 2$\sigma$ ellipses coming from experiment. It is shown below that the model is well consistent with precision electroweak data. Fourth, as an example (and simply as an example), we fix the values of some of the scalar masses and present a 3-dimensional plot of $\tilde{S}_{scalar}$ versus the mass splittings among members of the quintet and among members of the triplet.

\subsection{Unconstrained S and T parameters for the mirror fermion and scalar sectors}
\label{unconstrained}

The S and T parameters as shown in Section \ref{sec:obl} depend on a number of parameters such as the masses of the scalars as well as the mixing parameter $\sin \theta_H$ as defined in Section \ref{sec:obl}, and the masses of the new fermions from the model. For simplicity, we allow for the scalar masses to go from $M_Z$ to 650 GeV and for $\sin \theta_H$ to go from 0.1 to 0.89 as discussed in \cite{pqaranda} (we stretch the lower value to 0.1 for numerical purpose). The right handed neutrino masses are taken from $M_Z/2$ to 300 GeV, while the mirror charged lepton and the mirror quark masses vary from $M_Z$ to 600 GeV. The electroweak precision constraints which we will be using are given as $\tilde{S} = -0.02 \pm 0.14$; $\tilde{T} = 0.06 \pm 0.14$ \cite{pdg} for SM Higgs mass of 126 GeV. When about 10,000 different combinations of masses and mixings angles within these ranges are generated, and when the electroweak precision constraints are imposed the ranges of the scalar and the mirror fermion contributions to oblique parameters are seen to be:

\begin{itemize}

\item $\tilde{S}_{scalar}$ or $\tilde{S}_S$:
$\: -0.5\; \leq\; \tilde{S}_S\; \leq\; 0.5$

\item $\tilde{T}_{scalar}$ or $\tilde{T}_S$:
$\: -5\; \leq\; \tilde{T}_S\; \leq\; 22$

\item $\tilde{S}_{fermion}$ or $\tilde{S}_{MF}$:
$\: -0.1\; \leq\; \tilde{S}_{MF}\; \leq\; 1$

\item $\tilde{T}_{fermion}$ or $\tilde{T}_{MF}$:
$\: 0\; \leq\; \tilde{T}_{MF}\; \leq\; 32$

\end{itemize}

Before showing the combined scalar and mirror fermion contributions to $\tilde{T}$ and $\tilde{S}$, a few remarks are in order at this point. Let us look at the $\tilde{S}$ parameter. From the ranges given above one can see that the contribution to $\tilde{S}$ from the mirror fermion sector is almost always positive and can be quite large. This is to be expected since the addition of extra chiral doublets (the mirror fermions) always leads to such a phenomenon- a well-known fact. Cancellations from other contributions with the opposite sign will be needed in order to agree with the electroweak precision constraints. The range of $\tilde{S}_S$ shows that the contribution to $\tilde{S}$ coming from the scalar sector, in particular the Higgs triplet sectors, can be quite negative allowing for such cancellation to occur. This has been anticipated in \cite{pqnur} but this is the first detailed calculation of such a contribution to the electroweak precision parameters.

Now we show the comparison between the scalar and the mirror fermion contributions to the oblique parameters with the experimental constraints on the total $\tilde{S}$ and $\tilde{T}$. We will present the plots as follows:

\begin{itemize}

\item Scatter plot of $\tilde{T}$ versus $\tilde{S}$ for the scalar sector {\em with} the 1 and 2 $\sigma$ experimental contours (FIG.~\ref{TSscalar2}); 

\begin{figure}[H]
\centering
    \includegraphics[scale=0.35]{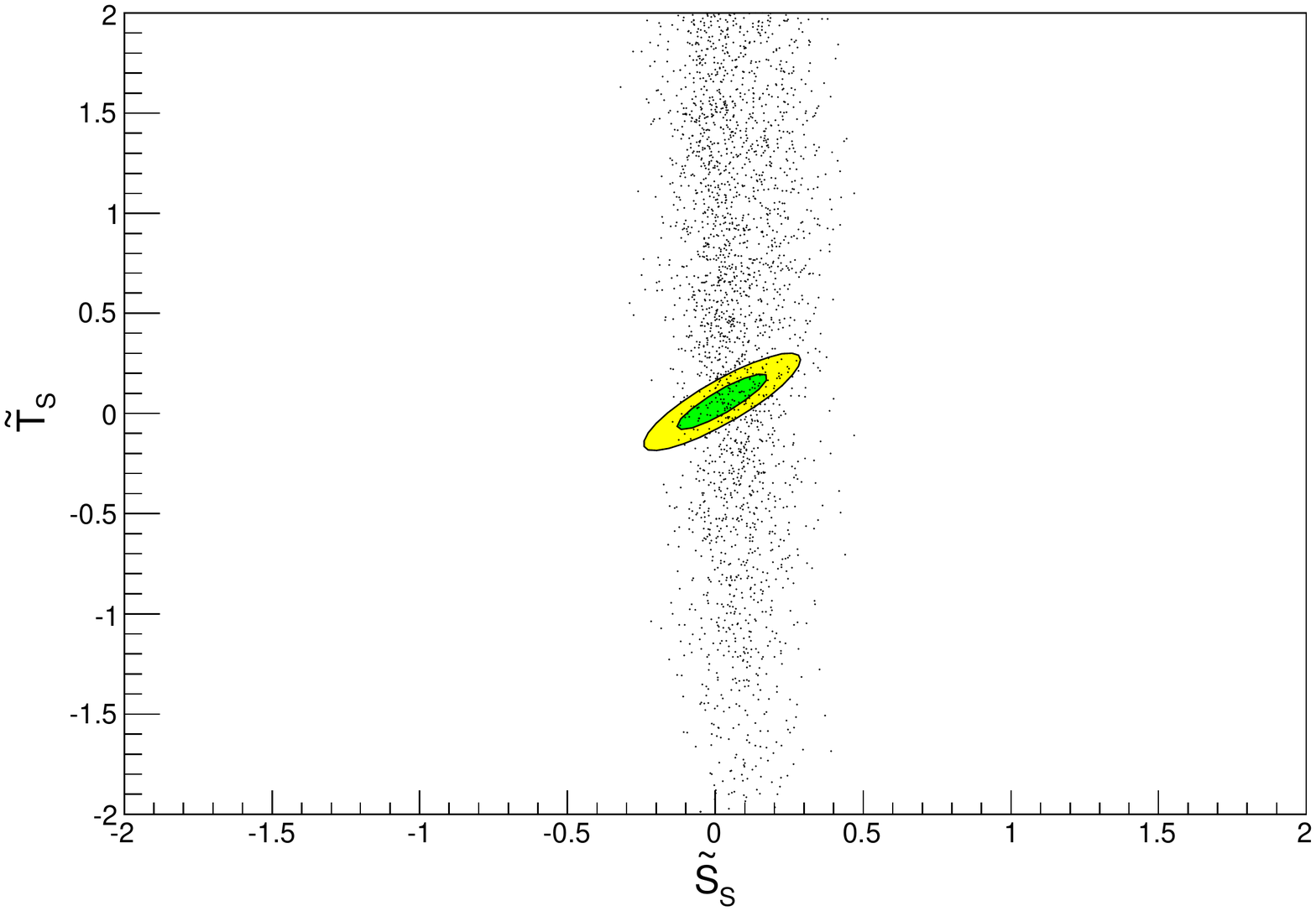} 
 \caption{\label{TSscalar2}{\small $\tilde{T}$ versus $\tilde{S}$ for the scalar sector with the 1 and 2 $\sigma$ experimental contours (about 500 points)}}
\end{figure}

\item Scatter plot of $\tilde{T}$ versus $\tilde{S}$ for the mirror fermion sector {\em with} the 1 and 2 $\sigma$ experimental contours (FIG.~\ref{TSMF2}). 

\begin{figure}[H]
\centering
    \includegraphics[scale=0.35]{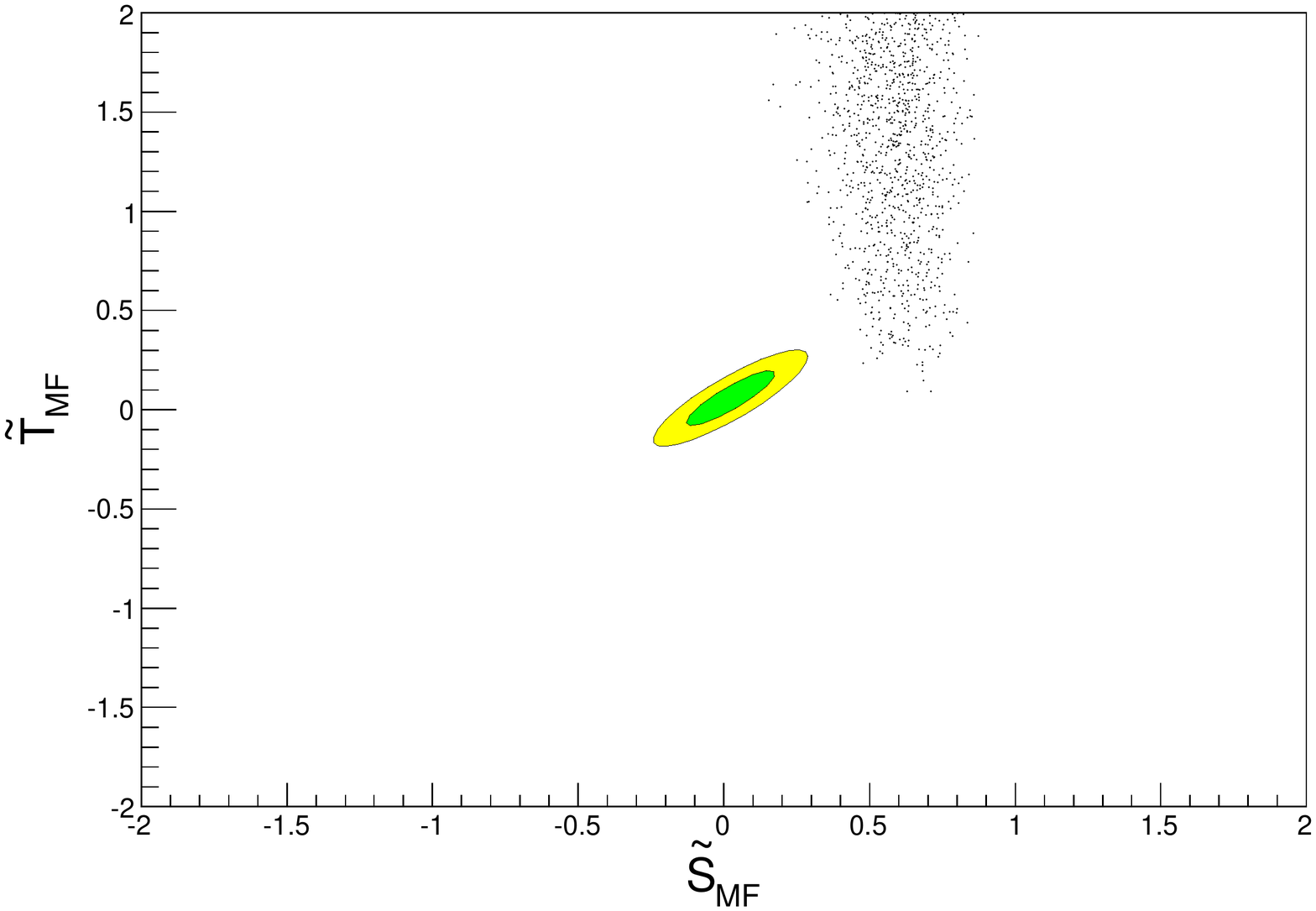} 
 \caption{\label{TSMF2}{\small $\tilde{T}$ versus $\tilde{S}$ for the mirror fermion sector with the 1 and 2 $\sigma$ experimental contours (about 500 points)}}
\end{figure}
\end{itemize}

To see a little more explicitly why the two sectors complement each other in such a way as to bring the EW$\nu_R$ model to be in agreement with the electroweak precision data, let us take a look at FIG.~\ref{TSscalar2} and FIG.~\ref{TSMF2}. From FIG.~\ref{TSscalar2}, we can see that the 1$\sigma$ and 2$\sigma$ experimental contours are well inside the region generated by the scalars of the EW$\nu_R$ model. However, these contours are way outside the region generated by the mirror fermions of the model. Again, one notices the importance of the scalar sector in bringing the EW$\nu_R$  into agreement with the electroweak precision data. 
\subsection{Constrained S and T parameters}
\label{constrained}

\begin{figure}[H]
\centering
    \includegraphics[scale=0.35]{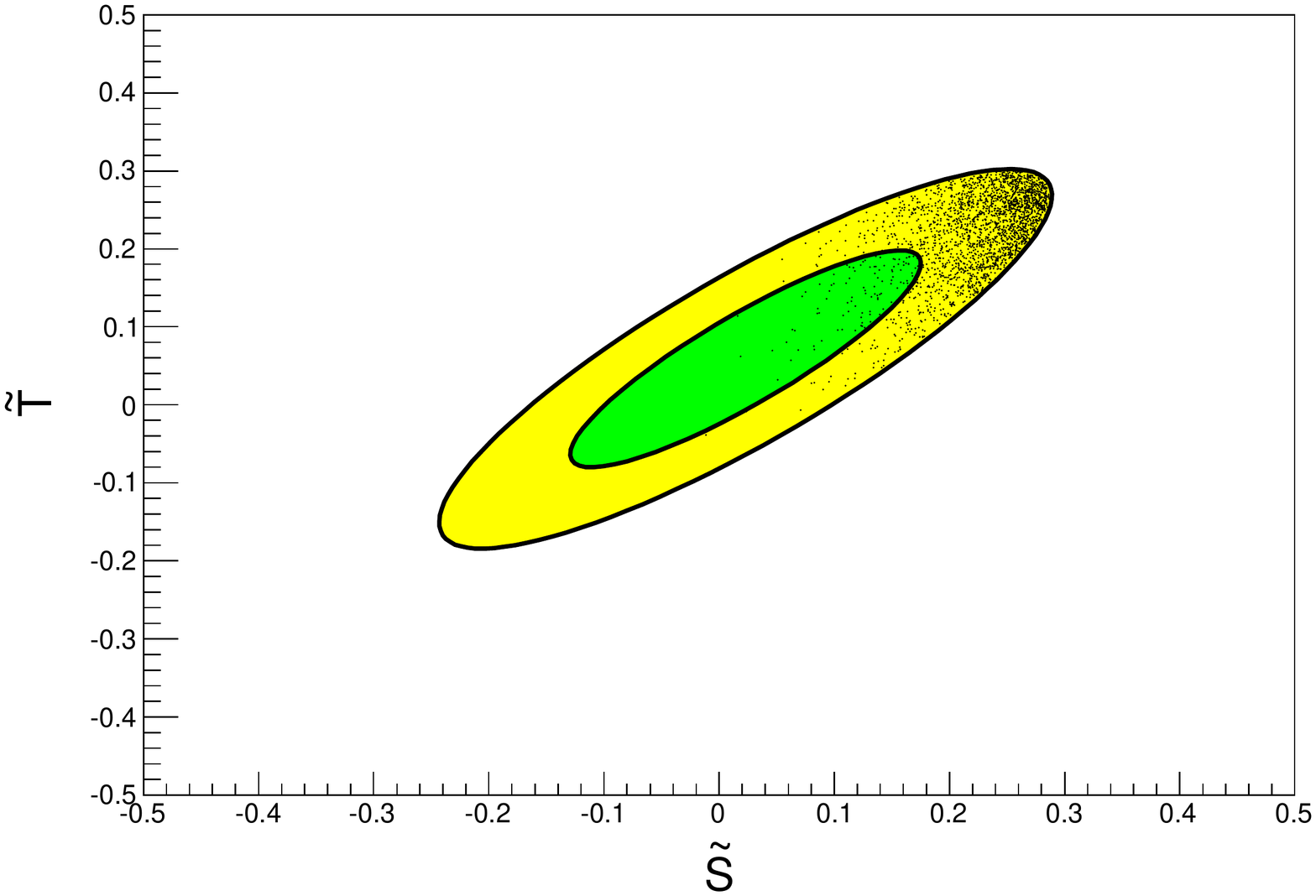} 
 \caption{\label{TStotal}{\small Total $\tilde{T}$ versus $\tilde{S}$ with the 1 and 2 $\sigma$ experimental contours}}
\end{figure}

\begin{figure}[H]
\centering
    \includegraphics[scale=0.35]{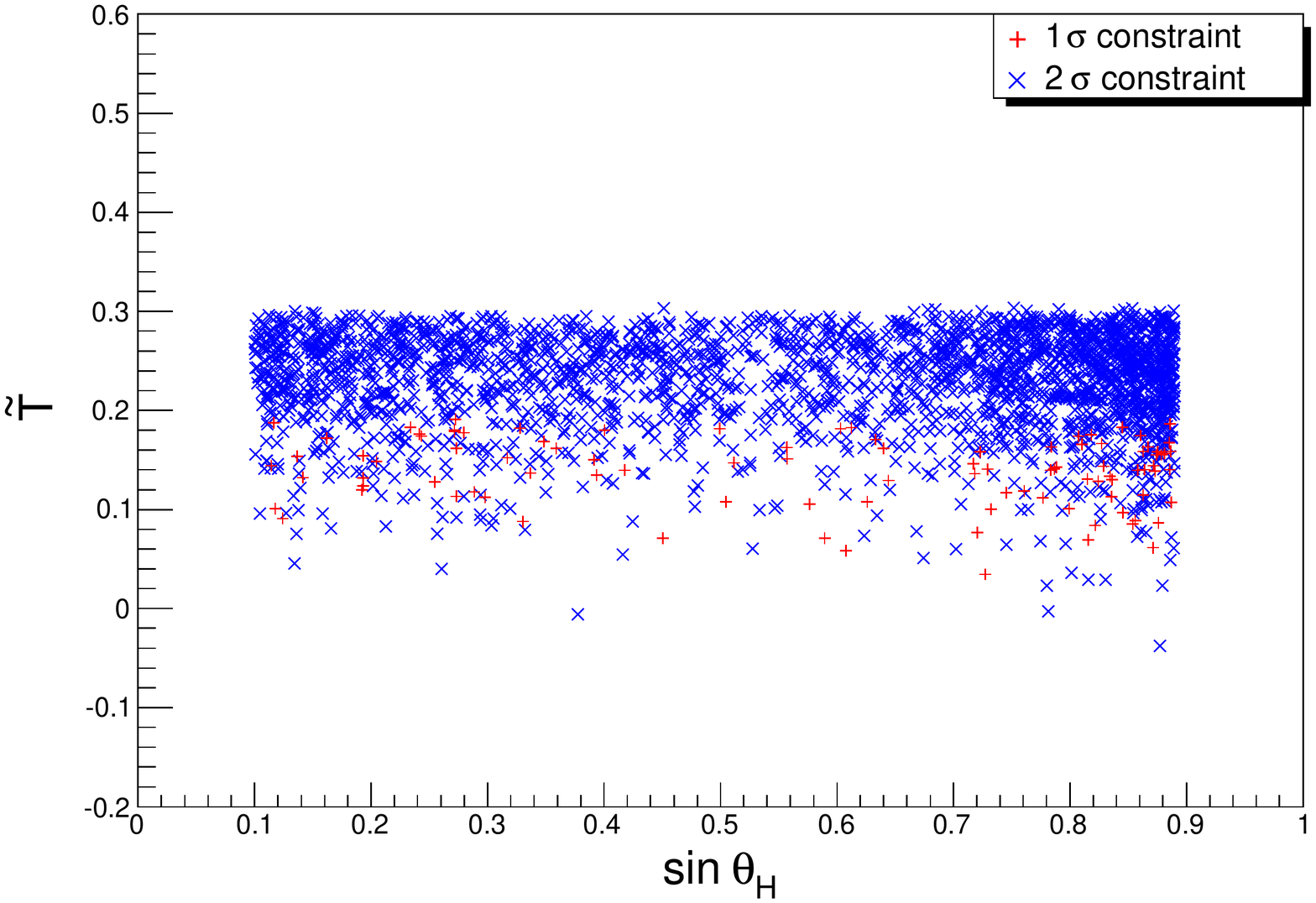} 
 \caption{\label{T_sh}{\small $\tilde{T}$ versus $\sin \theta_H$ for $\tilde{S}$ and $\tilde{T}$ satisfying $1\sigma$ and $2\sigma$ constraints of FIG.~\ref{TStotal} 
 }}
\end{figure}

To compare the model with data, we, of course, consider the total sum of the two contributions, namely $\tilde{S}=\tilde{S}_S + \tilde{S}_{MF}$ and $\tilde{T}=\tilde{T}_S + \tilde{T}_{MF}$. This is shown below in a plot which also includes the 1 and 2 $\sigma$ experimental contours. (The scatter plots in FIG.~\ref{TStotal} - FIG.~\ref{T_sh} show about 3,500 points each).

One can now see from FIG.~\ref{TStotal} that the sum of the scalar and mirror fermion contributions to  $\tilde{T}$ and $\tilde{S}$ generates ``data points'' {\em inside} the 1 and 2 $\sigma$ experimental contours. It implies that there is a non-negligible region of parameter space where the EW$\nu_R$ model is consistent with electroweak precision constraints. One notices again the crucial role played by the triplet scalars.

To understand better the ``data points'' in FIG.~\ref{TStotal}, we plot the constrained $\tilde{S}_S$ versus $\tilde{S}_{MF}$ as shown below.

\begin{figure}[H]
\centering
    \includegraphics[scale=0.35]{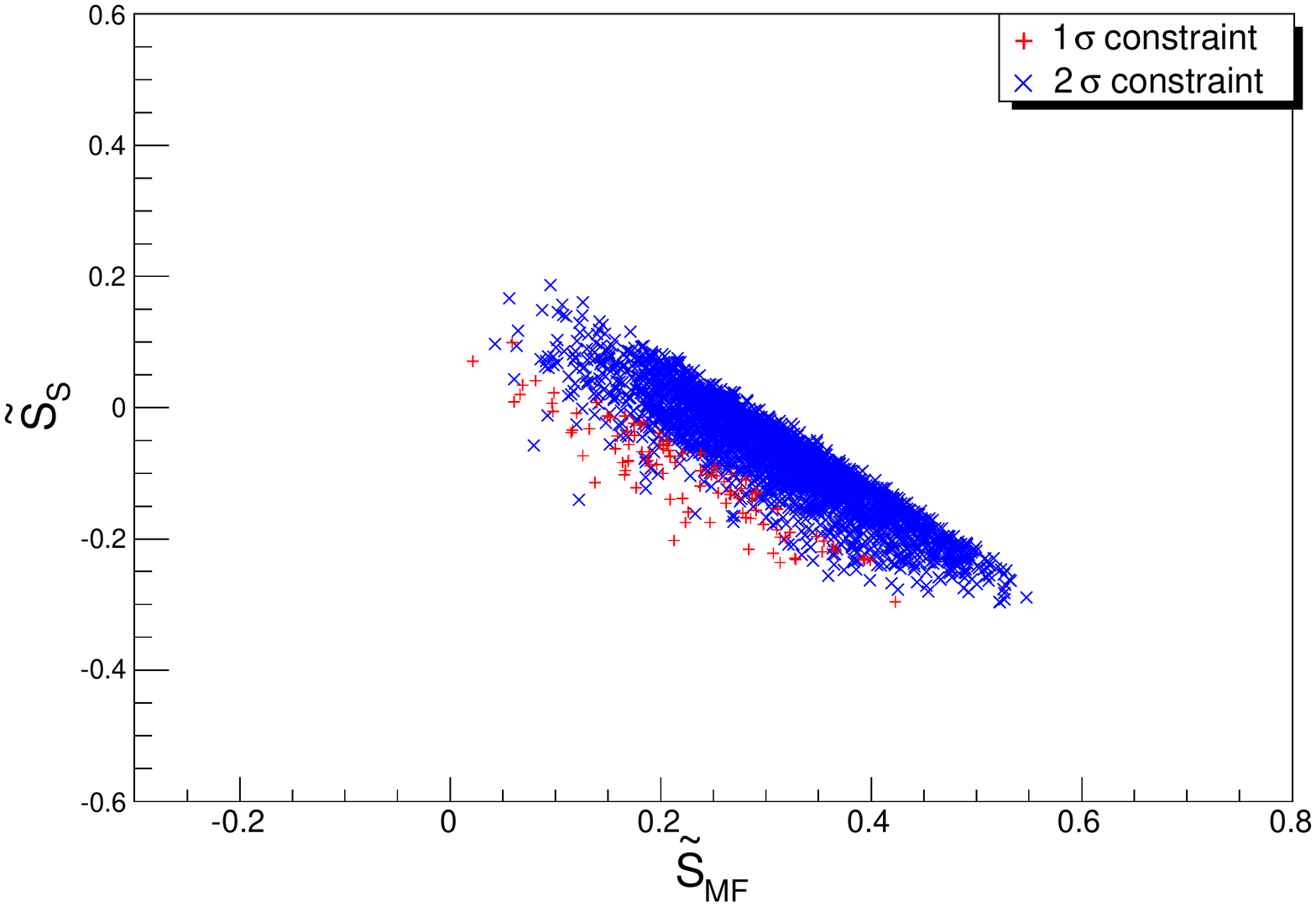} 
 \caption{\label{SvsS}{\small $\tilde{S}$ versus $\tilde{S}_{MF}$ for $\tilde{S}$ and $\tilde{T}$ satisfying $1\sigma$ and $2\sigma$ constraints of FIG.~\ref{TStotal}
 }}
\end{figure}

In the plot below, we also show the constrained $\tilde{T}_S$ versus $\tilde{T}_{MF}$.
\begin{figure}[H]
\centering
    \includegraphics[scale=0.35]{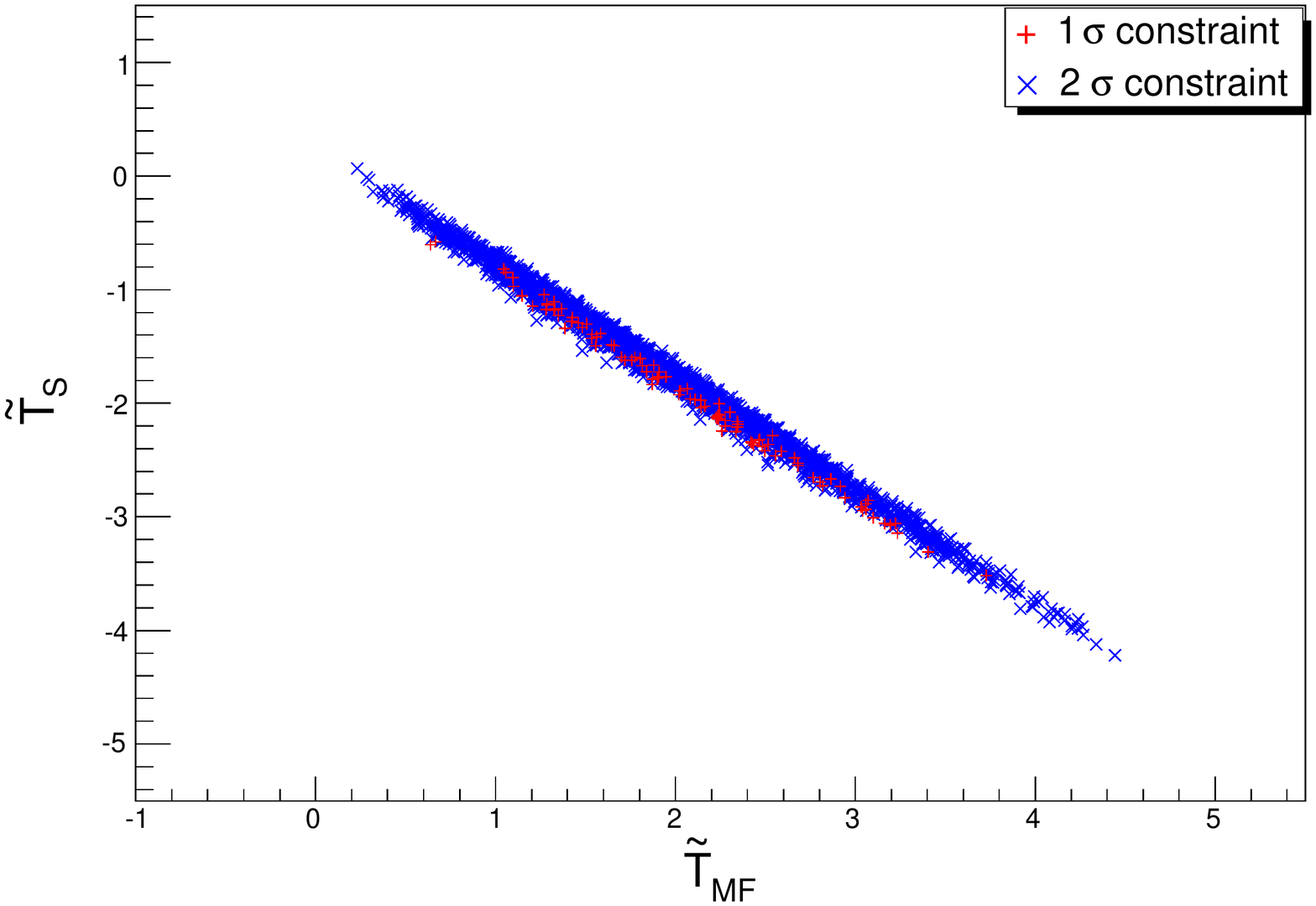} 
 \caption{\label{TvsT}{\small $\tilde{T}$ versus $\tilde{T}_{MF}$ for $\tilde{S}$ and $\tilde{T}$ satisfying $1\sigma$ and $2\sigma$ constraints of FIG.~\ref{TStotal}
 }}
\end{figure}

From  FIG.~\ref{SvsS} and  FIG.~\ref{TvsT}, one can clearly see the cancellation among the scalar and mirror fermion sectors in their contributions to $\tilde{T}$ and $\tilde{S}$. At the 1 $\sigma$ level, one can see from FIG.~\ref{SvsS} that the scalar contribution, $\tilde{S}_S$ , ranges roughly from 0 to -0.3. 

The last constrained plot we would like to show in this section is the variation of $\tilde{S}$ and $\tilde{T}$ as a function of $\sin \theta_H$.
\begin{figure}[H]
\centering
    \includegraphics[scale=0.35]{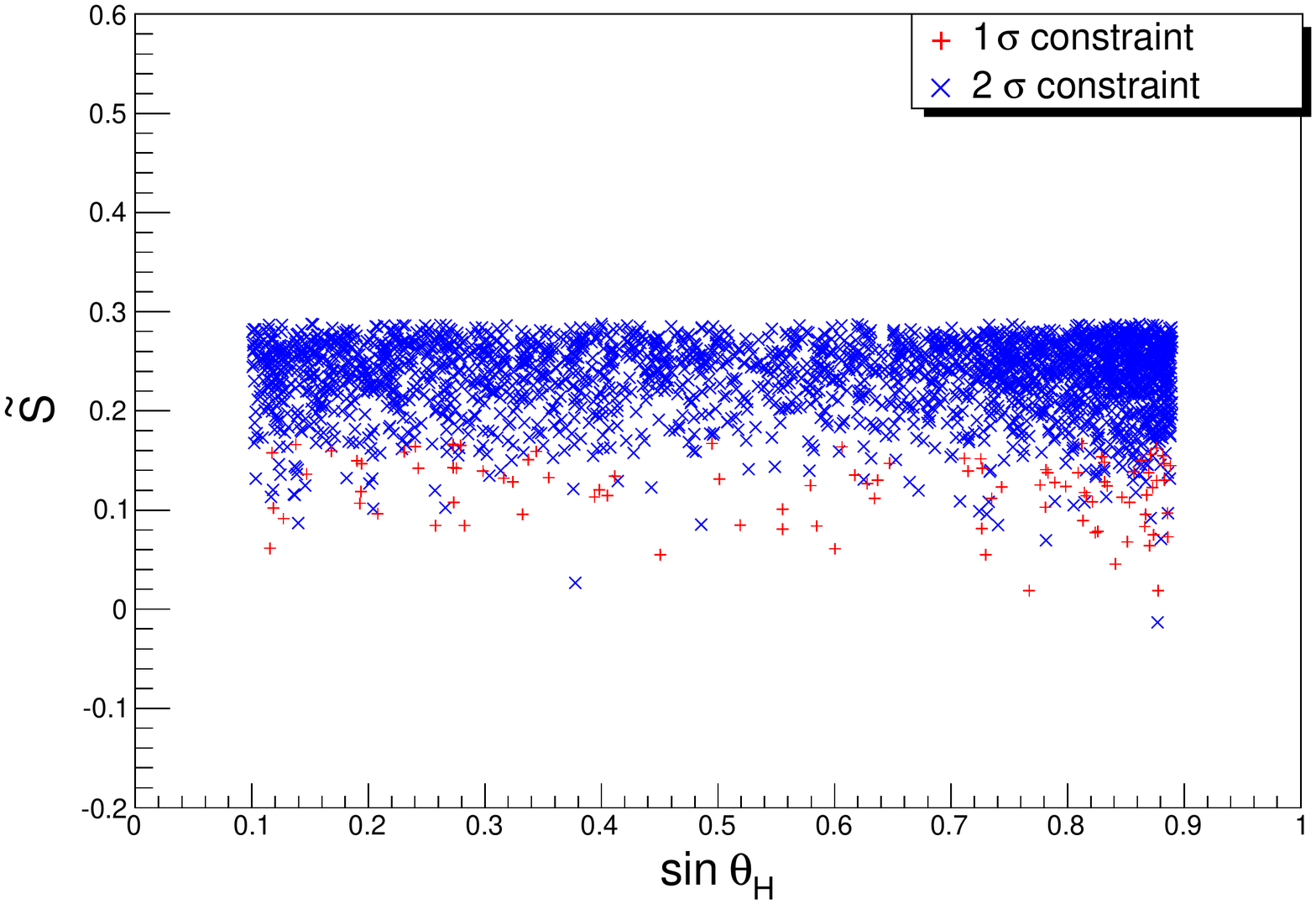} 
 \caption{\label{S_sh}{\small $\tilde{S}$ versus $\sin \theta_H$ for $\tilde{S}$ and $\tilde{T}$ satisfying $1\sigma$ and $2\sigma$ constraints of FIG.~\ref{TStotal}
 }}
\end{figure}

These plots should be considered, because in the \ewnur model various couplings and, hence, the branching ratios and the cross sections (e.g. Eq. (\ref{xsection})) depend on $\theta_H$ \cite{pqaranda,nurd}. Thus, it is necessary to take into account any constraints on $\sin \theta_H$ from the oblique parameters. Any restriction on the allowed range of $\sin \theta_H$, from the constraints on $\tilde{S}$, $\tilde{T}$, would surely affect the agreement of the model with the experimental data and the searches for experimental signals of this model. From FIG.~\ref{S_sh} and FIG.~\ref{T_sh}, we notice that the EW$\nu_R$ model agrees with electroweak precision data for the entire allowed range of $\sin \theta_H$. 

The next question that one might wish to ask is how the above informations influence the masses and mass splittings in the scalar sector.
In the next section, we will show some samples of three-dimensional plots of $\tilde{S}_S$ and $\tilde{T}_S$ versus the mass splittings in  the scalar quintet and triplet (i.e. $H_5$ and $H_3$). Some specific mass values are used in these plots for the purpose of illustration. An exhaustive study of a large range of masses is beyond the scope of this paper.

\subsection{$\tilde{S}_S$ and $\tilde{T}_S$ versus scalar mass splittings}
The experimental searches for the scalars presented in \cite{pqnur,pqaranda} will be guided partly by the mass splittings  among the scalars. These, in turn, are dictated by the electroweak precision constraints discussed above. In fact, the amount of mass splittings is constrained by e.g. the allowed ranges of $\tilde{S}_S$, $\tilde{T}_S$ which, at 1 $\sigma$ level, range approximately from 0 to -0.3 and from 0 to $-4$, respectively. 

A few remarks about the custodial symmetry and the mass splittings within a scalar or a fermion multiplet are in order here. The custodial symmetry is a global symmetry that is preserved whenever $M_Z \cos \theta_W = M_W$ is satisfied. In \ewnur model when the global $SU(2)_D$ symmetry is preserved after the electroweak symmetry breaking, all the members in a scalar or fermion multiplet (multiplet under the global $SU(2)_D$) are degenerate. This symmetry also makes sure that $M_Z \cos \theta_W = M_W$ is satisfied and hence it is the $SU(2)_D$ {\em custodial} symmetry. When the masses of scalars or fermions within a multiplet are non-degenerate, the global $SU(2)_D$ symmetry is {\em explicitly} broken. As a result, $M_W$ deviates from $M_Z \cos \theta_W$, but this deviation enters only at the loop level, when the loop-corrections to the self-energy diagrams of $W$ and $Z$ (with the {\em non-degenerate} members of a scalar or a fermion multiplet in the loop) are considered. This deviation from the 'custodiality' is restricted {\em only} by the experimental constraints on the $\widetilde{S}$ and $\widetilde{T}$ given above. Hence, the scalar- or fermion- mass splittings can be large as long as the total $\widetilde{S}$, $\widetilde{T}$ satisfy the experimental constraints. This is also observed within the Standard Model. In SM the custodial symmetry is explicitly broken, because e.g. the top- and the bottom- quark are non-degenerate with a large mass splitting ($|\Delta m_{tb}| \sim 40\, m_b$). However this results in only a small deviation from the custodial symmetry, but this deviation is seen only, when contributions of the top- and the bottom- quark loops to the self energy of the $W$ and $Z$ are considered. The effect of mass splittings within scalar and fermion multiplets on the $S$ parameter was discussed in \cite{RandallDugan}. And for the $T$ parameter, it can be realized in the same manner.

We will present a few of these plots and will comment on their implications.

\bi
\item {\bf Plots}:
The 3-D plots shown below will be for two definite values of $\sin \theta_H$, namely 0.1 and 0.89, in order to illustrate also the dependance of $\tilde{S}_S$ and $\tilde{T}_S$ on $\sin \theta_H$. As mentioned above, an exhaustive analysis for arbitrary scalar masses is beyond the scope of this paper. As a consequence, we will fix the values of $m_{H_1}$, $m_{H^{'}_1}$, $m_{H^{0}_3}$, $m_{H^{0}_5}$ and $m_{H^{++}_5}$ and will vary $m_{H^{+}_3}$ and $m_{H^{+}_5}$ (two of the three axes in the plots).

\begin{figure}[H]
\centering
    \includegraphics[scale=0.35]{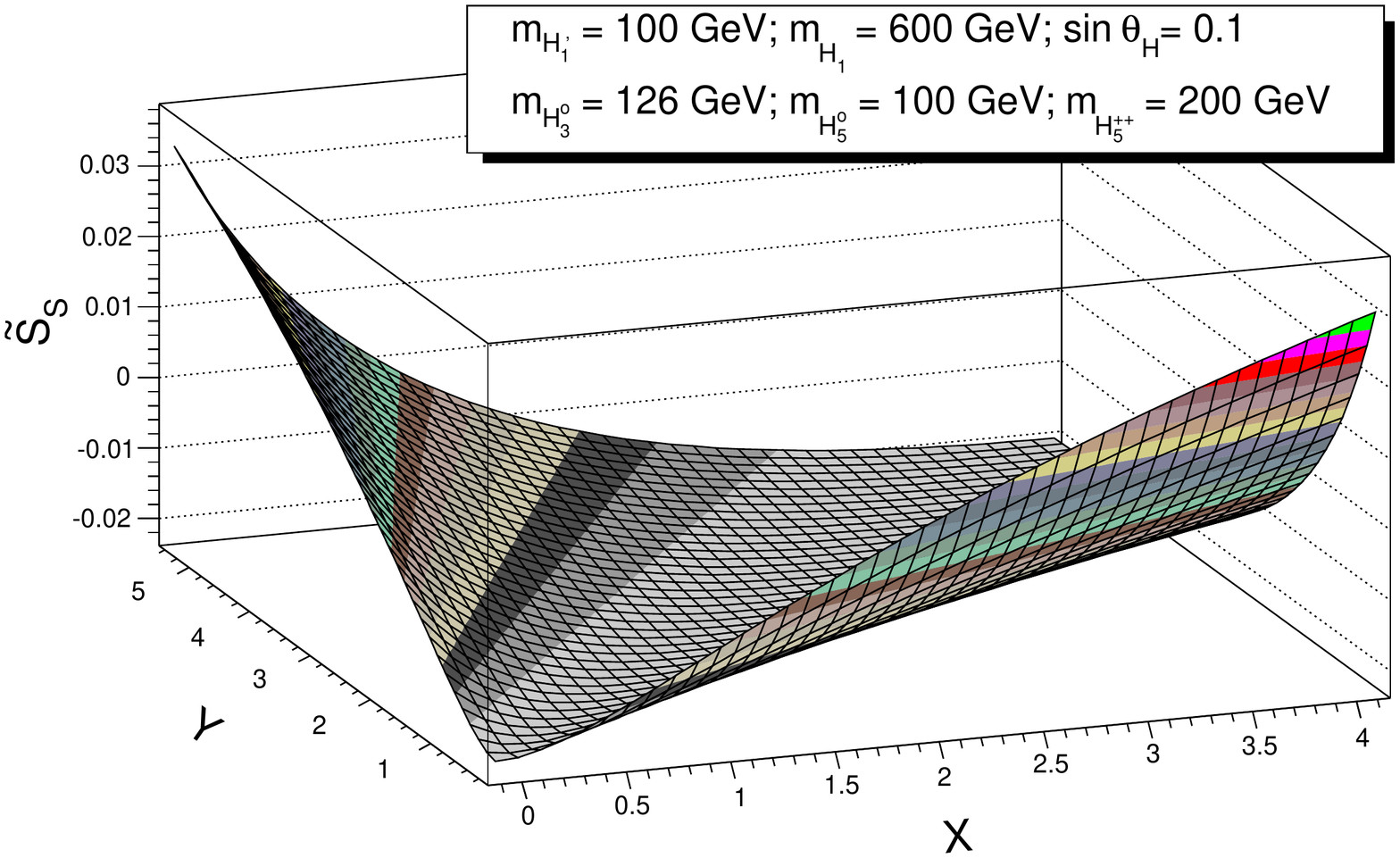} 
 \caption{\label{3DS_S1}{\small $\tilde{S}_S$  versus $Y \equiv \frac{m_{H^{+}_5}}{m_{H^{0}_5}}-1$  and $X \equiv \frac{m_{H^{+}_3}}{m_{H^{0}_3}}-1$  for $\sin \theta_H=0.1$ and $m_{H_1}=600\, GeV$}}
\end{figure}

\begin{figure}[H]
\centering
    \includegraphics[scale=0.35]{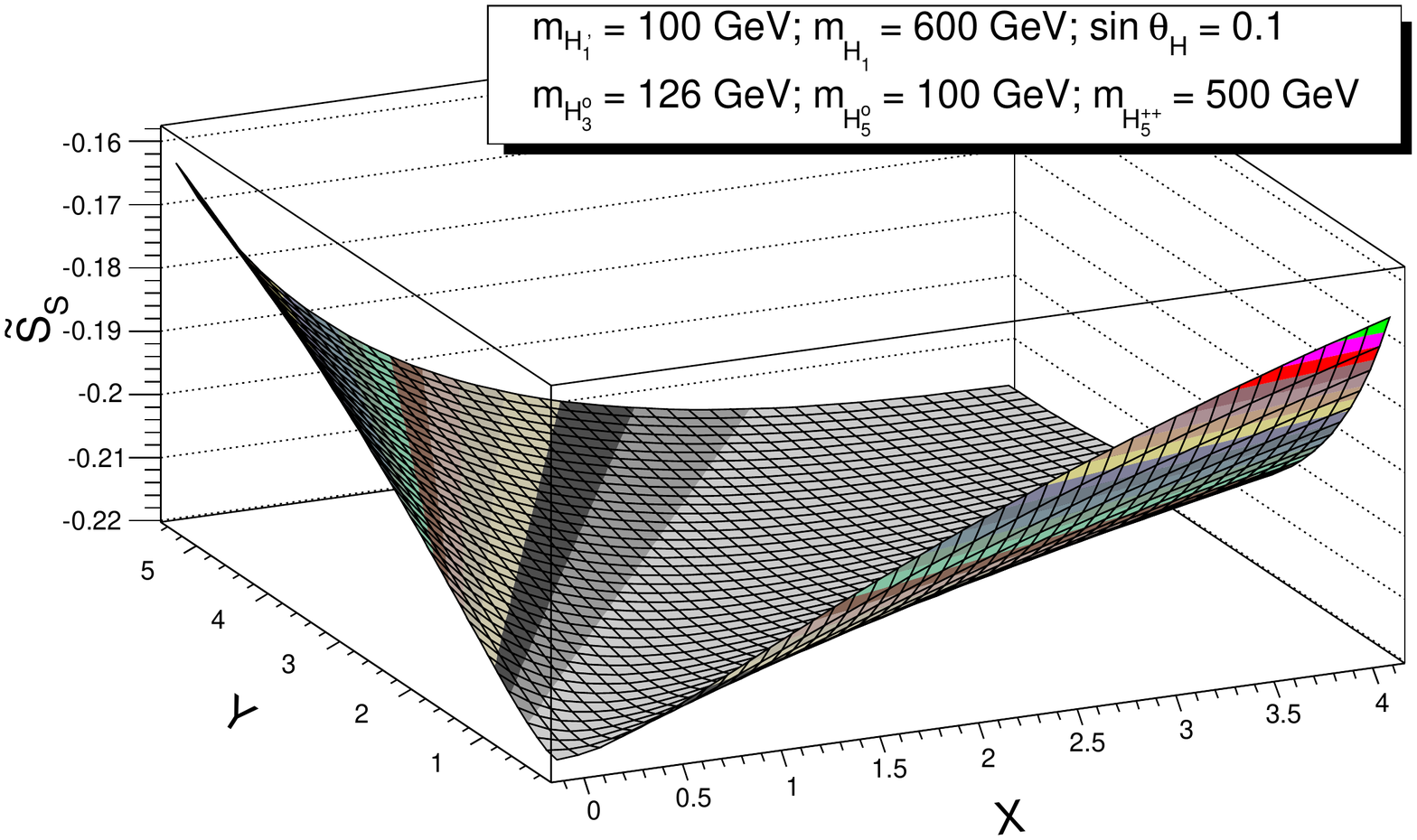} 
 \caption{\label{3DS_S2}{\small $\tilde{S}_S$  versus $Y \equiv \frac{m_{H^{+}_5}}{m_{H^{0}_5}}-1$  and $X \equiv \frac{m_{H^{+}_3}}{m_{H^{0}_3}}-1$  for $\sin \theta_H=0.1$ and $m_{H_1}=600\, GeV$}}
\end{figure}

\begin{figure}[H]
\centering
    \includegraphics[scale=0.35]{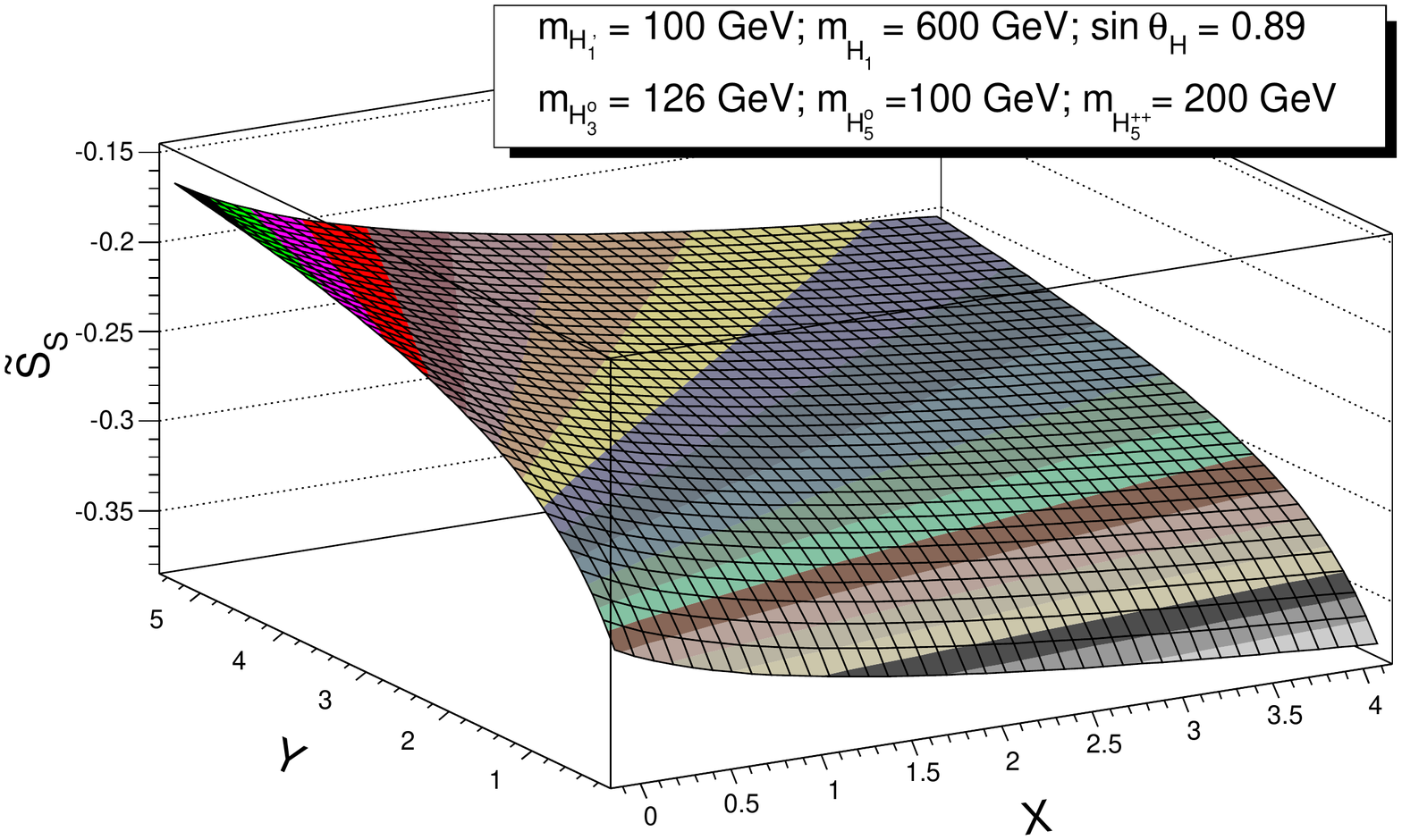} 
 \caption{\label{3DS_S3}{\small $\tilde{S}_S$  versus $Y \equiv \frac{m_{H^{+}_5}}{m_{H^{0}_5}}-1$  and $X \equiv \frac{m_{H^{+}_3}}{m_{H^{0}_3}}-1$  for $\sin \theta_H=0.89$ and $m_{H_1}=600\, GeV$}}
\end{figure}

\begin{figure}[H]
\centering
    \includegraphics[scale=0.35]{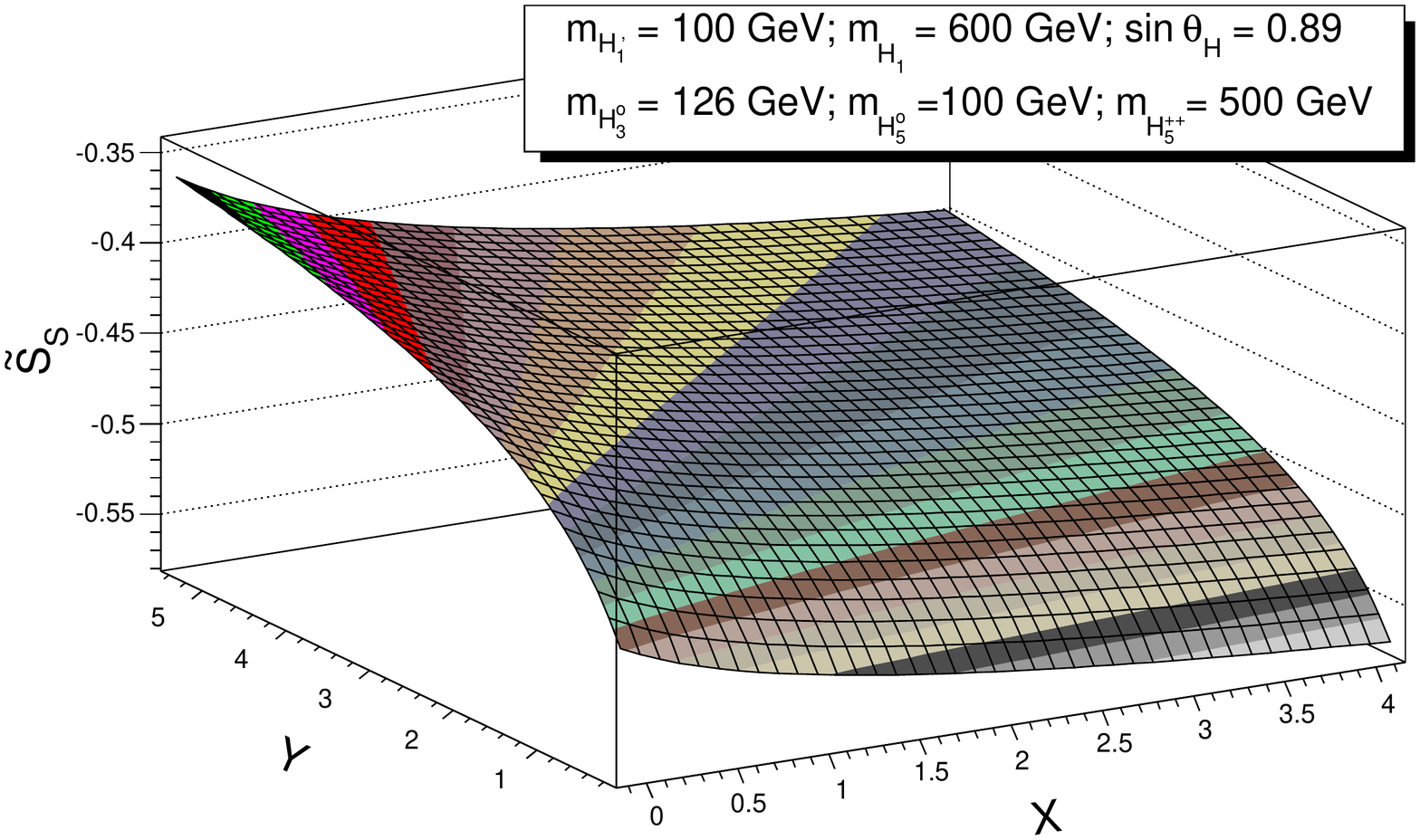} 
 \caption{\label{3DS_S4}{\small $\tilde{S}_S$  versus $Y \equiv \frac{m_{H^{+}_5}}{m_{H^{0}_5}}-1$  and $X \equiv \frac{m_{H^{+}_3}}{m_{H^{0}_3}}-1$  for $\sin \theta_H=0.89$ and $m_{H_1}=600\, GeV$}}
\end{figure}
\begin{figure}[H]
\centering
    \includegraphics[scale=0.35]{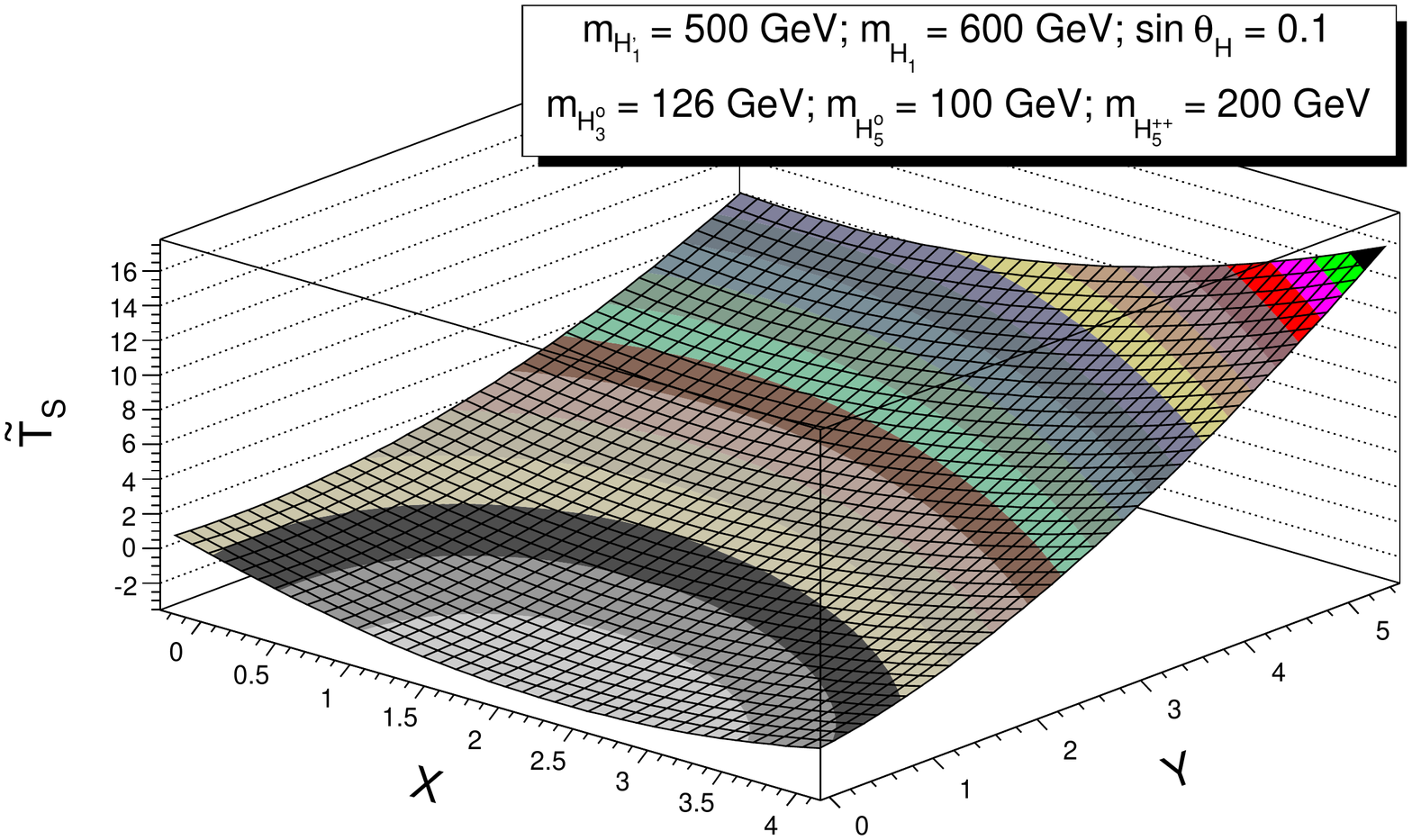} 
 \caption{\label{3DS_S5}{\small $\tilde{T}_S$  versus $Y \equiv \frac{m_{H^{+}_5}}{m_{H^{0}_5}}-1$  and $X \equiv \frac{m_{H^{+}_3}}{m_{H^{0}_3}}-1$  for $\sin \theta_H=0.1$ and $m_{H_1}=600\, GeV$}}
\end{figure}

\begin{figure}[H]
\centering
    \includegraphics[scale=0.35]{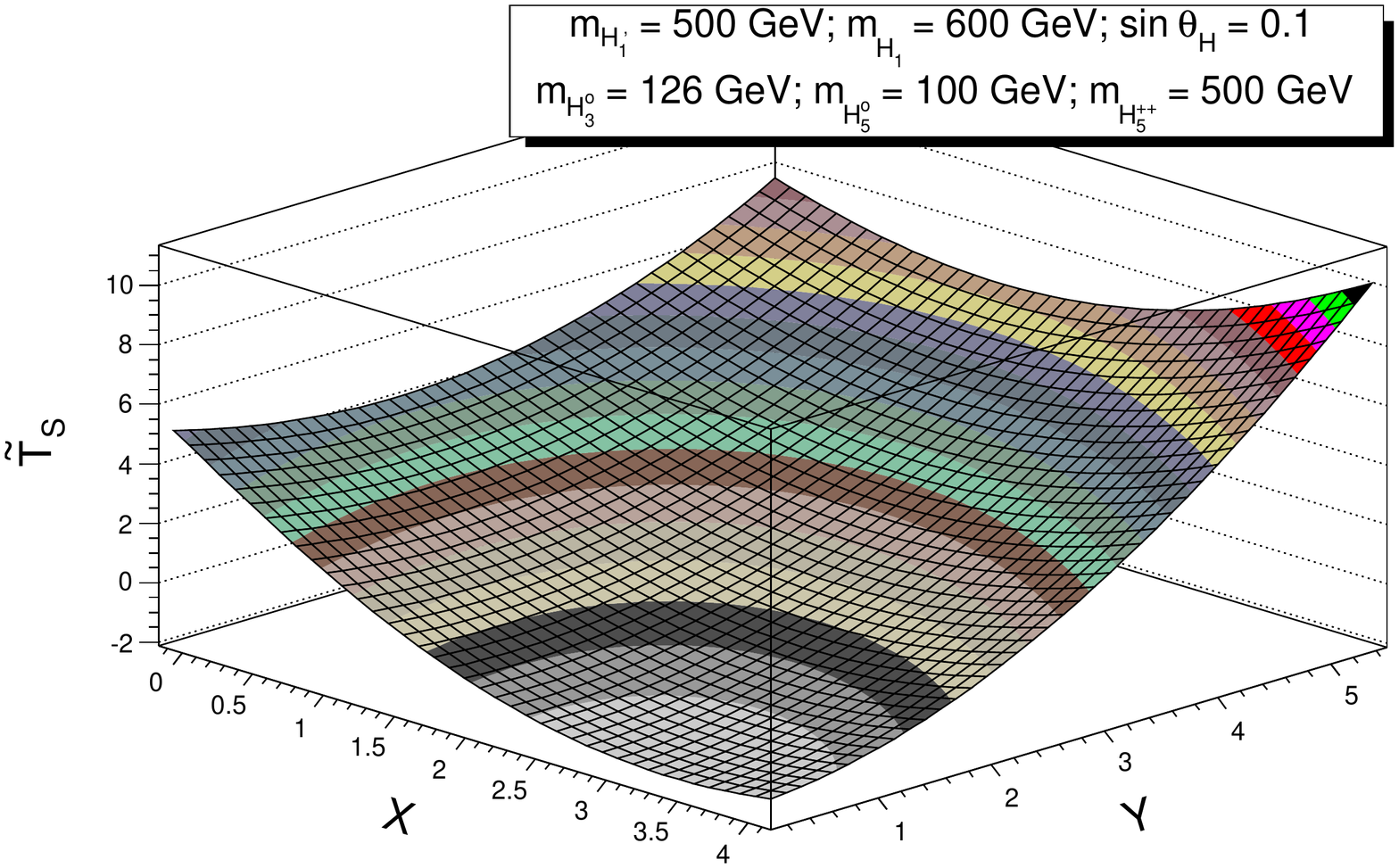} 
 \caption{\label{3DS_S6}{\small $\tilde{T}_S$  versus $Y \equiv \frac{m_{H^{+}_5}}{m_{H^{0}_5}}-1$  and $X \equiv \frac{m_{H^{+}_3}}{m_{H^{0}_3}}-1$  for $\sin \theta_H=0.1$ and $m_{H_1}=600\, GeV$}}
\end{figure}

\begin{figure}[H]
\centering
    \includegraphics[scale=0.35]{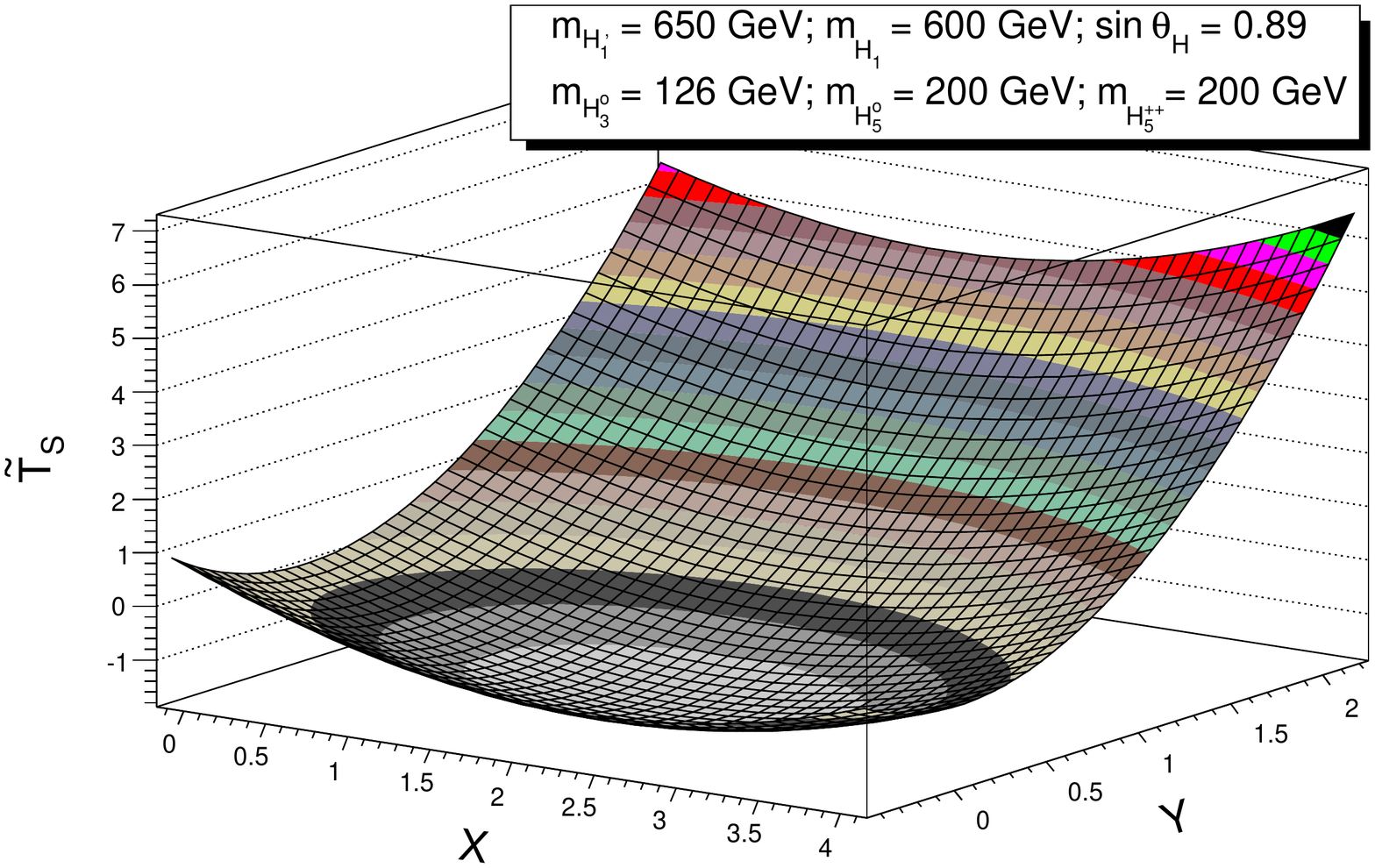} 
 \caption{\label{3DS_S7}{\small $\tilde{T}_S$  versus $Y \equiv \frac{m_{H^{+}_5}}{m_{H^{0}_5}}-1$  and $X \equiv \frac{m_{H^{+}_3}}{m_{H^{0}_3}}-1$  for $\sin \theta_H=0.89$ and $m_{H_1}=600\, GeV$}}
\end{figure}

\begin{figure}[H]
\centering
    \includegraphics[scale=0.35]{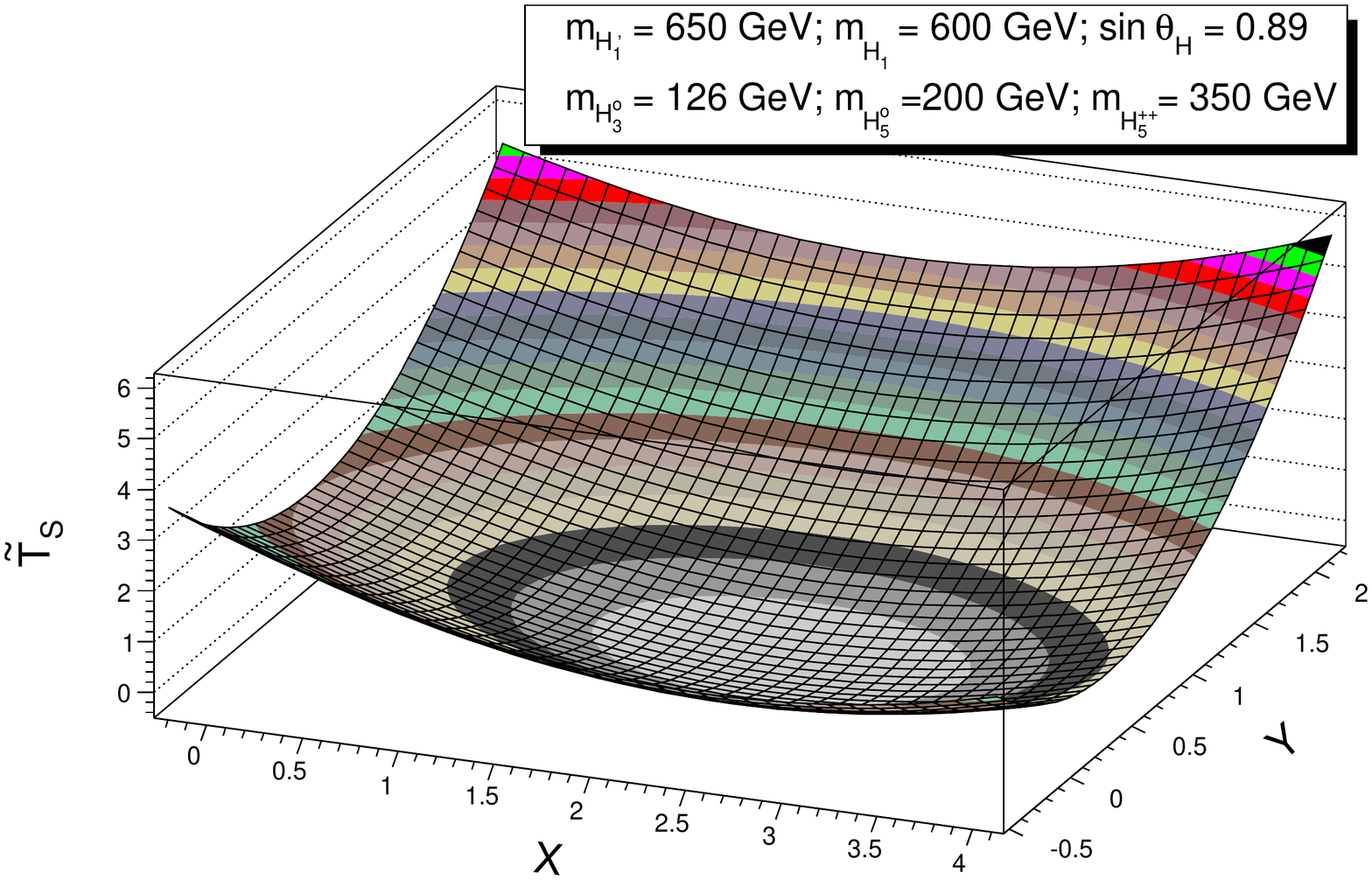} 
 \caption{\label{3DS_S8}{\small $\tilde{T}_S$  versus $Y \equiv \frac{m_{H^{+}_5}}{m_{H^{0}_5}}-1$  and $X \equiv \frac{m_{H^{+}_3}}{m_{H^{0}_3}}-1$  for $\sin \theta_H=0.89$ and $m_{H_1}=600\, GeV$}}
\end{figure}

\item {\bf Remarks }:
\bi
\item In the above figures, arbitrary values are chosen (for the purpose of illustration) for the two scalars $H_1$ and $H^{'}_1$ , namely $m_{H_1}=600\, GeV$ and $m_{H^{'}_1}=100, 500, 650\, GeV$. The reasons for these particular- albeit arbitrary- values will be given below. 

\item The value of 126 GeV was set for $H^{0}_3$ in the plots also for illustrative purpose. In the next section, we will discuss what the most recent LHC result on the spin-parity of the 126 GeV ``object'' implies on the minimal EW$\nu_R$ model and what extension is needed in the scalar sector. 

\item A look at FIG.~\ref{3DS_S1} - FIG.~\ref{3DS_S4} reveals the following pattern. $\tilde{S}_S$ becomes more negative as the mass of the doubly-charged scalar, $H^{++}_5$, goes from 200 GeV to 500 GeV. It also becomes more negative as one increases $\sin \theta_H$ from 0.1 to 0.89. 

\item For $\sin \theta_H=0.1$ (FIG.~\ref{3DS_S1} and FIG.~\ref{3DS_S2}) , we notice that $\tilde{S}_S$ {\em decreases}, becoming more negative, when mass splittings between $H^{+}_3$ and $H^{0}_3$ AND between $H^{+}_5$ and $H^{0}_5$ become similar. This feature persists until $\tilde{S}_S$ reaches its lowest value where it stays ``stable'' along the ``line'' $\frac{m_{H^{+}_5}}{m_{H^{0}_5}}-1 \sim \frac{m_{H^{+}_3}}{m_{H^{0}_3}}-1$. This implies that $\tilde{S}_S$ reaches its most negative value when the mass splittings among $H_5$'s and among $H_3$'s are {\em similar}. 

\item For $\sin \theta_H=0.89$, $\tilde{S}_S$ {\em decreases}, becoming more negative, when $H^{+}_5$ and $H^{0}_5$ become more and more {\em degenerate} {\em while} $H^{+}_3$ becomes heavier and heavier compared with $H^{0}_3$. This is in marked contrast with the $\sin \theta_H=0.1$ case.

\item If one restricts oneself to the 1 $\sigma$ constraint, namely $\tilde{S}_S \sim 0.0\; \text{to} \; -0.3$, then FIGs.~\ref{3DS_S1}, \ref{3DS_S2} and \ref{3DS_S3} seem to be favored. 

\item FIG.~\ref{3DS_S5} - FIG.~\ref{3DS_S8} reveal that the dependence of $\tilde{T}_S$ on the mass splittings is different from that of $\tilde{S}_S$.  

\item For $\sin \theta_H=0.1$ (FIG.~\ref{3DS_S5} and FIG.~\ref{3DS_S6}), with particular choice of $m_{H^\prime_1}=500\, GeV$, $\tilde{T}_S $ becomes more negative for large mass splitting within the triplet, and small mass splitting between $H_5^+$ and $H_5^0$. 

\item For $\sin \theta_H=0.89$ (FIG.~\ref{3DS_S7} and FIG.~\ref{3DS_S8}), we choose $m_{H^{'}_1}=650\, GeV$. The lowest value regions of  $\tilde{T}_S$ correspond to $1 < X < 4$ and $0 < Y < 1$.  And, noticeably, $\tilde{T}_S$ tends to be more negative as $m_{H^{++}_5}$ approaches $m_{H^0_5}$ .
\ei

The above observations are very useful in the search for signals of the EW$\nu_R$ model, in particular its scalar sector in light of recent results from the LHC. We shall discuss this aspect in the next section dealing with experimental implications.

\ei

\section{Some experimental implications}

As we have seen in the above discussions, the EW$\nu_R$ model contains a non-negligible region of parameter space which {\em agrees} with the electroweak precision data and thus has passed the first (indirect) test. The next test would be direct observations of the signatures coming from the new particles of the model: The mirror quarks and leptons and the scalars. Some of such signatures have been suggested in \cite{pqnur} such as like-sign dileptons as a sign of lepton number violation coming from the decay of the Majorana right-handed neutrinos. As mentioned in \cite{pqnur}, this would be the high-energy equivalent of neutrino-less double beta decay. This signal and those of other mirror quarks and leptons will be presented in a separate publication.

The scalar sector of the EW$\nu_R$ model has been studied in some details in \cite{pqaranda}. In light of the new LHC results, it is timely to update the status of this sector. In particular, the question that one may ask is the following: What are the implications of the above analysis on the masses of the scalars and their couplings to fermions? Several of these issues will be presented in a follow-up paper on the 126 GeV scalar but it is important to set the foundation for that paper here.

In terms of the minimal EW$\nu_R$ model discussed in this manuscript, one is most interested at this point in the neutral scalars given in Eq.~\ref{eq:higgs}. Furthermore, the 126-GeV object appears to be consistent, in terms production and decays, with the SM Higgs boson which is a $0^+$ particle \cite{LHC}. Although recent data on the spin-parity \cite{LHC2} seemed to disfavor the 126-GeV object as a $0^-$ particle and is more consistent with the $0^+$ interpretation, it did not completely rule out the $0^-$ possibility. In consequence, we will keep an open mind. As seen in Eq.~\ref{eq:higgs}, there are four neutral states: $H_5^0 $, $H_3^0$, $H_1^0$ and $H_1^{0\prime}$. Since the triplet scalars $\chi$ and $\xi$ do not couple to SM and mirror quarks while the doublet $\phi$ does, one can see from Eq.~\ref{eq:higgs} that only $H_3^0$ and $H_1^0$ could be candidates for the 126 GeV object. However, a close look at the production cross section reveals that, parity aside, only $H_3^0$ fits the bill. We summarize here some of the details which will be given in full in \cite{nurd}.

The dominant production mechanism for the aforementioned scalars is through gluon fusion. As a result, one should know the couplings of  $H_3^0$ and $H_1^0$ to the SM and mirror quarks. For $H_1^0$, the coupling to SM and mirror quarks is given generically as $g_{H_{1}^0\,q\bar{q}}=-\imath \frac{m_{q} g}{2m_{W} c_{H}}$ where $q$ represents SM and mirror quarks and $c_H$ is an abbreviation for $\cos \theta_H$. This is greater than the SM coupling by the factor $1/\cos \theta_H$. Furthermore, the gluon-fusion cross section is now proportional the square of the number of heavy quarks, namely $(7)^2=49$ where we count the top quark and the six mirror quarks. As a result, $\sigma_{EW\nu_R} \sim 49\, \sigma_{SM}$. This is evidently not acceptable. As a consequence, to be consistent with the LHC data, $H_1^0$ will have to be heavier than $\sim 600 GeV$. 

We are left with $H_3^0$. From \cite{gunion,pqaranda}, one can find its coupling to the SM and mirror quarks as follows: $g_{H_{3}^0\,t\bar{t}}=+\imath \frac{m_{t} g}{2m_{W} }\,\tan \theta_H \, \gamma_5$, $g_{H_{3}^0\,b\bar{b}}=-\imath \frac{m_{b} g}{2m_{W} }\,\tan \theta_H \, \gamma_5$, $g_{H_{3}^0\,u^{M}\bar{u}^{M}}=-\imath \frac{m_{u^{M}} g}{2m_{W} }\,\tan \theta_H \, \gamma_5$, $g_{H_{3}^0\,d^{M}\bar{d}^{M}}=+\imath \frac{m_{d^{M}} g}{2m_{W} }\,\tan \theta_H \, \gamma_5$. Notice that the pseudo scalar $H_3^0$ contains the imaginary part of $\phi^0$ which couples to the up and down quarks with opposite signs. This fact reflects in the above sign differences in the couplings. The amplitude for the gluon fusion production of $H_3^0$ involves a triangle loop denoted by $I$ which depends on $r_q=m_{q}^2/m_{H_3^0}^2$ ($q$ stands for any of the quarks). $I \rightarrow 1$ when $r_q \gg 1$ and $I \rightarrow 0$ when $r_q \ll 1$. It is well-known from the behavior of $I$ that the gluon fusion production of the SM Higgs boson is dominated by the top quark loop. In our case, in addition to the top quark, we have the mirror quarks which are assumed to be heavier than $H_3^0$ and hence $r_t\, ,\, r_{q^M} \gg 1$. However, $H_3^0$ is a pseudo scalar and, as we have seen above, the mirror up-quark loop cancels that of the mirror down-quark loop because both quarks are heavy so that $r_{q^M} \gg 1$ and because their couplings to $H_3^0$ have opposite signs. This means that $I_{u^M}+I_{d^M} \rightarrow 0$. As a consequence, again only the top quark loop ``contributes''. Details will be given in \cite{nurd}. Here we just quote the result. Because of the aforementioned cancellation in the mirror quark sector, the production cross section for $H_3^0$ is
\be
\label{xsection}
\sigma_{H_3^0} = \tan^{2}\theta_H\; \sigma_{H_{SM}}
\ee
If we assume that the various branching ratios for the $H_3^0$ decays are comparable to those of the SM, one can see from Eq.~\ref{xsection} that $\sigma_{H_3^0} \sim \sigma_{H_{SM}}$ if $\tan \theta_H \sim 1$ or $\sin \theta_H \sim 0.707$ which is well inside our allowed range as shown in FIG.~\ref{S_sh}. The only hitch is that the parity measurement seems to disfavor this interpretation at a $2\,\sigma-3\,\sigma$ level but does not rule out completely the pseudo scalar interpretation. As a consequence, we will keep an open mind regarding this possibility. In \cite{nurd} we also present a simple extension of the EW$\nu_R$ model which can accommodate the SM-like $0^+$ scalar as an interpretation of the 126 GeV object.

Last but not least are the direct searches for mirror fermions. In \cite{pqnur}, it was mentioned that  one of the most tell-tale signs of the EW$\nu_R$ is the production at the LHC and the decays of $\nu_R$'s which are Majorana particles and are their own antiparticles through the subprocess $q+\bar{q} \rightarrow \nu_R + \nu_R \rightarrow e_{M}^{-} + W^{+} + e_{M}^{-} + W^{+} \rightarrow e^{-} +\phi_S + W^{+} + e^{-} +\phi_S + W^{+}$, where $e$ stands for a generic charged lepton. These like-sign dileptons events would be the high-energy equivalent of the low-energy neutrino less double beta decay as emphasized in \cite{pqnur}. A detailed study of this and other processes involving mirror fermions is under investigation \cite{pqaranda2}.

\section{Conclusion}

The assumption that right-handed neutrinos are {\em non-singlets} under $SU(2) \times U(1)$ as proposed in \cite{pqnur} is a very reasonable one which can be tested experimentally. The EW$\nu_R$ model preserves the gauge structure of the SM but enriches it with mirror fermions and Higgs triplets. The price paid might be considered to be minimal considering the fact that the EW$\nu_R$ of \cite{pqnur} links the nature of right-handed neutrinos -and hence the energy scale of its Majorana mass- to details of the electroweak symmetry breaking. In addition, these aspects can be {\em tested} experimentally.

The first of such tests is the electroweak precision constraints. We have shown in this paper how the EW$\nu_R$ model has a non-negligible range of parameter space to fit the constraints ( see Fig.~\ref{TStotal}) on the oblique parameters S and T despite the presence of right-handed mirror quarks and leptons which by themselves alone would make a large positive contribution to the S parameter. We have shown in details how the scalar sector, in particular the Higgs triplet fields, dramatically avoids this potential disaster by making negative contributions which offset those of the mirror fermions and thus bringing the EW$\nu_R$ model in agreement with the electroweak precision data. We have shown also how mass splittings, in particular those of the scalar sector, affect the values of the oblique parameters such as S whose constraints in turn have interesting implications of those splittings themselves. This aspect would eventually be very useful in the search for the scalars of the model. The mass splittings of the mirror fermion sector can be straightforwardly computed as a function of the mass splittings in the scalar sector. 

The next test of the model would be signatures and searches for the mirror quarks and leptons and for the additional scalars. Of immediate interest for the EW$\nu_R$ (and for other BSM models as well) is the discovery of the SM-like boson with a mass of 126 GeV. This discovery puts a very strong constraint on any BSM model. What the EW$\nu_R$ model has to say about this 126 GeV object has been briefly discussed above and will be presented in detail in \cite{nurd}. Basically, the minimal EW$\nu_R$ model contains a pseudo-scalar, $H_3^0$, which could in principle be a candidate whose production cross section can be comparable to that for a SM Higgs boson with the same mass with a choice of the angle $\theta_H$ well within the allowed range discussed above. However, the spin-parity measurement \cite{LHC2} seemed to disfavor, but not completely ruling out, the interpretation of the 126 GeV object as a $0^-$ particle while the SM-like $0^+$ seems to be favored. Until more data come out to completely rule out the pseudo-scalar interpretation, we will keep an open mind however. Nevertheless, \cite{nurd} presents a minimal extension of the EW$\nu_R$ model where the presence of an additional $0^+$ state can act like a SM Higgs boson. Needless to say, one expects several scalars beyond the 126 GeV boson to be present in the model. A phenomenological study of the scalar sector of the EW$\nu_R$ model has been performed \cite{pqaranda} and it goes without saying that more studies of this sector are needed. The input from the electroweak precision constraints will be valuable in a new study of this sector.

One of the key points of the EW$\nu_R$ model was  the production and detection of electroweak-scale right-handed neutrinos through lepton-number violating signals such as like-sign dilepton events at the LHC \cite{pqnur} which represent the high-energy equivalent of the low-energy neutrino less double-beta decay. One could imagine that, after taking care of the SM background, it would be a ``much'' easier process to detect. This signal and others related to searches for mirror quarks and leptons will be presented in \cite{pqaranda2}.  

 \section{Acknowledgements}
 
 We would like to thank Goran Senjanovic for illuminating discussions. This work was supported by US DOE grant DE-FG02-97ER41027.

%
%
\appendix
%
%
\section{Loop Integrals and Functions}
\label{sec:appfunc}
%
Different contributions to the oblique parameters are expressed using loop integrals like $A_0$, $B_0$, $B_{22}$, $B_1$, $B_2$. and functions like $\mathcal{F}$, $G$, etc. Therefore, it is very important to define all the loop integrals and functions we have used in the calculations of different loop diagrams before listing contributions from loop diagrams and details of the calculations of the oblique parameters.
\par For calculation of oblique parameters we need the loop diagrams with two external vector bosons. These diagrams have a general form
	\begin{equation}
		\Pi_{\mu\nu} = \Pi_A\;g_{\mu\nu} + \Pi_B\; q_\mu q_\nu
	\end{equation}
For the purpose of oblique parameters we only need the '$\askGmu$' term in this equation. Hence, hereafter in this paper $\Pi_{\mu\nu}$ denotes only the first term on RHS above.
\par Loop diagrams involving one or two internal scalars or one or two internal fermions appear in the calculation of one-loop vector boson self-energy diagrams and $Z\gamma$- diagrams. Following loop integrals appear in the calculation of loops with scalar particles \cite{Holl}:\\
One-point integral:
	\begin{flalign}
		\int \frac{d^4k}{(2\pi)^4} \frac{1}{(k^2-m^2)} \equiv \frac{\askImath}{16\pi^2} A_0(m^2)
	\end{flalign}
Two-point integrals:
	\begin{flalign}\label{eq:B022int}
		\int \frac{d^4k}{(2\pi)^4} \frac{1}{(k^2-m_1^2)((k+q)^2-m_2^2)}&&\nonumber\\[2mm]
		\equiv \frac{\askImath}{16\pi^2}& B_0(q^2;m_1^2,m_2^2),&\\[4mm]
		\int \frac{d^4k}{(2\pi)^4} \frac{k_\mu k_\nu}{(k^2-m_1^2)((k+q)^2-m_2^2)}&&\nonumber\\[2mm]
		\equiv \frac{\askImath}{16\pi^2}g_{\mu\nu}& B_{22}(q^2;m_1^2,m_2^2)&
	\end{flalign}
The expansion of LHS in the latter equation also has term with $q_\mu q_\nu$ \cite{Holl}, but this term is omitted as it does not contribute to the oblique parameters \cite{PeskTak}.
\par Following \cite{Holl}, in the dimensional regularization these integrals can be simplified to
	\begin{flalign}
		\label{eq:A}
		A_0(m^2) &= m^2 \Big(\Delta + 1 - ln(m^2)\Big)&\\[2mm]
		\label{eq:B0}
		B_0(q^2;m_1^2,m_2^2) &= \Delta - \int_0^1 d\text{x}\; ln(X - \askImath \epsilon)&\\[2mm]
		\label{eq:B22}
		B_{22}(q^2;m_1^2,m_2^2) &= \frac{1}{4} (\Delta + 1) \Big(m_1^2 + m_2^2 - \frac{q^2}{3}\Big)&\nonumber\\[2mm]
		& - \frac{1}{2}\int_0^1 d\text{x}\; X\; ln(X - \askImath \epsilon)&
	\end{flalign}
where
  \begin{eqnarray}
  	\label{eq:X}
  	X &\equiv& m_1^2 \text{x} + m_2^2 (1 - \text{x}) - q^2 \text{x} (1 - \text{x}),\\[3mm]
	\label{eq:D}
	\Delta &\equiv& \dfrac{2}{4-d} + ln(4\pi) - \gamma.
  \end{eqnarray}
in $d$ space-time dimensions with $\gamma= 0.577216...$, the Euler's constant \cite{PeskSchro}. The integrals in eqns. (\ref{eq:B0}), (\ref{eq:B22}) can be calculated numerically up to desired accuracy. Note that these equations involve the logarithm of a dimensionful quantity, $X$ and the scale of this logarithm is hidden in the $2/(4 - d)$ term in $\Delta$ (refer to section 7.5 of \cite{PeskSchro}). It is useful, especially in deriving $\widetilde{T}_{scalar}$ in Eq. (\ref{eq:expts}), to note that \cite{HaberOneil}
	\begin{flalign}
		B_0(0;m_1^2,m_2^2) &= \frac{A_0(m_1^2) - A_0(m_2^2)}{m_1^2 - m_2^2},&\\[2mm]
		\label{eq:B22FAA}
		4 B_{22}(0;m_1^2,m_2^2) &= \mathcal{F}(m_1^2,m_2^2) + A_0(m_1^2) + A_0(m_2^2),&
	\end{flalign}
where
	\begin{eqnarray}\label{eq:calF}
  		\mathcal{F}(m_1^2, m_2^2) &=& \dfrac{m1^2 + m_2^2}{2} - \dfrac{m_1^2 m_2^2}{m_1^2 - m_2^2}\; ln\left(\dfrac{m_1^2}{m_2^2}\right),\nonumber\\
		&&\hspace{2em}\text{if $m_1 \neq m_2$},\nonumber\\[2mm]
		&=& 0 \hspace{1.5em} \text{if $m_1 = m_2$}.
	\end{eqnarray}
Note that 
  \begin{equation}
  	\mathcal{F}(m_1^2, m_2^2)=\mathcal{F}(m_2^2, m_1^2)\,.
  \end{equation}
Also notice that
  \begin{eqnarray}
  	B_{22}(q^2; m_1^2, m_2^2) &=& B_{22}(q^2; m_2^2, m_1^2)\nonumber\\[2mm]
	B_0(q^2; m_1^2, m_2^2) &=& B_0(q^2; m_2^2, m_1^2).
  \end{eqnarray}
\par While evaluating the fermion loops which contribute to the oblique parameters following two-point loop integrals are useful (refer section 21.3 of \cite{PeskSchro}):
	\begin{flalign}\label{eq:B1B2}
		B_1(q^2;m_1^2,m_2^2) &= \int_0^1 d\text{x}\; (1 - \text{x})\; ln\Big(\frac{X - \askImath \epsilon}{M^2}\Big),&\\[2mm]
		B_2(q^2;m_1^2,m_2^2) &= \int_0^1 d\text{x}\; \text{x}(1 - \text{x})\; ln\Big(\frac{X - \askImath \epsilon}{M^2}\Big),&
	\end{flalign}
where $X$ is as defined in Eq. (\ref{eq:X}). The logarithms in these integrals involve a mass scale $M$. All the terms, which depend on this scale cancel while evaluating the final expressions for oblique parameters. For $m_1 = m_2 = m$ and $q^2 = M_Z^2$,
	\begin{flalign}
		B_1(M_Z^2;m^2,m^2) &= -1 - \frac{G(\text{x})}{4} + ln\Big(\frac{m^2}{M^2}\Big),&\\[2mm]
		B_2(M_Z^2;m^2,m^2) &= \frac{1}{18} \Bigg[-\frac{3}{2}\; G(x) \big(2\; x + 1\big)&\nonumber\\[2mm]
		& + \Bigg(-12\; x - 5 + 3\; ln\Big(\frac{m^2}{M^2}\Big) \Bigg) \Bigg],&
	\end{flalign}
where
	\begin{equation}\label{eq:G}
		  	G(x) =\; -4\; \sqrt{4 x - 1}\;Arctan\Big(\frac{1}{\sqrt{4 x - 1}}\Big)\;.
	\end{equation}
While deriving $\widetilde{T}_{fermion}$ in Eq.(\ref{eq:ttt}) we need to evaluate integrals in Eq. (\ref{eq:B1B2}) for $q=0$ and $m_1 \neq m_2$. One of the integrals, which appear in this calculation is
	\begin{flalign}
		&\int_0^1 dx\; \big(m_1^2 x + m_2^2 (1 - x) \big)\; ln\Big(\frac{m_1^2 x + m_2^2 (1 - x)}{M^2} \Big)&\nonumber\\[2mm]
		&=\frac{ \big(m_2^4 - m_1^4\big) + 2\; m_1^4\; ln\Big(\frac{m_1^2}{M^2}\Big) - 2\; m_2^4\; ln\Big(\frac{m_1^2}{M^2}\Big)}{4 \big(m_1^2-m_2^2\big)}.&
	\end{flalign}
\par Using the loop integrals and functions defined and enlisted in this appendix we can derive the expressions for the oblique parameters, which are suitable for the numerical analysis.
%
%
\section{Gauge Couplings of Higgs' in \ewnur model}
\label{sec:appfeyns}
%
In this appendix we derive the cubic and quartic couplings of the Higgs' in \ewnur model with the electroweak gauge bosons. We start with the scalar fields $\Phi$ and $\chi$ in the \ewnur model, get the physical scalar states from a generic potential with a global $SU(2)_L \times SU(2)_R$ symmetry, and which after spontaneous symmetry breaking preserves $SU(2)_D$ custodial symmetry. Then we derive the gauge couplings of the physical scalar states from the kinetic part of the scalar Lagrangian in \ewnur model. We work in the 't Hooft Feynman gauge (gauge parameter, $\xi_{gauge}=1$) throughout the calculations in this appendix and all the appendices, which follow. To calculate the new Physics contributions due to \ewnur model to the oblique parameters we also need the corresponding contributions from SM (refer to equations (\ref{eq:s}, \ref{eq:t}, \ref{eq:u})). Therefore, in this section we also list the related SM couplings.
\par The most general scalar potential for $\Phi$ and $\chi$ that preserves global $SU(2)_L \times SU(2)_R$ is given by \cite{pqaranda, ChanoGold}:
	\begin{flalign}\label{eq:pot}
		V(\Phi,\chi) &= \lambda_1 \Big(Tr\Phi^\dagger \Phi - v_2^2 \Big)^2 + \lambda_2 \Big(Tr\chi^\dagger \chi - 3 v_M^2\Big)^2&\nonumber\\[2mm]
		&+\lambda_3 \Big(Tr\Phi^\dagger \Phi - v_2^2 + Tr\chi^\dagger \chi - 3 v_M^2 \Big)^2&\nonumber\\[2mm]
		&+\lambda_4 \Big((Tr\Phi^\dagger \Phi)\; (Tr\chi^\dagger \chi) &\nonumber\\[2mm]
		&\hspace{3em}- 2\; (Tr\Phi^\dagger \frac{\tau^a}{2} \Phi \frac{\tau^b}{2})\; (Tr\chi^\dagger T^a \chi T^b) \Big)&\nonumber\\[2mm]
		&+\lambda_5 \Big(3\; Tr\chi^\dagger \chi \chi^\dagger \chi - (Tr\chi^\dagger \chi)^2 \Big)&
	\end{flalign}
where repeated indices $a$, $b$ are summed over. Note that this potential is invariant under $\chi \rightarrow -\chi$ so that the cubic terms in the potential are eliminated. In order for this potential to be positive semidefinite the following conditions must be imposed: $\lambda_1 + \lambda_2 + 2\lambda_3 > 0$, $\lambda_1\lambda_2 + \lambda_1\lambda_3 + \lambda_2\lambda_3 > 0$, $\lambda_4 > 0$, $\lambda_5 > 0$.
\par At the minimum of this potential, the vacuum alignment in eqs. (\ref{chivev}, \ref{phivev}) breaks the global $SU(2)_L \times SU(2)_R$ down to the custodial $SU(2)_D$. One obtains $M_W = g\,v/2$ and $M_Z = M_W/\cos \theta_W$, with $v= \sqrt{v_2^2+ 8\,v_M^2} \approx 246 \, GeV$ and, at tree level, $\rho= M_W/M_Z \, \cos \theta_W=1$ as desired (this confirms that $SU(2)_D$ custodial is, indeed, preserved at the tree level).
\par After spontaneous breaking of $SU(2)_L \times U(1)_Y$, besides the three Nambu-Goldstone bosons which are absorbed by $W$ and $Z$, there are {\em ten} physical scalars which are grouped into {\bf 5} + {\bf 3} + {\bf 1} (2 singlets) of the custodial $SU(2)_D$. These Nambu-Goldstone bosons and physical scalars are given in eqns. (\ref{eq:goldstone}), (\ref{eq:higgs}) respectively. The masses of the physical scalars are given as:
\begin{eqnarray}
	m_{H_5^{\pm\pm,\pm,0}}^2 &=& m_5^2 = 3\;( \lambda_4 c_H^2 + \lambda_5 s_H^2 )\nonumber\\[2mm]
	m_{H_3^{\pm,0}}^2 &=& m_3^2 = \lambda_4\; v^2\;.
\end{eqnarray}
The two singlets $H_1^0$ and $H_1^{0\prime}$ can mix according to the mass-squared matrix given as:
\begin{equation}
	\mathcal{M}_{H_1^0,\;H_1^{0\prime}}^2 = v^2 \left[
	\begin{array}{cc}
		8 c_H^2\; (\lambda_1 + \lambda_3)	&	2 \sqrt{6} s_H c_H \lambda_3 \\[2mm]
		2 \sqrt{6} s_H c_H \lambda_3		&	3 s_H^2\; (\lambda_2 + \lambda_3)
	\end{array} \right]
\end{equation}
The oblique parameters, the Feynman rules and the loop diagrams are expressed in terms of the VEVs of the doublet and triplets, and the masses of the physical scalars- $m_{H_5^{\pm\pm,\pm,0}}$, $m_{H_3^{\pm,0}}$, $m_{1}$, $m_{H_1^\prime}$.
\par The gauge couplings of the physical scalars can be obtained from the kinetic part of the scalar Lagrangian in \ewnur model \cite{GeorgMach, gunion, pqaranda}:

	\begin{flalign}\label{eq:lskin}
		\left(\mathcal{L}_{S^{EW\nu_R}}\right)_{kin} =& \frac{1}{2}\; Tr\left[(D_\mu \Phi)^\dagger (D_\mu \Phi)\right]&\\\nonumber
		& + \frac{1}{2}\; Tr\left[(D_\mu \chi)^\dagger (D_\mu \chi)\right] + |\partial_\mu \phi_S |^2&
	\end{flalign}
The notation $\left(\mathcal{L}_{S^{EW\nu_R}}\right)_{kin}$ is used to denote the kinetic part (denoted by subscript '$kin$') of the Higgs Lagrangian (denoted by subscript '$S$' for Scalar) in \ewnur model (denoted by '$EW\nu_R$' in the subscript). Here, $\Phi$ and $\chi$ are used in their $2 \times 2$ and $3 \times 3$ representations respectively, as given in equations (\ref{chi}, \ref{Phi}); $\phi_S$ is the neutral $SU(2)$ singlet scalar in \ewnur model and
	\begin{equation}
		D_\mu \Phi \equiv \partial_\mu \Phi + \frac{1}{2}\;\askImath g (\mathbf{W_\mu \cdot\tau}) \Phi - \frac{1}{2}\;\askImath g^\prime \Phi B_\mu \tau_3\; ;
	\end{equation}
	\begin{equation}
		D_\mu \chi \equiv \partial_\mu \chi + \askImath g (\mathbf{W_\mu \cdot t}) \chi - \askImath g^\prime \chi B_\mu t_3
	\end{equation}
The $\tau_i/2$ and $t_i$ are the $2 \times 2$ and $3 \times 3$ representation matrices of the $SU(2)$ generators respectively, following reference \cite{gunion}.
\par We work under the premise that the hierarchy in neutrino masses in \ewnur model comes from the VEV of $\phi_S$. Thus, $v_S \sim 10^5$ eV and as a result the mixing between $\phi_S$ and other scalars in negligible. Hence, hereafter  in the related calculations we neglect this mixing, unless otherwise is stated.
After the spontaneous breaking of $SU(2)_L \times U(1)_Y$ to $U(1)_{EM}$, expanding the Lagrangian in equation (\ref{eq:lskin}), one can find the Feynman rules for the three point and four point interactions between physical scalars, Nambu-Goldstone bosons and electroweak gauge bosons $W$, $Z$ and $\gamma$. For the corresponding SM Feynman rules it is useful to recall the kinetic part of the SM-Higgs Lagrangian:
	\begin{equation}\label{eq:lkinsm}
		\left(\mathcal{L}_{S^{SM}}\right)_{kin} = \frac{1}{2} Tr\left[(D_\mu \Phi)^\dagger (D_\mu \Phi)\right]
	\end{equation}
The resulting Feynman rules in \ewnur model and SM are listed in tables (\ref{table:feyn21}, \ref{table:feyn12}, \ref{table:feyn22} and \ref{table:feyn22n}) below.
  \begin{table*}[!htbp]
  \renewcommand{\arraystretch}{2}
  \centering
  \caption{\label{table:feyn21}$S_1S_2V$ type couplings($V$ is a vector gauge boson and $S_1$, $S_2$ are Higgs/ Goldstone bosons), which contribute to Oblique Corrections. Common factor: $\askImath g \askPmu$, where $p (p')$ is the \textit{incoming} momentum of the $S_1 (S_2)$.}
  \bigskip
  \begin{tabular}{|ll||ll|}
  	\hline
  	$g_{H_5^0H_5^-W^+}$		& $-\frac{\sqrt{3}}{2}$ 	& $g_{H_5^{++}H_5^{--}Z}$	& $-\frac{(1 - 2 s_W^2)}{2 c_W}$\\
  	$g_{H_5^+H_5^{--}W^+}$	& $-\frac{1}{\sqrt{2}}$ 	& $g_{H_3^+H_3^-Z}$		& $\frac{(1 - 2 s_W^2)}{2 c_W}$\\
  	$g_{H_3^0H_3^-W^+}$		& $-\frac{1}{2}$ 		& $g_{H_3^+H_5^-Z}$		& $-\frac{1}{2 c_W}$\\
  	$g_{H_3^+H_5^{--}W^+}$	& $-\frac{1}{\sqrt{2}} c_H$ & $g_{H_3^0H_5^0Z}$		& $\frac{1}{\sqrt{3}} \frac{c_H}{c_W}$\\
  	$g_{H_3^0H_5^{-}W^+}$	& $-\frac{1}{2} c_H$ 		& $g_{G_3^+G_3^-Z}$		& $\frac{(1 - 2 s_W^2)}{2 c_W}$\\
  	$g_{H_5^0H_3^{-}W^+}$	& $-\frac{1}{2\sqrt{3}} c_H$ & $g_{G_3^0H_5^0Z}$		& $\frac{1}{\sqrt{3}} \frac{s_H}{c_W}$\\
  	$g_{G_3^0G_3^-W^+}$		& $-\frac{1}{2}$ 		& $g_{G_3^+H_5^-Z}$		& $-\frac{1}{2 c_W} s_H$\\
  	$g_{G_3^+H_5^{--}W^+}$	& $-\frac{1}{\sqrt{2}} s_H$ & $g_{H_1^0G_3^0Z}$		& $\frac{c_H}{c_W}$\\
  	$g_{G_3^0H_5^-W^+}$		& $-\frac{1}{2} s_H$ 		& $g_{H_1^{0\prime}G_3^0Z}$	& $\sqrt{\frac{2}{3}} \frac{s_H}{c_W}$\\
  	$g_{H_5^0G_3^-W^+}$		& $\frac{1}{2\sqrt{3}} s_H$ & $g_{H_1^0H_3^0Z}$		& $-\frac{s_H}{2 c_W}$\\
  	$g_{H_1^0G_3^-W^+}$		& $\frac{1}{2} c_H$ 		& $g_{H_1^{0\prime}H_3^0Z}$	& $\sqrt{\frac{2}{3}} \frac{c_H}{c_W}$\\
  	$g_{H_1^{0\prime}G_3^-W^+}$		& $\sqrt{\frac{2}{3}} s_H$ & $g_{H_5^+H_5^-\gamma}$	& $-s_W$\\
  	$g_{H_1^0H_3^-W^+}$		& $-\frac{1}{2} s_H$ 		& $g_{H_5^{++}H_5^{--}\gamma}$	& $-2 s_W$\\
  	$g_{H_1^{0\prime}H_3^-W^+}$		& $\sqrt{\frac{2}{3}} c_H$ & $g_{H_3^+H_3^-\gamma}$		& $s_W$\\
  	$g_{H_5^+H_5^-Z}$		& $\frac{(1 - 2 s_W^2)}{2 c_W}$ & $g_{G_3^+G_3^-\gamma}$		& $s_W$\\
  	\hline
  \end{tabular}
  \end{table*}
  \begin{table*}[!htbp]
  \renewcommand{\arraystretch}{2}
  \centering
  \caption{\label{table:feyn12}$SV_1V_2$ type couplings($V_1$ and $V_2^\prime$ are vector gauge bosons and $S$ is a Higgs boson), which contribute to Oblique Corrections. Common factor: $\askImath g M_W \askGmu$}
  \bigskip
  \begin{tabular}{|lc||lc|}
  	\hline
  	$g_{H_5^0W^+W^-}$		& $\frac{s_H}{\sqrt{3}}$	& $g_{H_5^0ZZ}$		& $-\frac{2}{\sqrt{3}} \frac{s_H}{c_W^2}$\\
  	$g_{H_5^{++}W^-W^-}$		& $\sqrt{2} s_H$		& $g_{H_5^{+}W^-Z}$	& $-\frac{s_H}{c_W}$\\
  	$g_{H_1^0W^+W^-}$		& $c_H$				& $g_{H_1^0ZZ}$		& $\frac{c_H}{c_W^2}$\\
  	$g_{H_1^{0\prime}W^+W^-}$ & $\frac{2 \sqrt{2}}{\sqrt{3}} s_H$	& $g_{H_1^{0\prime}ZZ}$	& $\frac{2 \sqrt{2}}{\sqrt{3}} \frac{s_H}{c_W^2}$\\
  	\hline
  \end{tabular}
  \end{table*}
  \begin{table*}[!htbp]
  \renewcommand{\arraystretch}{2}
  \centering
  \caption{\label{table:feyn22}$H_1H_2V_1V_2$ type couplings, which contribute to Oblique Corrections. Common factor: $\askImath g^2 \askGmu$}
  \bigskip
  \begin{tabular}{|lc||lc|}
  	\hline
  	$g_{H_5^0H_5^0W^+W^-}$		& $\frac{5}{3}$				& $g_{H_5^0H_5^0ZZ}$		& $\frac{2}{3} \frac{1}{c_W^2}$\\
  	$g_{H_5^+H_5^-W^+W^-}$		& $-\frac{3}{2}$			& $g_{H_5^+H_5^-ZZ}$		& $-\frac{(c_W^4 + s_W^4)}{c_W^2}$\\
  	$g_{H_5^{++}H_5^{--}W^+W^-}$	& $1$					& $g_{H_5^{++}H_5^{--}ZZ}$		& $2 \frac{(1 - 2 s_W^2)^2}{c_W^2}$\\
  	$g_{H_3^0H_3^0W^+W^-}$		& $-(c_H^2 + \frac{s_H^2}{2})$		& $g_{H_3^0H_3^0ZZ}$	& $-\frac{(1 + c_H^2)}{2 c_W^2}$\\
  	$g_{H_3^+H_3^-W^+W^-}$		& $-(\frac{1}{2} + c_H^2)$		& $g_{H_3^+H_3^-ZZ}$	& $-\left[ \frac{s_H^2}{2} \frac{(1 - s_W^2)^2}{c_W^2} + c_H^2\frac{(c_W^4 + s_W^4)}{c_W^2}\right]$\\
  	$g_{G_3^0G_3^0W^+W^-}$		& $-\frac{(1 + s_H^2)}{2}$		& $g_{G_3^0G_3^0ZZ}$		& $-\frac{1}{2 c_W^2} (1 + 3 s_H^2)$\\
  	$g_{G_3^+G_3^-W^+W^-}$		& $-(\frac{1}{2} + s_H^2)$	& $g_{G_3^+G_3^-ZZ}$		& $-\left[\frac{1}{2} c_H^2 (1 - 2 s_W^2)^2 + s_H^2 (c_W^4 + s_W^4)\right]$\\
  	$g_{H_1^0H_1^0W^+W^-}$		& $\frac{1}{2}$				& $g_{H_1^0H_1^0ZZ}$		& $\frac{1}{2 c_W^2}$\\
  	$g_{H_1^{0\prime}H_1^{0\prime}W^+W^-}$	& $\frac{4}{3}$		& $g_{H_1^{0\prime}H_1^{0\prime}ZZ}$	& $\frac{4}{3 c_W^2}$\\
  	$g_{H_5^+H_5^-\gamma\gamma}$		& $-2 s_W^2$			& $g_{H_5^+H_5^-Z\gamma}$		& $-\frac{s_W}{c_W} (1 - 2 s_W^2)$\\
  	$g_{H_5^{++}H_5^{--}\gamma\gamma}$		& $8 s_W^2$		& $g_{H_5^{++}H_5^{--}Z\gamma}$	& $4 \frac{s_W}{c_W} (1 - 2 s_W^2)$\\
  	$g_{H_3^+H_3^-\gamma\gamma}$		& $-2 s_W^2$			& $g_{H_3^+H_3^-Z\gamma}$		& $-\frac{s_W}{c_W} (1 - 2 s_W^2)$\\
  	$g_{G_3^+G_3^-\gamma\gamma}$		& $-2 s_W^2$			& $g_{G_3^+G_3^-Z\gamma}$		& $-\frac{s_W}{c_W} (1 - 2 s_W^2)$\\
  	\hline
  \end{tabular}
  \end{table*}
  \begin{table*}[!htbp]
  \renewcommand{\arraystretch}{2}
  \centering
  \caption{\label{table:feyn22n}$H_1H_2V_1V_2$ type couplings, which \textit{do not} contribute to Oblique Corrections. Common factor: $\askImath g^2 \askGmu$}
  \bigskip
  \begin{tabular}{|lc||lc|}
  	\hline
  	$g_{H_1^{0\prime}H_5^0W^+W^-}$		& $\frac{\sqrt{2}}{3}$		& $g_{H_1^{0\prime}H_5^0ZZ}$		& $-\frac{2 \sqrt{2}}{3 c_W^2}$\\
  	$g_{H_3^+H_5^-W^+W^-}$		& $-\frac{c_H}{2}$			& $g_{H_3^+H_5^-ZZ}$			& $c_H \frac{(1 - 2 s_W^2)}{c_W^2}$\\
  	$g_{H_3^0G_3^0W^+W^-}$		& $-\frac{c_H s_H}{2}$		& $g_{H_3^0G_3^0ZZ}$			& $-\frac{3}{2} \frac{c_H s_H}{c_W^2}$\\
  	$g_{H_3^+G_3^-W^+W^-}$		& $-c_H s_H$				& $g_{H_3^+G_3^-ZZ}$			& $-\frac{c_H s_H}{2 c_W^2}$\\
  	$g_{H_5^+G_3^-W^+W^-}$		& $-\frac{s_H}{2}$			& $g_{H_5^+G_3^-ZZ}$			& $s_H \frac{(1 - 2 s_W^2)}{c_W^2}$\\
  	~ & ~ & $g_{H_3^+H_5^-Z\gamma}$		& $c_H \frac{s_W}{c_W}$\\
  	\hline
  \end{tabular}
  \end{table*}
%
%
\section{Gauge Couplings of mirror fermions in \ewnur model}
\label{sec:appfeynf}
%
%
%
\par In this appendix we start with the fermion content of the \ewnur model and derive the electroweak gauge couplings of these fermions from the Lagrangian. The $SU(2)_L \times U(1)_Y$ fermion content of the \ewnur model of \cite{pqnur} is given in eqns. (\ref{ldoublet}), (\ref{lsinglet}), (\ref{qdoublet}), (\ref{qsinglet}).%
\par The interaction of mirror leptons with the $SU(2)_L \times U(1)_Y$ gauge bosons are found in the terms
\begin{equation}
	\bar{l}^{M}_{R} \slashed{D} l^{M}_{R}\, ; \, \bar{e}^{M}_{L} \slashed{D} e^{M}_{L} \, ,
\end{equation}
where
\begin{eqnarray}
	\slashed{D}l^M_R &\equiv& \gamma^\mu(\partial_\mu - \frac{1}{2}\;\askImath g (\mathbf{W_\mu \cdot\tau}) + \frac{1}{2}\;\askImath g^\prime B_\mu) l^M_R\,,\nonumber\\[2mm]
	\slashed{D}e^M_e &\equiv& \gamma^\mu(\partial_\mu + \askImath g^\prime B_\mu) e^M_L\,.
\end{eqnarray}
The gauge interactions for mirror quarks can similarly be found. Thus, the Feynman rules for the gauge interactions for fermions (Sm fermions and mirror fermions) in the \ewnur model can be evaluated from
	\begin{flalign}
		\left(\mathcal{L}_{F^{EW\nu_R}}\right)_{int} = \left(\mathcal{L}_{F^{SM}}\right)_{int} + \left(\mathcal{L}_{F^M}\right)_{int}\;,
	\end{flalign}
where $\left(\mathcal{L}_{F^{SM}}\right)_{int}$ comes from the fermion-sector in the Standard Model (and is well known) and $\left(\mathcal{L}_{F^M}\right)_{int}$ includes interaction terms arising due to the mirror fermion-sector in \ewnur model. $\left(\mathcal{L}_{F^{SM}}\right)_{int}$ is well known to be \cite{PeskSchro}
%
%
To write the mirror fermions part $\left(\mathcal{L}_{F^M}\right)_{int}$ remember that the $W$ bosons couple only to $SU(2)$ doublets of fermions. Thus only right-handed mirror fermions couple to the $W^\pm$, as opposed to $\left(\mathcal{L}_{F^{SM}}\right)_{int}$, where only left-handed SM fermions interact with the $W^\pm$ bosons. Similarly the three-point couplings of the right-handed mirror fermions with $Z$ and $\gamma$ bosons at the tree-level are same as those for the left-handed SM fermions. Hence, $\left(\mathcal{L}_{F^M}\right)_{int}$, is given by
	\begin{flalign}\label{eq:lintm}
		&\left(\mathcal{L}_{F^M}\right)_{int} = \;\frac{g}{\sqrt{2}}\left[\left(\;\overline{u}_R^{Mi} \;\gamma^\mu d_{Ri}^M + \overline{\nu}_{R}^{i} \;\gamma^\mu e_{Ri}^M\;\right) W_\mu^+\right.&\nonumber\\[3mm]
		&\left.+\;\left(\;\overline{d}_R^{M\;i} \;\gamma^\mu u_{R\;i}^M + \overline{e}_{R}^{M\;i} \;\gamma^\mu \nu_{R\;i}^M\;\right) W_\mu^-\right]&\nonumber\\[3mm]
		&+\frac{g}{c_W} \left[\sum_{f^M=\;u^M,d^M, \nu^M, e^M}\left(T_3^{f^M}- s_W^2 Q_{f^M}\right) \overline{f}_R^{M\;i}\; \gamma^\mu f_{R\;i}^M \right.&\nonumber\\[3mm]
		&\left. -\sum_{f^M=\;u^M,d^M, e^M}\; s_W^2 Q_{f^M}\; \overline{f}_L^{M\;i}\; \gamma^\mu f_{L\;i}^M\; \right] Z_\mu&\nonumber\\[3mm]
		&+ e\; \sum_{f^M=\;u^M,d^M, e^M}Q_{f^M} \left(\overline{f}_R^{M\;i}\; \gamma^\mu f_{R\;i}^M + \overline{f}_L^{M\;i}\; \gamma^\mu f_{L\;i}^M\;\right) A_\mu&
	\end{flalign}
  \begin{table*}[!htbp]
  \renewcommand{\arraystretch}{2}
  \centering
  \caption{\label{table:feyn21f}$f_1^M f_2^M V$ type couplings, which contribute to the Oblique Corrections. For each Feynman rule the charge conservation is implicit. $f_{1R}^M$ and $f_{2R}^M$ are members of the same mirror fermion doublet with isospins $\dfrac{1}{2}$ and $-\dfrac{1}{2}$ respectively (ref. \cite{pqnur}, \cite{PeskSchro} \& \cite{HaberMartin}). Common factor for all couplings: $\askImath g \gamma_\mu$}
  \bigskip
  \begin{tabular}{|lc|}
  	\hline
  	$g_{\overline{f}_{1R}^M f_{2R}^M W^+}$	& $\frac{1}{\sqrt{2}}$\\
	$g_{\overline{f}_R^M f_R^{M}Z}$			& $\frac{1}{c_W} (T_3^f - s_W^2 Q_f)$\\
	$g_{\overline{f}_R^M f_R^{M}\gamma}$		& $s_W Q_f$\\
	$g_{\overline{f}_L^M f_L^{M}Z}$			& $-\frac{s_W^2}{c_W}Q_f$\\
	$g_{\overline{f}_L^M f_L^{M}\gamma}$		& $s_W Q_f$\\[1mm]
  	\hline
  \end{tabular}
  \end{table*}
In equation (\ref{eq:lintm}) $i,j=1,2,3$, where $i$ denotes fermions in the $i^{th}$ mirror-quark or mirror-lepton generation. Sums over $i$ are implicit, when an index is contracted. ($u^M_i$ and $d^M_i$) denote the (up- and the down-) members of a mirror-quark generation respectively. Following a similar notation ($\nu_{Ri}$ and $e^M_i$) denote (the neutrino and the 'electron') members of a mirror-lepton generation respectively.
\par All the tree-level interactions calculated from equation (\ref{eq:lintm}) can be tabulated in a compact form as given in table (\ref{table:feyn21f}). Corresponding SM interactions can be similarly calculated from equation $\left(\mathcal{L}_{F^{SM}}\right)_{int}$.
%
\FloatBarrier
%
%
\section{Calculation of One Loop Contributions to Oblique Parameters in \ewnur model}
\label{sec:appst}
%
The one loop contributions to the oblique parameters in \ewnur model can be calculated from the cubic and quartic couplings listed in Appendix \ref{sec:appfeyns} and \ref{sec:appfeynf} and using the loop integral functions illustrated in Appendix \ref{sec:appfunc}. The SM loop diagrams contributing to $S$, $T$, $U$ can be similarly obtained from the SM cubic and quartic couplings in equations (\ref{eq:lkinsm}), $\left(\mathcal{L}_{F^{SM}}\right)_{int}$ and using loop integrals from Appendix \ref{sec:appfunc}. Hereafter, the focus of calculations will be on $S$ and $T$ parameters. The new Physics contributions to $S$ from the scalar sector and mirror fermion sector in \ewnur model will be calculated separately and then added to find the total contribution $\widetilde{S}$ (Eq. (\ref{eq:s})). Similary procedure will be followed to calculate $\widetilde{T}$ (Eq. (\ref{eq:t})). Thus, as in eqns. (\ref{eq:sss}), (\ref{eq:ttt}),
	\begin{eqnarray*}
		\widetilde{S} &=& \widetilde{S}_{scalar} + \widetilde{S}_{fermion},\\[1mm]
		\widetilde{T} &=& \widetilde{T}_{scalar} + \widetilde{T}_{fermion}.
	\end{eqnarray*}
Recall (Eq. (\ref{eq:s})) that the contributions to $\widetilde{S}$ come from $Z$ and $\gamma$ self-energies, $Z\gamma$ mixing, each calculated up to one-loop level. To evaluate $\widetilde{T}$ using equation (\ref{eq:t}) the isospin current $\Pi_{11}$ and electromagnetic current $\Pi_{33}$ are used. The $W$ and $Z$ self-energies are related to these isospin currents by \cite{PeskTak},
	\begin{flalign}\label{eq:13wz}
		\hspace{2em}\Pi_{WW} &= \frac{e^2}{s_W^2} \Pi_{11};&\nonumber\\[2mm]
		\hspace{2em}\Pi_{ZZ} &= \frac{e^2}{s_W^2 c_W^2} (\Pi_{33} - 2s_W^2\Pi_{3Q} + s_W^4\Pi_{QQ})&
	\end{flalign}
Using these relations $\Pi_{11}$ can be obtained from the loop contributions to $\Pi_{WW}$ listed in tables \ref{table:loopss}, \ref{table:loopsv}, \ref{table:loopgss}. From equation (\ref{eq:13wz}) the one-loop contributions to $\Pi_{33}$ can be obtained using $\displaystyle \lim_{g^\prime \rightarrow 0} (\Pi_{ZZ})$. These contributions are listed below, separately from $\Pi_{ZZ}$ for scalar as well as fermion sectors in \ewnur model.
%
%
\subsection{One Loop Contributions to $\widetilde{S}_{scalar}$ and $\widetilde{T}_{scalar}$}
\label{sec:appsts}
%
In this subsection the one-loop contributions to $\widetilde{S}_{scalar}$ and $\widetilde{T}_{scalar}$ are listed. In every table the loop contributions in \ewnur model are listed first and then the corresponding contributions in SM are also listed. The one-loop diagrams, which contribute to $\widetilde{S}_{scalar}$ can be found in tables \ref{table:loopss}, \ref{table:loopsv}, \ref{table:loopgss}, \ref{table:loopgsv}, \ref{table:loops}, \ref{table:loopgs} below. To calculate $\widetilde{T}_{scalar}$, $\Pi_{11}$ contributions from scalar sector in \ewnur model can be obtained from contributions to $\Pi_{WW}$ listed in tables \ref{table:loopss}, \ref{table:loopsv}, \ref{table:loopgss}. The scalar-loop diagrams contributing to $\Pi_{33}$ are listed in tables \ref{table:loopsst}, \ref{table:loopsvt}, \ref{table:loopst}.
  	\renewcommand{\arraystretch}{3}
	\begin{longtable*}{|p{1cm}p{1cm}p{4.5cm}||p{1cm}p{1cm}p{4.5cm}|}
	\caption{\label{table:loopss}One-loop diagrams with two internal scalar ($S$) (Higgs or Goldstone boson) lines, which contribute to $W^+$ and $Z$ self-energies. Common factor: $\dfrac{g^2}{16\pi^2}$}\\[3em]
	\hline
	\multicolumn{3}{|c||}{Contributions to $\Pi_{WW}(q^2)$}
	& \multicolumn{3}{c|}{Contributions to $\Pi_{ZZ}(q^2)$} \\
	\multicolumn{3}{|c||}%
	{
	\includegraphics{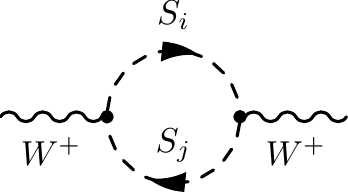}
	}
	&
	\multicolumn{3}{c|}%
	{
	\includegraphics{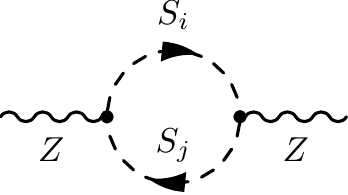}
	}\\
	\hline
	\multicolumn{1}{|l}{$S_i$}	& \multicolumn{1}{l}{$S_j$}	& \multicolumn{1}{c||}{~} &
	\multicolumn{1}{l}{$S_i$} & \multicolumn{1}{l}{$S_j$}	& \multicolumn{1}{c|}{~} \\
	\hline
	\endfirsthead
	\multicolumn{6}{c}%
	{{\tablename\ \thetable{} -- continued from previous page}} \\
	\hline
	\multicolumn{1}{|l}{$S_i$}	& \multicolumn{1}{l}{$S_j$}	& \multicolumn{1}{c||}{~} &
	\multicolumn{1}{l}{$S_i$} & \multicolumn{1}{l}{$S_j$}	& \multicolumn{1}{c|}{~} \\
	\hline
	\endhead
	\hline \multicolumn{6}{|c|}{{Continued on next page...}} \\ \hline
	\endfoot
	\hline
	\endlastfoot
	$H_5^+$		& $H_5^0$			& $3 B_{22}(q^2;\;m_{H_5^+}^2,\;m_{H_5^0}^2)$
	& $H_5^+$	& $H_5^+$			& $\dfrac{c_{2W}^2}{c_W^2} B_{22}(q^2;\;m_{H_5^+}^2,\;m_{H_5^+}^2)$ \\
	$H_5^{++}$	& $H_5^+$			& $2 B_{22}(q^2;\;m_{H_5^{++}}^2,\;m_{H_5^+}^2)$
	& $H_5^{++}$	& $H_5^{++}$			& $4\dfrac{c_{2W}^2}{c_W^2} B_{22}(q^2;\;m_{H_5^{++}}^2,\;m_{H_5^{++}}^2)$ \\
	$H_3^+$		& $H_3^0$			& $B_{22}(q^2;\;m_{H_3^+}^2,\;m_{H_3^0}^2)$
	& $H_3^+$	& $H_3^+$			& $\dfrac{c_{2W}^2}{c_W^2} B_{22}(q^2;\;m_{H_3^+}^2,\;m_{H_3^+}^2)$ \\
	$H_5^{++}$	& $H_3^+$			& $2 c_H^2 B_{22}(q^2;\;m_{H_5^{++}}^2,\;m_{H_3^+}^2)$
	& $H_5^+$	& $H_3^+$			& $\dfrac{c_H^2}{c_W^2} B_{22}(q^2;\;m_{H_5^+}^2,\;m_{H_3^+}^2)$ \\
	$H_5^+$		& $H_3^0$ 			& $c_H^2 B_{22}(q^2;\;m_{H_5^+}^2,\;m_{H_3^0}^2)$
	& $H_5^-$	& $H_3^-$			& $\dfrac{c_H^2}{c_W^2} B_{22}(q^2;\;m_{H_5^+}^2,\;m_{H_3^+}^2)$ \\
	$H_5^0$		& $H_3^+$ 			& $\dfrac{c_H^2}{3} B_{22}(q^2;\;m_{H_5^0}^2,\;m_{H_3^+}^2)$
	& $H_5^0$	& $H_3^0$			& $\dfrac{4}{3}\;\dfrac{c_H^2}{c_W^2} B_{22}(q^2;\;m_{H_5^0}^2,\;m_{H_3^0}^2)$ \\
  	$H_3^+$		& $H_1^0$			& $s_H^2 B_{22}(q^2;\;m_{H_3^+}^2,\;m_{H_1}^2)$
	& $H_3^0$	& $H_1^0$			& $\dfrac{s_H^2}{c_W^2} B_{22}(q^2;\;m_{H_3^0}^2,\;m_{H_1}^2)$ \\
	$H_3^+$		& $H_1^{0\prime}$	& $\dfrac{8}{3}\;c_H^2 B_{22}(q^2;\;m_{H_3^+}^2,\;m_{H_1^\prime}^2)$
	& $H_3^0$	& $H_1^{0\prime}$	& $\dfrac{8}{3}\;\dfrac{c_H^2}{c_W^2} B_{22}(q^2;\;m_{H_3^0}^2,\;m_{H_1^\prime}^2)$ \\
	$G_3^+$		& $H_5^{++}$			& $2 s_H^2 B_{22}(q^2;\;M_W^2,\;m_{H_5^{++}}^2)$
	& $G_3^+$	& $H_5^+$			& $\dfrac{s_H^2}{c_W^2} B_{22}(q^2;\;M_W^2,\;m_{H_5^+}^2)$ \\
	$G_3^0$		& $H_5^+$			& $s_H^2 B_{22}(q^2;\;M_Z^2,\;m_{H_5^+}^2)$
	& $G_3^-$	& $H_5^-$			& $\dfrac{s_H^2}{c_W^2} B_{22}(q^2;\;M_W^2,\;m_{H_5^+}^2)$ \\
	$G_3^+$		& $H_5^0$			& $\dfrac{s_H^2}{3} B_{22}(q^2;\;M_W^2,\;m_{H_5^0}^2)$
	& $G_3^0$	& $H_5^0$			& $\dfrac{4}{3}\;\dfrac{s_H^2}{c_W^2} B_{22}(q^2;\;M_Z^2,\;m_{H_5^0}^2)$ \\
	$G_3^+$		& $H_1^0$			& $c_H^2 B_{22}(q^2;\;M_W^2,\;m_{H_1}^2)$
	& $G_3^0$	& $H_1^0$			& $\dfrac{c_H^2}{c_W^2} B_{22}(q^2;\;M_Z^2,\;m_{H_1}^2)$ \\	
	$G_3^+$		& $H_1^{0\prime}$	& $\dfrac{8}{3}\;s_H^2 B_{22}(q^2;\;M_W^2,\;m_{H_1^\prime}^2)$
	& $G_3^0$	& $H_1^{0\prime}$	& $\dfrac{8}{3}\;\dfrac{s_H^2}{c_W^2} B_{22}(q^2;\;M_Z^2,\;m_{H_1^\prime}^2)$ \\
	$G_3^+$		& $G_3^0$			& $B_{22}(q^2;\;M_W^2,\;M_Z^2)$
	& $G_3^+$	& $G_3^+$			& $\dfrac{c_{2W}^2}{c_W^2} B_{22}(q^2;\;M_W^2,\;M_W^2)$ \\
	\hline
	\multicolumn{6}{|c|}{Standard Model contributions} \\
	\hline
	$H$			& $G_{SM}^+$			& $B_{22}(q^2;\;M_W^2,\;m_H^2)$
	& $H$		& $G_{SM}^0$			& $\dfrac{1}{c_W^2} B_{22}(q^2;\;M_Z^2,\;m_H^2)$ \\
	$G_{SM}^+$	& $G_{SM}^0$			& $B_{22}(q^2;\;M_W^2,\;M_Z^2)$
	& $G_{SM}^+$	& $G_{SM}^+$			& $\dfrac{c_{2W}^2}{c_W^2} B_{22}(q^2;\;M_W^2,\;M_W^2)$ \\
	\end{longtable*}
  	\renewcommand{\arraystretch}{3}
	\begin{longtable*}{|p{1.2cm}p{4cm}||p{1.2cm}p{4cm}|}
	\caption{\label{table:loops}Tadpole diagrams with one internal scalar ($S$) (Higgs or Goldstone boson) line, which contribute to $W^+$ and $Z$ self-energies.
	Common factor: $\dfrac{g^2}{16\pi^2}$}\\[3em]
	\hline
	\multicolumn{2}{|c||}{Contributions to $\Pi_{WW}(q^2)$}
	& \multicolumn{2}{c|}{Contributions to $\Pi_{ZZ}(q^2)$} \\
	\multicolumn{2}{|c||}%
	{
	\includegraphics{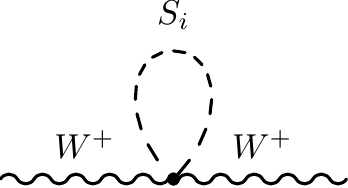}
	}
	&
	\multicolumn{2}{c|}%
	{
	\includegraphics{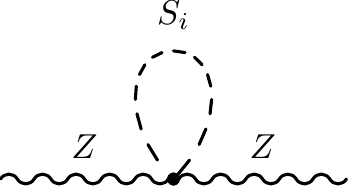}
	}\\
	\hline
	\multicolumn{1}{|l}{$S_i$}	& \multicolumn{1}{c||}{~} &
	\multicolumn{1}{l}{$S_i$} & \multicolumn{1}{c|}{~} \\
	\hline
	\endfirsthead
	\multicolumn{4}{c}%
	{{\tablename\ \thetable{} -- continued from previous page}} \\
	\hline
	\multicolumn{1}{|l}{$S_i$}	& \multicolumn{1}{c||}{~} &
	\multicolumn{1}{l}{$S_i$} & \multicolumn{1}{c|}{~} \\
	\hline
	\endhead
	\hline \multicolumn{4}{|c|}{{Continued on next page...}} \\ \hline
	\endfoot
	\hline
	\endlastfoot
	$H_5^0$			& $-\dfrac{5}{6} A_0(m_{H_5^0}^2)$
	& $H_5^0$		& $-\dfrac{2}{6 c_W^2} A_0(m_{H_5^0}^2)$ \\
	$H_5^+$			& $-\dfrac{3}{2} A_0(m_{H_5^+}^2)$
	& $H_5^+$		& $-\dfrac{c_W^4 + s_W^4}{c_W^2} A_0(m_{H_5^+}^2)$ \\
	$H_5^{++}$		& $-A_0(m_{H_5^{++}}^2)$
	& $H_5^{++}$		& $-2 \dfrac{c_{2W}^2}{c_W^2}A_0(m_{H_5^{++}}^2)$ \\
	$H_3^0$			& $-\dfrac{1}{4}(1 + c_H^2) A_0(m_{H_3^0}^2)$
	& $H_3^0$		& $-\dfrac{1}{4 c_W^2}(1 + 3 c_H^2) A_0(m_{H_3^0}^2)$ \\
	$H_3^+$			& $-\dfrac{1}{2}(1 + 2 c_H^2) A_0(m_{H_3^+}^2)$
	& $H_3^+$		& $-\dfrac{c_{2W}^2}{2 c_W^2} (1 + c_H^2) A_0(m_{H_3^+}^2)$ \\
	$H_1^0$			& $-\dfrac{1}{4} A_0(m_{H_1}^2)$
	& $H_1^0$		& $-\dfrac{1}{4 c_W^2} A_0(m_{H_1}^2)$ \\
	$H_1^{0\prime}$	& $-\dfrac{2}{3} A_0(m_{H_1^\prime}^2)$
	& $H_1^{0\prime}$	& $-\dfrac{2}{3 c_W^2} A_0(m_{H_1^\prime}^2)$ \\
	$G_3^0$			& $-\dfrac{1}{4}(1 + s_H^2) A_0(M_Z^2)$
	& $G_3^0$		& $-\dfrac{1}{4 c_W^2}(1 + 3 s_H^2) A_0(M_Z^2)$ \\
	$G_3^+$			& $-\dfrac{1}{2}(1 + 2 s_H^2) A_0(M_W^2)$
	& $G_3^+$		& $-\dfrac{c_{2W}^2}{2 c_W^2} (1 + s_H^2) A_0(M_W^2)$ \\
	\hline
	\multicolumn{4}{|c|}{Standard Model contributions} \\
	\hline
	$H$				& $-\dfrac{1}{4} A_0(m_H^2)$
	& $H$			& $-\dfrac{1}{4 c_W^2} A_0(m_H^2)$ \\
	$G_{SM}^0$			& $-\dfrac{1}{4} A_0(M_Z^2)$
	& $G_{SM}^0$		& $-\dfrac{1}{4 c_W^2} A_0(M_Z^2)$ \\
	$G_{SM}^+$			& $-\dfrac{1}{2} A_0(M_W^2)$
	& $G_{SM}^+$		& $-\dfrac{c_{2W}^2}{2 c_W^2} A_0(M_W^2)$ \\
    \end{longtable*}
  \begin{table*}[!htb]
  	\renewcommand{\arraystretch}{3}
	\caption{\label{table:loopsv}One-loop diagrams with one internal scalar ($S$) (Higgs or Goldstone boson) line and one internal vector boson line, which contribute to $W^+$ and $Z$ self-energies. Common factor: $\dfrac{g^2}{16\pi^2}$}
	\begin{center}
	\begin{tabular}{|p{1.2cm}p{1.2cm}p{4.5cm}||p{1.2cm}p{1.2cm}p{4.5cm}|}
	\hline
	\multicolumn{3}{|c||}{Contributions to $\Pi_{WW}(q^2)$}
	& \multicolumn{3}{c|}{Contributions to $\Pi_{ZZ}(q^2)$} \\
	\multicolumn{3}{|c||}%
	{
	\includegraphics{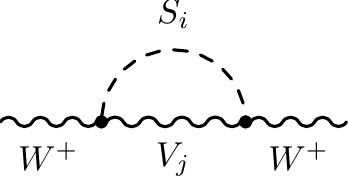}
	}
	&
	\multicolumn{3}{c|}%
	{
	\includegraphics{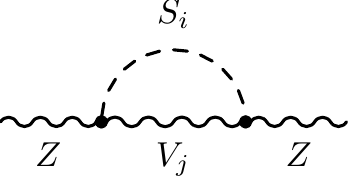}
	}\\
	\hline
	$S_i$ & $V_j$ & ~ & $S_i$ & $V_j$ & ~ \\
	\hline
	$H_5^0$			& $W^+$		& $-\dfrac{s_H^2}{3} M_W^2 B_0(q^2;\;M_W^2,\;m_{H_5^0}^2)$
	& $H_5^0$		& $Z$		& $-\dfrac{4}{3}\;\dfrac{s_H^2}{c_W^2} M_Z^2 B_0(q^2;\;M_Z^2,\;m_{H_5^0}^2)$ \\
	$H_1^0$			& $W^+$		& $-c_H^2 M_W^2 B_0(q^2;\;M_W^2,\;m_{H_1}^2)$	
	& $H_1^0$		& $Z$		& $-\dfrac{c_H^2}{c_W^2} M_Z^2 B_0(q^2;\;M_Z^2,\;m_{H_1}^2)$ \\
	$H_1^{0\prime}$	& $W^+$		& $-\dfrac{8}{3}\;s_H^2 M_W^2 B_0(q^2;\;M_W^2,\;m_{H_1^\prime}^2)$
	& $H_1^{0\prime}$	& $Z$	& $-\dfrac{8}{3}\;\dfrac{s_H^2}{c_W^2} M_Z^2 B_0(q^2;\;M_Z^2,\;m_{H_1^\prime}^2)$ \\
	$H_5^+$			& $Z$		& $-\dfrac{s_H^2}{c_W^2} M_W^2 B_0(q^2;\;M_Z^2,\;m_{H_5^+}^2)$
	& $H_5^+$		& $W^-$		& $-\dfrac{s_H^2}{c_W^2} M_W^2 B_0(q^2;\;M_W^2,\;m_{H_5^+}^2)$ \\
	$H_5^{++}$		& $W^-$		& $-2 s_H^2 M_W^2 B_0(q^2;\;M_W^2,\;m_{H_5^{++}}^2)$
	& $H_5^-$		& $W^+$		& $-\dfrac{s_H^2}{c_W^2} M_W^2 B_0(q^2;\;M_W^2,\;m_{H_5^+}^2)$ \\
	\hline
	\multicolumn{6}{|c|}{Standard Model contributions} \\
	\hline
	$H$				& $W^+$		& $-M_W^2 B_0(q^2;\;M_W^2,\;m_H^2)$
	& $H$			& $Z$		& $-\dfrac{M_Z^2}{c_W^2} B_0(q^2;\;M_Z^2,\;m_H^2)$ \\
	$G_{SM}^+$			& $Z$		& $-\dfrac{s_W^4}{c_W^2} M_W^2 B_0(q^2;\;M_Z^2,\;M_W^2)$
	& $G_{SM}^+$		& $W^-$		& $-2 \dfrac{s_W^4}{c_W^2} M_W^2 B_0(q^2;\;M_W^2,\;M_W^2)$ \\
	$G_{SM}^+$			& $\gamma$	& $-s_W^2 M_W^2 B_0(q^2;\;0,\;M_W^2)$
	& $G_{SM}^-$		& $W^+$		& $-2 \dfrac{s_W^4}{c_W^2} M_W^2 B_0(q^2;\;M_W^2,\;M_W^2)$ \\
	\hline
	\end{tabular}
	\end{center}
  \end{table*}
  \begin{table*}[!htb]
  	\renewcommand{\arraystretch}{3}
	\caption{\label{table:loopgss}One-loop diagrams with two internal scalar ($S$) (Higgs or Goldstone boson) lines, which contribute to photon ($\gamma$) self-energy and $Z-\gamma$ transition amplitude. Common factor: $\dfrac{g^2}{16\pi^2}$}
	\begin{center}
	\begin{tabular}{|p{1.2cm}p{1.2cm}p{5cm}||p{1.2cm}p{1.2cm}p{5cm}|}
	\hline
	\multicolumn{3}{|c||}{Contributions to $\Pi_{\gamma\gamma}(q^2)$}
	& \multicolumn{3}{c|}{Contributions to $\Pi_{Z\gamma}(q^2)$} \\
	\multicolumn{3}{|c||}%
	{
	\includegraphics{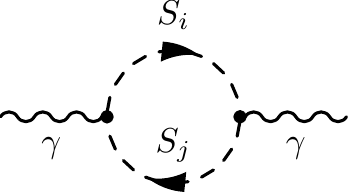}
	}
	&
	\multicolumn{3}{c|}%
	{
	\includegraphics{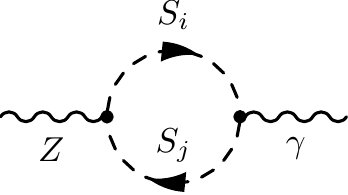}
	}\\
	\hline
	$S_i$	& $S_j$	& ~ & $S_i$	& $S_j$	& ~ \\
	\hline
	$H_5^+$			& $H_5^+$		& $4 s_W^2 B_{22}(q^2;\;m_{H_5^+}^2,\;m_{H_5^+}^2)$
	& $H_5^+$		& $H_5^+$		& $2 \dfrac{s_W}{c_W} c_{2W} B_{22}(q^2;\;m_{H_5^+}^2,\;m_{H_5^+}^2)$ \\
	$H_5^{++}$		& $H_5^{++}$		& $16 s_W^2 B_{22}(q^2;\;m_{H_5^{++}}^2,\;m_{H_5^{++}}^2)$
	& $H_5^{++}$		& $H_5^{++}$		& $8 \dfrac{s_W}{c_W} c_{2W} B_{22}(q^2;\;m_{H_5^{++}}^2,\;m_{H_5^{++}}^2)$ \\
	$H_3^+$			& $H_3^+$		& $4 s_W^2 B_{22}(q^2;\;m_{H_3^+}^2,\;m_{H_3^+}^2)$
	& $H_3^+$		& $H_3^+$		& $2 \dfrac{s_W}{c_W} c_{2W} B_{22}(q^2;\;m_{H_3^+}^2,\;m_{H_3^+}^2)$ \\
	$G_3^+$			& $G_3^+$		& $4 s_W^2 B_{22}(q^2;\;M_W^2,\;M_W^2)$
	& $G_3^+$		& $G_3^+$		& $2 \dfrac{s_W}{c_W} c_{2W} B_{22}(q^2;\;M_W^2,\;M_W^2)$ \\
	\hline
	\multicolumn{6}{|c|}{Standard Model contributions} \\
	\hline
	$G_{SM}^+$			& $G_{SM}^+$		& $4 s_W^2 B_{22}(q^2;\;M_W^2,\;M_W^2)$
	& $G_{SM}^+$		& $G_{SM}^+$		& $2 \dfrac{s_W}{c_W} c_{2W} B_{22}(q^2;\;M_W^2,\;M_W^2)$ \\	
	\hline
	\end{tabular}
	\end{center}
  \end{table*}
  \begin{table*}[!htb]
  	\renewcommand{\arraystretch}{3}
	\caption{\label{table:loopgs}Tadpole diagrams with one internal scalar ($S$) (Higgs or Goldstone boson) line, which contribute to photon ($\gamma$) self-energy and $Z-\gamma$ transition amplitude. Common factor: $\dfrac{g^2}{16\pi^2}$}
	\begin{center}
	\begin{tabular}{|p{1.2cm}p{3.5cm}||p{1.2cm}p{3.5cm}|}
	\hline
	\multicolumn{2}{|c||}{Contributions to $\Pi_{\gamma\gamma}(q^2)$}
	& \multicolumn{2}{c|}{Contributions to $\Pi_{Z\gamma}(q^2)$} \\
	\multicolumn{2}{|c||}%
	{
	\includegraphics{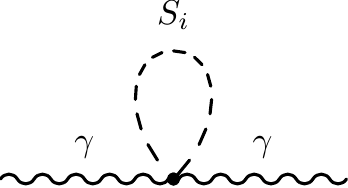}
	}
	&
	\multicolumn{2}{c|}%
	{
	\includegraphics{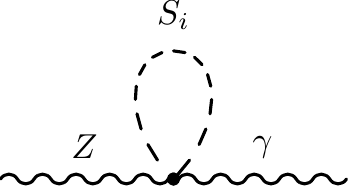}
	}\\
	\hline
	$S_i$	& ~ & $S_i$	& ~ \\
	\hline
	$H_5^+$			& $-2 s_W^2 A_0(m_{H_5^+}^2)$
	& $H_5^+$		& $-\dfrac{s_W}{c_W} c_{2W} A_0(m_{H_5^+}^2)$ \\
	$H_5^{++}$		& $-8 s_W^2 A_0(m_{H_5^{++}}^2)$
	& $H_5^{++}$		& $-4 \dfrac{s_W}{c_W} c_{2W} A_0(m_{H_5^{++}}^2)$ \\
	$H_3^+$			& $-2 s_W^2 A_0(m_{H_3^+}^2)$
	& $H_3^+$		& $-\dfrac{s_W}{c_W} c_{2W} A_0(m_{H_3^+}^2)$ \\
	$G_3^+$			& $-2 s_W^2 A_0(M_W^2)$
	& $G_3^+$		& $-\dfrac{s_W}{c_W} c_{2W} A_0(M_W^2)$ \\
	\hline
	\multicolumn{4}{|c|}{Standard Model contributions} \\
	\hline
	$G_{SM}^+$			& $-2 s_W^2 A_0(M_W^2)$
	& $G_{SM}^+$		& $-\dfrac{s_W}{c_W} c_{2W} A_0(M_W^2)$ \\
	\hline
	\end{tabular}
	\end{center}
  \end{table*}
  \begin{table*}[!htb]
  	\renewcommand{\arraystretch}{3}
	\caption{\label{table:loopgsv}One-loop diagrams with one internal scalar ($S$) (Higgs or Goldstone boson) line and one internal vector boson line, which contribute to photon ($\gamma$) self-energy and $Z-\gamma$ transition amplitude. Common factor: $\dfrac{g^2}{16\pi^2}$}
	\begin{center}
	\begin{tabular}{|p{1.2cm}p{1.2cm}p{4.5cm}||p{1.2cm}p{1.2cm}p{4.5cm}|}
	\hline
	\multicolumn{3}{|c||}{Contributions to $\Pi_{\gamma\gamma}(q^2)$}
	& \multicolumn{3}{c|}{Contributions to $\Pi_{Z\gamma}(q^2)$} \\
	\multicolumn{3}{|c||}%
	{
	\includegraphics{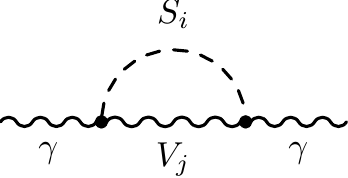}
	}
	&
	\multicolumn{3}{c|}%
	{
	\includegraphics{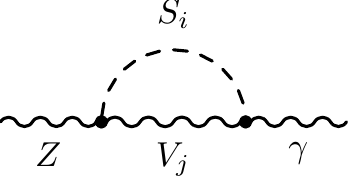}
	}\\
	\hline
	$S_i$ & $V_j$ & ~ & $S_i$ & $V_j$ & ~ \\
	\hline
	$G_3^+$			& $W^-$			& $-s_W^2 M_W^2 B_0(M_Z^2;\;M_W^2,\;M_W^2)$
	& $G_3^+$		& $W^-$			& $\dfrac{s_W^3}{c_W} M_W^2 B_0(M_Z^2;\;M_W^2,\;M_W^2)$ \\
	$G_3^-$			& $W^+$			& $-s_W^2 M_W^2 B_0(M_Z^2;\;M_W^2,\;M_W^2)$
	& $G_3^-$		& $W^+$			& $\dfrac{s_W^3}{c_W} M_W^2 B_0(M_Z^2;\;M_W^2,\;M_W^2)$ \\
	\hline
	\multicolumn{6}{|c|}{Standard Model contributions} \\
	\hline
	$G_{SM}^+$			& $W^-$			& $-s_W^2 M_W^2 B_0(M_Z^2;\;M_W^2,\;M_W^2)$
	& $G_{SM}^+$		& $W^-$			& $\dfrac{s_W^3}{c_W} M_W^2 B_0(M_Z^2;\;M_W^2,\;M_W^2)$ \\
	$G_{SM}^-$			& $W^+$			& $-s_W^2 M_W^2 B_0(M_Z^2;\;M_W^2,\;M_W^2)$
	& $G_{SM}^-$		& $W^+$			& $\dfrac{s_W^3}{c_W} M_W^2 B_0(M_Z^2;\;M_W^2,\;M_W^2)$ \\
	\hline
	\end{tabular}
	\end{center}
  \end{table*}
  	\begin{table*}[!htb]
	\renewcommand{\arraystretch}{3}
	\caption{\label{table:loopsst}One-loop diagrams with two internal scalar ($S$) (Higgs or Goldstone boson) lines, which contribute to $\Pi_{33}(q^2)$ in $T$.
	Common factor: $\dfrac{g^2}{16\pi^2}$}
	\begin{center}
	\begin{tabular}{|p{1cm}p{1cm}p{4cm}||p{1cm}p{1cm}p{4cm}|}
	\hline
	\multicolumn{6}{|c|}{Contributions to $\Pi_{33}(q^2)$} \\
	\hline
	\multicolumn{6}{|c|}%
	{
	\raisebox{5ex}{$\lim_{g^\prime \rightarrow 0}$} \includegraphics{loopzssz.pdf}}\\
	\hline
	\multicolumn{1}{|l}{$S_i$} & \multicolumn{1}{l}{$S_j$}	& \multicolumn{1}{c||}{~} &
	\multicolumn{1}{l}{$S_i$} & \multicolumn{1}{l}{$S_j$}	& \multicolumn{1}{c|}{~} \\
	\hline
	$H_5^+$			& $H_5^+$			& $B_{22}(q^2;\;m_{H_5^+}^2,\;m_{H_5^+}^2)$ &
	$H_5^{++}$		& $H_5^{++}$			& $4 B_{22}(q^2;\;m_{H_5^{++}}^2,\;m_{H_5^{++}}^2)$ \\
	$H_3^+$			& $H_3^+$			& $B_{22}(q^2;\;m_{H_3^+}^2,\;m_{H_3^+}^2)$ &
	$H_5^+$			& $H_3^+$			& $c_H^2 B_{22}(q^2;\;m_{H_5^+}^2,\;m_{H_3^+}^2)$ \\
	$H_5^-$			& $H_3^-$			& $c_H^2 B_{22}(q^2;\;m_{H_5^+}^2,\;m_{H_3^+}^2)$ &
	$H_5^0$			& $H_3^0$			& $\dfrac{4}{3}\;c_H^2 B_{22}(q^2;\;m_{H_5^0}^2,\;m_{H_3^0}^2)$ \\
  	$H_3^0$			& $H_1^0$			& $s_H^2 B_{22}(q^2;\;m_{H_3^0}^2,\;m_{H_1}^2)$ &
	$H_3^0$			& $H_1^{0\prime}$	& $\dfrac{8}{3}\;c_H^2 B_{22}(q^2;\;m_{H_3^0}^2,\;m_{H_1^\prime}^2)$ \\
	$G_3^+$			& $H_5^+$			& $s_H^2 B_{22}(q^2;\;M_W^2,\;m_{H_5^+}^2)$ &
	$G_3^-$			& $H_5^-$			& $s_H^2 B_{22}(q^2;\;M_W^2,\;m_{H_5^+}^2)$ \\
	$G_3^0$			& $H_5^0$			& $\dfrac{4}{3}\;s_H^2 B_{22}(q^2;\;M_W^2,\;m_{H_5^0}^2)$ &
	$G_3^0$			& $H_1^0$			& $c_H^2 B_{22}(q^2;\;M_W^2,\;m_{H_1}^2)$ \\	
	$G_3^0$			& $H_1^{0\prime}$	& $\dfrac{8}{3}\;s_H^2 B_{22}(q^2;\;M_W^2,\;m_{H_1^\prime}^2)$ &
	$G_3^+$			& $G_3^+$			& $B_{22}(q^2;\;M_W^2,\;M_W^2)$ \\
	\hline
	\multicolumn{6}{|c|}{Standard Model contributions} \\
	\hline
	$H$				& $G_{SM}^0$			& $B_{22}(q^2;\;M_W^2,\;m_H^2)$ &
	$G_{SM}^+$		& $G_{SM}^+$			& $B_{22}(q^2;\;M_W^2,\;M_W^2)$ \\
	\hline
	\end{tabular}
	\end{center}
	\end{table*}
	\begin{table*}[!htb]
  	\renewcommand{\arraystretch}{3}
	\caption{\label{table:loopst}Tadpole diagrams with one internal scalar ($S$) (Higgs or Goldstone boson) line, which contribute to $\Pi_{33}(q^2)$ in $T$.
	Common factor: $\dfrac{g^2}{16\pi^2}$}
	\begin{center}
	\begin{tabular}{|p{1.2cm}p{3.5cm}|p{1.2cm}p{3.5cm}|}
	\hline
	\multicolumn{4}{|c|}{Contributions to $\Pi_{33}(q^2)$} \\
	\hline
	\multicolumn{4}{|c|}%
	{
	\raisebox{5ex}{$\lim_{g^\prime \rightarrow 0}$} \includegraphics{tadpolezsz.pdf}}\\
	\hline
	$S_i$ & ~ & $S_i$ & ~\\
	\hline
	$H_5^0$		& $-\dfrac{2}{6} A_0(m_{H_5^0}^2)$ &
	$H_5^+$		& $-A_0(m_{H_5^+}^2)$ \\
	$H_5^{++}$	& $-2 A_0(m_{H_5^{++}}^2)$ &
	$H_3^0$		& $-\dfrac{1}{4}(1 + 3 c_H^2) A_0(m_{H_3^0}^2)$ \\
	$H_3^+$		& $-\dfrac{1}{2}(1 + c_H^2) A_0(m_{H_3^+}^2)$ &
	$H_1^0$		& $-\dfrac{1}{4} A_0(m_{H_1}^2)$ \\
	$H_1^{0\prime}$	& $-\dfrac{2}{3} A_0(m_{H_1^\prime}^2)$ &
	$G_3^0$		& $-\dfrac{1}{4}(1 + 3 s_H^2) A_0(M_W^2)$ \\
	$G_3^+$		& $-\dfrac{1}{2}(1 + s_H^2) A_0(M_W^2)$ & ~ & ~\\
	\hline
	\multicolumn{4}{|c|}{Standard Model contributions} \\
	\hline
	$H$			& $-\dfrac{1}{4} A_0(m_H^2)$ &
	$G_{SM}^0$	& $-\dfrac{1}{4} A_0(M_W^2)$ \\
	$G_{SM}^+$	& $-\dfrac{1}{2} A_0(M_W^2)$ & ~ & ~\\
	\hline
	\end{tabular}
	\end{center}
	\end{table*}
    \begin{table*}[!htb]
  	\renewcommand{\arraystretch}{3}
	\caption{\label{table:loopsvt}One-loop diagrams with one internal scalar ($S$) (Higgs or Goldstone boson) line and one internal vector boson line, which contribute to $\Pi_{33}(q^2)$ in $T$. Common factor: $\dfrac{g^2}{16\pi^2}$}
	\begin{center}
	\begin{tabular}{|p{1cm}p{1cm}p{5cm}|p{1cm}p{1cm}p{5cm}|}
	\hline
	\multicolumn{6}{|c|}{Contributions to $\Pi_{33}(q^2)$} \\
	\multicolumn{6}{|c|}%
	{
	\raisebox{5ex}{$\lim_{g^\prime \rightarrow 0}$} \includegraphics{loopzsvz.pdf}}\\
	\hline
	$S_i$ & $V_j$ & ~ & $S_i$ & $V_j$ & ~ \\
	\hline
	$H_5^0$			& $Z$		& $-\dfrac{4}{3}\;s_H^2 M_W^2 B_0(q^2;\;M_W^2,\;m_{H_5^0}^2)$ &
	$H_1^0$			& $Z$		& $-c_H^2 M_W^2 B_0(q^2;\;M_W^2,\;m_{H_1}^2)$ \\
	$H_1^{0\prime}$	& $Z$		& $-\dfrac{8}{3}\;s_H^2 M_W^2 B_0(q^2;\;M_W^2,\;m_{H_1^\prime}^2)$ &
	$H_5^+$			& $W^-$		& $-s_H^2 M_W^2 B_0(q^2;\;M_W^2,\;m_{H_5^+}^2)$ \\
	$H_5^-$			& $W^+$		& $-s_H^2 M_W^2 B_0(q^2;\;M_W^2,\;m_{H_5^+}^2)$ &~&~&~\\
	\hline
	\multicolumn{6}{|c|}{Standard Model contributions} \\
	\hline
	$H$				& $Z$		& $-M_W^2 B_0(q^2;\;M_W^2,\;m_H^2)$ &~&~&~\\
	\hline
	\end{tabular}
	\end{center}
  \end{table*}
\FloatBarrier
Using tables \ref{table:loopss}, \ref{table:loops}, \ref{table:loopsv}, \ref{table:loopgss}, \ref{table:loopgs}, \ref{table:loopgsv} above and Eq. (\ref{eq:s}) the new Physics contribution, $\widetilde{S}_{scalar}$ is given by (as in Eq. (\ref{eq:expss}))
  \begin{widetext}
  \begin{flalign}\label{eq:expss}
  	\widetilde{S}_{scalar} =\;& S_{scalar}^{EW\nu_R} - S_{scalar}^{SM}&\nonumber \\[3mm]
  	=\; &\dfrac{1}{M_Z^2 \pi} \Bigg\{\;
  	\frac{4}{3}\; s_H^2 \left[\; \askBbar_{22}(M_Z^2; M_Z^2, m_{H_5^0}^2) - M_Z^2\; \askBbar_0(M_Z^2; M_Z^2, m_{H_5^0}^2)\;\right]
  	+ 2\; s_H^2 \left[\; \askBbar_{22}(M_Z^2; M_Z^2, m_{H_5^+}^2) \right.\nonumber&\\[3mm]
  	&\left. -\; M_W^2\; \askBbar_0(M_Z^2; M_Z^2, m_{H_5^+}^2)\;\right] + c_H^2 \left[\; \askBbar_{22}(M_Z^2; M_Z^2, m_{H_1}^2) - M_Z^2\; \askBbar_0(M_Z^2; M_Z^2, m_{H_1}^2)\;\right] \nonumber&\\[3mm]
  	&+\; \frac{8}{3}\; s_H^2 \left[\; \askBbar_{22}(M_Z^2; M_Z^2, m_{H_1^\prime}^2) - M_Z^2\; \askBbar_0(M_Z^2; M_Z^2, m_{H_1^\prime}^2)\;\right] + \frac{4}{3} c_H^2\; \askBbar_{22}(M_Z^2; m_{H_5^0}^2, m_{H_3^0}^2) &\nonumber\\[3mm]
  	&+ 2\; c_H^2\; \askBbar_{22}(M_Z^2; m_{H_5^+}^2, m_{H_3^+}^2) + s_H^2\; \askBbar_{22}(M_Z^2; m_{H_3^0}^2, m_{H_1}^2) + \frac{8}{3}\; c_H^2\; \askBbar_{22}(M_Z^2; m_{H_3^0}^2, m_{H_1^\prime}^2) \nonumber&\\[3mm]
  	&-\; 4\; \askBbar_{22}(M_Z^2; m_{H_5^{++}}^2, m_{H_5^{++}}^2) -\; \askBbar_{22}(M_Z^2; m_{H_5^+}^2, m_{H_5^+}^2)
  	- \;\askBbar_{22}(M_Z^2; m_{H_3^+}^2, m_{H_3^+}^2) \nonumber&\\[3mm]
  	&-\; \frac{}{}\left[\askBbar_{22}(M_Z^2; M_Z^2, m_H^2) - M_Z^2\; \askBbar_0(M_Z^2; M_Z^2, m_H^2)\right]\Bigg\},&
  \end{flalign}
	\end{widetext}
\par Although it is not apparent from Eq. (\ref{eq:expss}), $\widetilde{S}_{scalar}$ decreases with increasing mass-splitting within a $SU(2)_D$ scalar multiplet and between two $SU(2)_D$ scalar singlets of \ewnur model. For large enough splitting(s) it becomes negative, which is desired to compensate the large positive contribution from mirror fermions (refer to section \ref{numerical}).
\par To obtain $\widetilde{T}_{scalar}$ the contributions to $\Pi_{11}$ in Eq. (\ref{eq:t}) are obtained from the $\Pi_{WW}$ contributions in tables \ref{table:loopss}, \ref{table:loops}, \ref{table:loopsv}, \ref{table:loopgss}, \ref{table:loopgs}, \ref{table:loopgsv} and using Eq. (\ref{eq:13wz}). The corresponding $\Pi_{ZZ}$ contributions are obtained using Eq. (\ref{eq:13wz}) and tables \ref{table:loopsst}, \ref{table:loopst}, \ref{table:loopsvt}. Thus, we get
	\begin{widetext}
	\begin{flalign}
		\widetilde{T}_{scalar} =\;& T_{scalar}^{EW\nu_R} - T_{scalar}^{SM}&\nonumber \\[3mm]
		=\;&\dfrac{1}{4 \pi s_W^2 M_W^2} \Bigg\{ 2\; B_{22}(0;m_{H_5^{++}}^2;m_{H_5^+}^2) + 3\; B_{22}(0;m_{H_5^+}^2;m_{H_5^0}^2) + B_{22}(0;m_{H_3^+}^2;m_{H_3^0}^2)+ c_H^2\; \Big[\; 2\; B_{22}(0;m_{H_5^{++}}^2;m_{H_3^+}^2) &\nonumber \\[3mm]
		& + B_{22}(0;m_{H_5^+}^2;m_{H_3^0}^2) + \frac{1}{3}\; B_{22}(0;m_{H_3^+}^2;m_{H_5^0}^2) + \frac{8}{3}\; B_{22}(0;m_{H_3^+}^2;m_{H_1^\prime}^2) - \frac{8}{3}\; B_{22}(0;m_{H_3^0}^2;m_{H_1^\prime}^2)\; \Big]&\nonumber \\[3mm]
		& + s_H^2\; \Big[\; 2\; B_{22}(0;M_W^2;m_{H_5^{++}}^2) - B_{22}(0;M_W^2;m_{H_5^+}^2) - B_{22}(0;M_W^2;m_{H_5^0}^2) + B_{22}(0;m_{H_3^+}^2;m_{H_1}^2) - B_{22}(0;m_{H_3^0}^2;m_{H_1}^2)&\nonumber \\[3mm]
		& + M_W^2\; \Big( B_0(0;M_W^2,m_{H_5^0}^2) + B_0(0;M_W^2,m_{H_5^+}^2) - 2\; B_0(0;M_W^2,m_{H_5^{++}}^2) \Big)\; \Big] &\nonumber \\[3mm]
		& + A_0(m_{H_5^{++}}^2) - \frac{1}{2}\; A_0(m_{H_5^+}^2) - \frac{1}{2}\; A_0(m_{H_5^0}^2) - \Big( \frac{1}{2} - s_H^2 \Big) A_0(m_{H_3^+}^2) + \frac{1}{2}\; A_0(m_{H_3^0}^2) - \Big( \frac{1}{2} + s_H^2 \Big) A_0(M_W^2)\;\Bigg\}&
	\end{flalign}
%
Hence, using Eq. (\ref{eq:B22FAA}), 
%
  \begin{flalign}\label{eq:expts}
  	\widetilde{T}_{scalar} =\;&\dfrac{1}{4 \pi s_W^2 M_W^2} \Bigg\{ \frac{1}{2} \mathcal{F}(m_{H_5^{++}}^2, m_{H_5^0}^2) + \frac{3}{4} \mathcal{F}(m_{H_5^+}^2, m_{H_5^0}^2) + \frac{1}{4} \mathcal{F}(m_{H_3^+}^2, m_{H_3^0}^2) + \frac{c_H^2}{2} \mathcal{F}(m_{H_5^{++}}^2, m_{H_3^+}^2)&\nonumber\\[3mm]
  	& +\; \frac{c_H^2}{4} \mathcal{F}(m_{H_5^+}^2, m_{H_3^0}^2) + \frac{c_H^2}{12} \mathcal{F}(m_{H_5^0}^2, m_{H_3^+}^2) - \frac{c_H^2}{2} \mathcal{F}(m_{H_5^+}^2, m_{H_3^+}^2) - \frac{c_H^2}{3} \mathcal{F}(m_{H_5^0}^2, m_{H_3^0}^2) &\nonumber\\[3mm]
  	&+\; \frac{s_H^2}{4} \left[ \mathcal{F}(m_{H_3^+}^2, m_{H_1}^2) - \mathcal{F}(m_{H_3^0}^2, m_{H_1}^2) \right] + \frac{2}{3}\; c_H^2 \left[ \mathcal{F}(m_{H_3^+}^2, m_{H_1^\prime}^2) - \mathcal{F}(m_{H_3^0}^2, m_{H_1^\prime}^2)\right]&\nonumber\\[3mm]
  	&+\;\frac{s_H^2}{2} \mathcal{F}(M_W^2, m_{H_5^{++}}^2) - \frac{s_H^2}{4} \mathcal{F}(M_W^2, m_{H_5^+}^2) - \frac{s_H^2}{4} \mathcal{F}(M_W^2, m_{H_5^0}^2) &\nonumber\\[3mm]
  	&+\;  \frac{}{} M_W^2 s_H^2 B_0 (0; M_W^2, m_{H_5^0}^2) + M_W^2 s_H^2 B_0 (0; M_W^2, m_{H_5^+}^2) - M_W^2 s_H^2 B_0 (0; M_W^2, m_{H_5^{++}}^2)  \Bigg\}\,.&
  \end{flalign}
  \end{widetext}
It should be noted that individual loop integral functions on RHS of eqns. (\ref{eq:expss}), (\ref{eq:expts}) do contain divergences by definition, but these divergences cancel as expected resulting in finite $\widetilde{S}_{scalar}$ and $\widetilde{T}_{scalar}$ respectively. Similar cancellations ensure that $\widetilde{S}_{lepton}$, $\widetilde{T}_{lepton}$ and $\widetilde{S}_{quark}$, $\widetilde{T}_{quark}$ are all separately finite.
%
%
\subsection{One Loop Contributions to $\widetilde{S}_{fermion}$ and $\widetilde{T}_{fermion}$}
\label{sec:appstf}
\begin{figure*}[!htb]
	\begin{flushleft}
		\subfigure[\hspace{13em}]	{
		\label{fig:loopexrr}
		\includegraphics{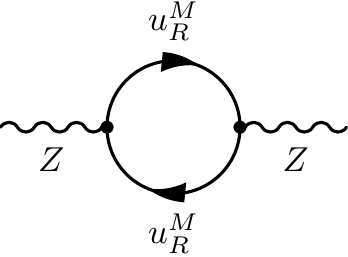}
		\raisebox{3.5em}{$\hspace{2em}=\;-\dfrac{4}{c_W^2} (T_3^{u^M} - s_W^2 Q_{u^M})^2 \left[(\dfrac{q^2}{6}-\dfrac{m_{u^M}^2}{2})\Delta - q^2 B_2(q^2;\;m_{u^M}^2,\;m_{u^M}^2) + m_{u^M}^2 B_1(q^2;\;m_{u^M}^2,\;m_{u^M}^2) \right]$}
		}\\
		\subfigure[]	{
		\label{fig:loopexrl}
		\includegraphics{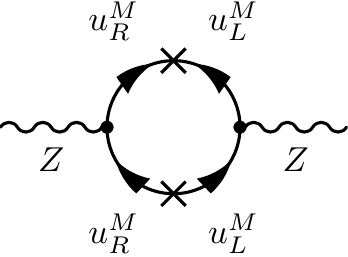}
		\raisebox{3.5em}{$\hspace{2em}=\;-\dfrac{2}{c_W^2} m_{u^M}^2 (T_3^{u^M} - s_W^2 Q_{u^M}) s_W^2 Q_{u^M} \left[\Delta - 2 B_1(q^2;\;m_{u^M}^2,\;m_{u^M}^2) \right]$}
		}
		\caption{\ewnur model mirror fermion loop examples}
		\label{fig:loopex}
	\end{flushleft}
	\end{figure*}

\begin{figure*}[!htb]
	\begin{flushleft}
		\subfigure[]	{
		\label{fig:loopexllsm}
		\includegraphics{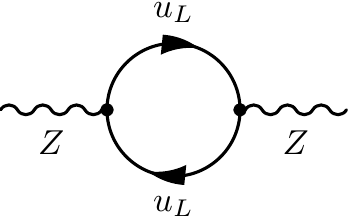}
		\raisebox{3.5em}{$\hspace{2em}=\;-\dfrac{4}{c_W^2} (T_3^u - s_W^2 Q_u)^2 \left[(\dfrac{q^2}{6}-\dfrac{m_u^2}{2})\Delta - q^2 B_2(q^2;\;m_u^2,\;m_u^2) + m_u^2 B_1(q^2;\;m_u^2,\;m_u^2) \right]$}
		}\\
		\subfigure[]	{
		\label{fig:loopexlrsm}
		\includegraphics{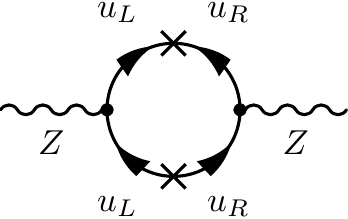}
		\raisebox{3.5em}{$\hspace{2em}=\;-\dfrac{2}{c_W^2} m_u^2 (T_3^u - s_W^2 Q_u) s_W^2 Q_u \left[\Delta - 2 B_1(q^2;\;m_u^2,\;m_u^2) \right]$}
		}
		\caption{Standard Model fermion loop examples}
		\label{fig:loopexsm}
	\end{flushleft}
	\end{figure*}

%
The new Physics contributions, $\widetilde{S}_{fermion}$ and $\widetilde{T}_{fermion}$, due to fermion sector in \ewnur model can be calculated by adding the respective contributions due to the lepton- and quark-sectors in \ewnur model that is,
	\begin{flalign}
		\hspace{2em}\widetilde{S}_{fermion} &= \widetilde{S}_{lepton} + \widetilde{S}_{quark}& \\
		\hspace{2em}\widetilde{T}_{fermion} &= \widetilde{T}_{lepton} + \widetilde{T}_{quark}&
	\end{flalign}
In this subsection the one-loop contributions to $\widetilde{S}_{fermion}$ and $\widetilde{T}_{fermion}$ are listed in tables \ref{table:floopww}, \ref{table:floopzz}, \ref{table:floopzg}, \ref{table:floop33}. In each of these tables only the loop contributions due to the mirror fermions in \ewnur model are listed. The same expressions for the loop contributions can be used to calculate the lepton loop diagrams and the quark loop diagrams. The fermion loop contributions in SM can be obtained from the mirror fermion loop having fermions with the opposite chirality going in the loop. Consider, for example, the mirror-up-quark-loop diagrams in FIG. \ref{fig:loopex} and SM-up-quark-loop diagrams in FIG. \ref{fig:loopexsm}.
  \begin{table*}[!htb]
  	\renewcommand{\arraystretch}{1}
	\caption{\label{table:floopww}Fermion loop diagrams with two internal mirror fermion lines, which contribute to $\Pi_{WW}(q^2)$. Here $f_{1R}^M$'s and $f_{2R}^M$'s are members of a mirror fermion doublet with isospins ($T_3^f$) equal to $\dfrac{1}{2}$ and $-\dfrac{1}{2}$ respectively. Common factor: $\dfrac{g^2 N_c}{16\pi^2}$}
	\begin{centering}
	\begin{tabular}{|p{4cm}c|}
	\hline
	\multicolumn{2}{|c|}{}\\
	\multicolumn{2}{|c|}{Contributions to $\Pi_{WW}(q^2)$} \\[4mm]
	\hline
	~&~\\
	\includegraphics{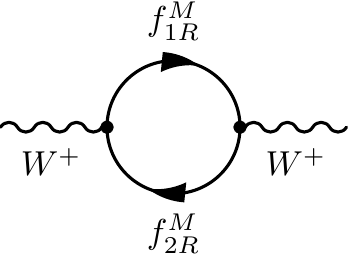}		& \raisebox{3em}{$-2\Bigg[\Bigg(\dfrac{q^2}{6}-\dfrac{1}{4}(m_{1f}^2+m_{2f}^2)\Bigg)\Delta - q^2 B_2(q^2;\;m_{1f}^2,\;m_{2f}^2)$}\\
	~& \raisebox{2em}{\hspace{2em}$ + \dfrac{1}{2}(m_{1f}^2 B_1(q^2;\;m_{1f}^2,\;m_{2f}^2) + m_{2f}^2 B_1(q^2;\;m_{2f}^2,\;m_{1f}^2))\Bigg]$}\\[3mm]
	\hline
	\end{tabular}
	\end{centering}
  \end{table*}
  \begin{table*}[!htb]
  	\renewcommand{\arraystretch}{1}
	\caption{\label{table:floopzz}Fermion loop diagrams with two internal mirror fermion lines, which contribute to $\Pi_{ZZ}(q^2)$. Common factor: $\dfrac{g^2 N_c}{16\pi^2}$}
	\begin{center}
	\begin{tabular}{|p{4cm}c|}
	\hline
	\multicolumn{2}{|c|}{}\\
	\multicolumn{2}{|c|}{Contributions to $\Pi_{ZZ}(q^2)$} \\[4mm]
	\hline
	~&~\\
	\includegraphics{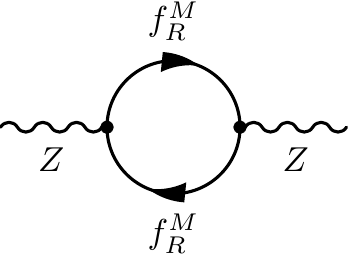}			& \raisebox{3em}{$-\dfrac{4}{c_W^2} (T_3^f - s_W^2 Q_f)^2 \Bigg[\Bigg(\dfrac{q^2}{6}-\dfrac{m_f^2}{2}\Bigg)\Delta$}\\
	~& \raisebox{2em}{$\hspace{2em}- q^2 B_2(q^2;\;m_f^2,\;m_f^2) + m_f^2 B_1(q^2;\;m_f^2,\;m_f^2) \Bigg]$}\\
	\hline
	~&~\\
	\includegraphics{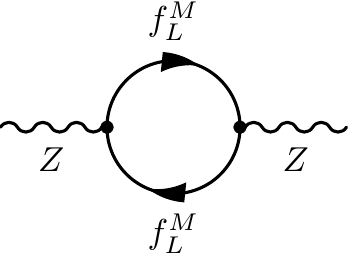}			& \raisebox{3em}{$-\dfrac{4}{c_W^2} s_W^4 Q_f^2 \Bigg[\Bigg(\dfrac{q^2}{6}-\dfrac{m_f^2}{2}\Bigg)\Delta$}\\
	~	& \raisebox{2em}{$\hspace{2em}- q^2 B_2(q^2;\;m_f^2,\;m_f^2) + m_f^2 B_1(q^2;\;m_f^2,\;m_f^2) \Bigg]$}\\
	\hline
	~&~\\
	\includegraphics{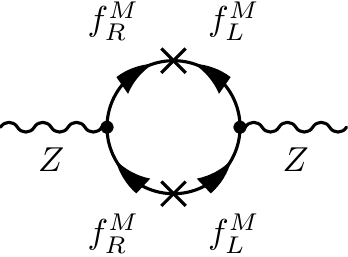}			& \raisebox{3em}{$-\dfrac{2}{c_W^2} m_f^2 (T_3^f - s_W^2 Q_f) s_W^2 Q_f \Big[\Delta - 2 B_1(q^2;\;m_f^2,\;m_f^2) \Big]$}\\[3mm]
	\hline
	~&~\\
	\includegraphics{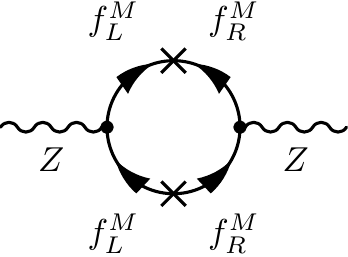}			& \raisebox{3em}{$-\dfrac{2}{c_W^2} m_f^2 (T_3^f - s_W^2 Q_f) s_W^2 Q_f \Big[\Delta - 2 B_1(q^2;\;m_f^2,\;m_f^2) \Big]$}\\[3mm]
	\hline
	\end{tabular}
	\end{center}
  \end{table*}
  \begin{table*}[!htb]
  	\renewcommand{\arraystretch}{1}
	\caption{\label{table:floopzg}Fermion loop diagrams with two internal mirror fermion lines, which contribute to $\Pi_{Z\gamma}(q^2)$. Common factor: $\dfrac{g^2 N_c}{16\pi^2}$}
	\begin{center}
	\begin{tabular}{|p{4cm}c|}
	\hline
	\multicolumn{2}{|c|}{}\\
	\multicolumn{2}{|c|}{Contributions to $\Pi_{Z\gamma}(q^2)$} \\[4mm]
	\hline
	~&~\\
	\includegraphics{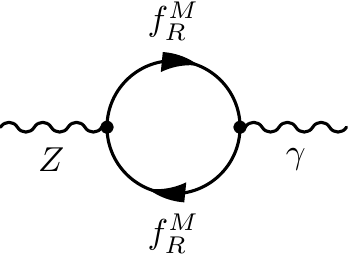}		& \raisebox{3em}{$-\dfrac{4}{c_W} (T_3^f - s_W^2 Q_f) s_W Q_f \Bigg[\Bigg(\dfrac{q^2}{6}-\dfrac{m_f^2}{2}\Bigg)\Delta$}\\
	~& \raisebox{2em}{$\hspace{2em}- q^2 B_2(q^2;\;m_f^2,\;m_f^2) + m_f^2 B_1(q^2;\;m_f^2,\;m_f^2) \Bigg]$}\\
	\hline
	~&~\\
	\includegraphics{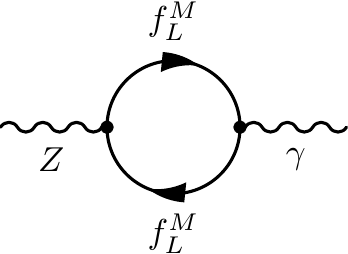}	& \raisebox{3em}{$\dfrac{4}{c_W} s_W^3 Q_f^2 \Bigg[\Bigg(\dfrac{q^2}{6}-\dfrac{m_f^2}{2}\Bigg)\Delta$}\\
	~& \raisebox{2em}{$\hspace{2em}- q^2 B_2(q^2;\;m_f^2,\;m_f^2) + m_f^2 B_1(q^2;\;m_f^2,\;m_f^2) \Bigg]$}\\
	\hline
	~&~\\
	\includegraphics{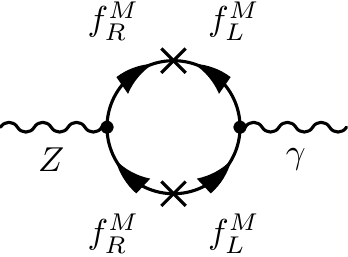}		& \raisebox{3em}{$\dfrac{2}{c_W} m_f^2 (T_3^f - s_W^2 Q_f) s_W Q_f \Big[\Delta - 2 B_1(q^2;\;m_f^2,\;m_f^2) \Big]$}\\[3mm]
	\hline
	~&~\\
	\includegraphics{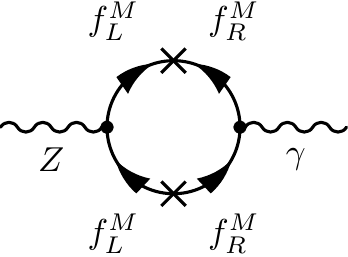}	& \raisebox{3em}{$-\dfrac{2}{c_W} m_f^2 s_W^3 Q_f^2 \Big[\Delta - 2 B_1(q^2;\;m_f^2,\;m_f^2) \Big]$}\\[3mm]
	\hline
	\end{tabular}
	\end{center}
  \end{table*}
  \begin{table*}[!htb]
  	\renewcommand{\arraystretch}{1}
	\caption{\label{table:floopgg}Fermion loop diagrams with two internal mirror fermion lines, which contribute to $\Pi_{\gamma\gamma}(q^2)$. Common factor: $\dfrac{g^2 N_c}{16\pi^2}$}
	\begin{center}
	\begin{tabular}{|p{4cm}c|}
	\hline
	\multicolumn{2}{|c|}{}\\
	\multicolumn{2}{|c|}{Contributions to $\Pi_{\gamma\gamma}(q^2)$} \\[4mm]
	\hline
	~&~\\
	\includegraphics{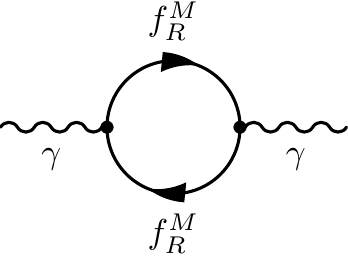}		& \raisebox{3em}{$-4 s_W^2 Q_f^2 \Bigg[\Bigg(\dfrac{q^2}{6}-\dfrac{m_f^2}{2}\Bigg)\Delta$}\\
	~& \raisebox{2em}{$\hspace{2em}- q^2 B_2(q^2;\;m_f^2,\;m_f^2) + m_f^2 B_1(q^2;\;m_f^2,\;m_f^2) \Bigg]$}\\
	\hline
	~&~\\
	\includegraphics{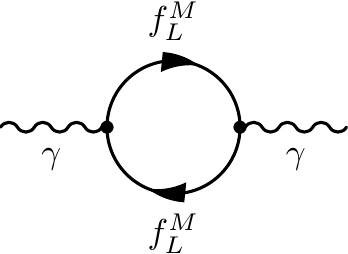}	& \raisebox{3em}{$-4 s_W^2 Q_f^2 \Bigg[\Bigg(\dfrac{q^2}{6}-\dfrac{m_f^2}{2}\Bigg)\Delta$}\\
	~& \raisebox{2em}{$\hspace{2em}- q^2 B_2(q^2;\;m_f^2,\;m_f^2) + m_f^2 B_1(q^2;\;m_f^2,\;m_f^2) \Bigg]$}\\
	\hline
	~&~\\
	\includegraphics{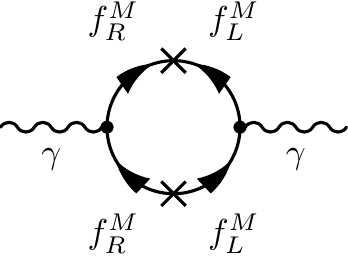}		& \raisebox{3em}{$2 m_f^2 s_W^2 Q_f^2 \Big[\Delta - 2 B_1(q^2;\;m_f^2,\;m_f^2) \Big]$}\\[3mm]
	\hline
	~&~\\
	\includegraphics{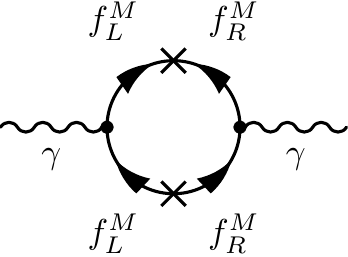}	& \raisebox{3em}{$2 m_f^2 s_W^2 Q_f^2 \Big[\Delta - 2 B_1(q^2;\;m_f^2,\;m_f^2) \Big]$}\\[3mm]	
	\hline
	\end{tabular}
	\end{center}
  \end{table*}
  \begin{table*}[!htb]
  	\renewcommand{\arraystretch}{1}
	\caption{\label{table:floop33}Fermion loop diagrams with two internal mirror fermion lines, which contribute to $\Pi_{33}(q^2)$. Common factor: $\dfrac{g^2 N_c}{16\pi^2}$}
	\begin{center}
	\begin{tabular}{|p{4cm}c|}
	\hline
	\multicolumn{2}{|c|}{}\\
	\multicolumn{2}{|c|}{Contributions to $\Pi_{33}(q^2)$} \\[4mm]
	\hline
	~&~\\
	\raisebox{2em}{$\displaystyle\lim_{g^\prime \rightarrow 0}\;$}\includegraphics{loopzfrfrz.pdf}	& \raisebox{3em}{$-4\; \left(T_3^f\right)^2 \Bigg[\Bigg(\dfrac{q^2}{6}-\dfrac{m_f^2}{2}\Bigg)\Delta$}\\
	~& \raisebox{2em}{$\hspace{2em}- q^2 B_2(q^2;\;m_f^2,\;m_f^2) + m_f^2 B_1(q^2;\;m_f^2,\;m_f^2) \Bigg]$}\\[3mm]	
	\hline
	\end{tabular}
	\end{center}
  \end{table*}
The contribution due to the loop diagram in FIG. \ref{fig:loopexllsm} (with two left-handed SM up quarks in the loop) has similar form of expression as the loop diagram in FIG. \ref{fig:loopexrr} with two right-handed mirror-up-quarks in the loop. Also, if the SM up quark loop diagram has mass-insertion propagators as in FIG. \ref{fig:loopexlrsm}, then it has similar form of expression as the loop diagram with mass-insertion propagators of mirror up quarks, FIG. \ref{fig:loopexrl}, when the left-handed-up-quarks-side of the loop is replaced by the right-handed-mirror-up-quark side of the loop and vice versa. The same correspondence exists between other one loop diagrams involving mirror fermions listed in tables \ref{table:floopww}, \ref{table:floopzz}, \ref{table:floopzg}, \ref{table:floop33} and the diagrams involving SM fermions. Therefore, we have not listed separately the SM fermion loop diagrams in this paper.

\par The definitions of the loop functions used in these tables are given in Appendix \ref{sec:appfunc}. Using these loop diagrams and the definitions of $S$, $T$ in eqns. (\ref{eq:s}), (\ref{eq:t}), we obtain the new Physics contributions, $\widetilde{S}_{lepton}$, $\widetilde{T}_{lepton}$ and $\widetilde{S}_{quark}$, $\widetilde{T}_{quark}$.  We also use $Q_f = T_3^f + \dfrac{Y_f}{2}$. Thus, for $\widetilde{S}_{lepton}$ we get (as given in Eq. (\ref{eq:expsl})):
  \begin{flalign}\label{eq:expsl}
  	\widetilde{S}_{lepton} =& S_{lepton}^{EW\nu_R} - S_{lepton}^{SM}&\nonumber \\[3mm]
  	=&\;\frac{(N_{C})_{lepton}}{6\pi} \sum_{i=1}^{3} \Bigg\{ -2\; Y_{lepton}\; x_{\nu i}&\nonumber\\[3mm]
	 &+  2 \left(-4\frac{Y_{lepton}}{2} + 3\right) x_{ei} - Y_{lepton}\; ln\left(\frac{x_{\nu i}}{x_{ei}}\right)&\nonumber\\[3mm]
  	&+\; \left(1 - x_{\nu i}\right) \frac{Y_{lepton}}{2} G(x_{\nu i})&\nonumber\\[3mm]
	&+\; \left[\left(\frac{3}{2} - \frac{Y_{lepton}}{2}\right) x_{ei} - \frac{Y_{lepton}}{2}\right] G(x_{ei})\Bigg\}&
  \end{flalign}
For $\widetilde{T}_{lepton}$ we obtain,
	\begin{flalign}
		&\widetilde{T}_{lepton} =\; T_{lepton}^{EW\nu_R} - T_{lepton}^{SM}&\nonumber \\[3mm]
		& =\; \frac{(N_{C})_{lepton}}{4 \pi s_W^2 M_W^2}\times&\nonumber \\[3mm]
		& \sum_{i=1}^{3} \Bigg[ m_{\nu i}^2 \Big(B_1(0;m_{\nu i}^2,m_{\nu i}^2) - B_1(0;m_{\nu i}^2,m_{ei}^2)\Big)&\nonumber \\[3mm]
		& + m_{ei}^2 \Big(B_1(0;m_{ei}^2,m_{ei}^2) - B_1(0;m_{ei}^2,m_{\nu i}^2)\Big) \Bigg]&
	\end{flalign}
Hence, as given in Eq. (\ref{eq:exptl}):
  \begin{flalign}\label{eq:exptl}
  	\widetilde{T}_{lepton} &=\; \frac{(N_{C})_{lepton}}{8 \pi s_W^2 M_W^2} \sum_{i=1}^{3} \mathcal{F}(m_{\nu i}^2, m_{ei}^2).&
  \end{flalign}
Here, because we have subtracted the contribution from three generations of SM leptons, the summation is over three generations of mirror leptons only. Subscripts $\nu i$ and $ei$ represent the mass eigenstates, right-handed neutrino ($\nu_{Ri}$) and mirror electron ($e^M_i$) member, of the $i^{th}$ mirror lepton generation respectively. $(N_{C})_{lepton} = 1$ is the lepton color factor and $Y_{lepton} = -1$ is  hypercharge for mirror leptons. $x_{\nu i,\;ei} = (m_{\nu i,\;ei} / M_Z)^2$, where $m_{\nu i,\;ei}$ are masses of $\nu_{Ri}$ and $e^M_i$ respectively. And $G(x)$ is given by Eq. (\ref{eq:G}).
\par The new Physics contributions to $S$ and $T$ from the quark sector in \ewnur model are given by
  \begin{flalign}\label{eq:expsq}
  	\widetilde{S}_{quark} =& S_{quark}^{EW\nu_R} - S_{quark}^{SM}&\nonumber \\[3mm]
  	=&\;\frac{(N_{C})_{quark}}{6\pi} \sum_{i=1}^{3} \Bigg\{ 2 \left( 4 \frac{Y_{quark}}{2} + 3\right) x_{ui}&\nonumber\\[3mm]
	&+ 2 \left(-4\frac{Y_{quark}}{2} + 3\right) x_{di} - Y_{quark}\; ln\left(\frac{x_{ui}}{x_{di}}\right)&\nonumber\\[3mm]
  	&+\;\left[\left(\frac{3}{2} + Y_{quark}\right) x_{ui} + \frac{Y_{quark}}{2}\right] G(x_{ui})&\nonumber\\[3mm]
	&+ \left[\left(\frac{3}{2} - Y_{quark}\right) x_{di} - \frac{Y_{quark}}{2}\right] G(x_{di}) \Bigg\}&
  \end{flalign}
and
  \begin{flalign}\label{eq:exptq}
  	\widetilde{T}_{quark} &= \; T_{quark}^{EW\nu_R} - T_{quark}^{SM}&\nonumber \\[3mm]
		& =\; \frac{(N_{C})_{quark}}{4 \pi s_W^2 M_W^2}\times&\nonumber \\[3mm]
		& \sum_{i=1}^{3} \Bigg[ m_{ui}^2 \Big(B_1(0;m_{ui}^2,m_{ui}^2) - B_1(0;m_{ui}^2,m_{di}^2)\Big)&\nonumber \\[3mm]
		& + m_{di}^2 \Big(B_1(0;m_{di}^2,m_{di}^2) - B_1(0;m_{di}^2,m_{ui}^2)\Big) \Bigg]&\nonumber \\[3mm]
  	&=\; \frac{(N_{C})_{quark}}{8 \pi s_W^2 M_W^2} \sum_{i=1}^{3} \mathcal{F}(m_{ui}^2, m_{di}^2),&
  \end{flalign}
respectively. Once again, because we have subtracted the contribution from three generations of SM quarks, the summation is over three generations of mirror quarks only. Subscripts $ui$ and $di$ represent the mass eigenstates of the mirror up- ($u^M_i$) and the mirror down- ($d^M_i$) member of the $i^{th}$ mirror-quark generation respectively (refer to the arguments about negligible mirror fermion mixings given after Eq. (\ref{eq:exptl}) ). $(N_{C})_{quark} = 3$ is the quark color factor and $Y_{quark} = -1/3$ is  hypercharge for mirror quarks. $x_{ui,\;di} = (m_{ui,\;di} / M_Z)^2$, where $m_{ui,\;di}$ are masses of $u^M_i$ and $d^M_i$ respectively. Refer to Appendix \ref{sec:appstf} for the mirror fermion loop diagrams contributing to $S$ and $T$.
%
\par As in in section \ref{unconstrained}, both $\widetilde{S}_{lepton}$ and $\widetilde{S}_{quark}$ favor positive values more than the negative values, although this trend is not apparent in eqns. (\ref{eq:expsl}), (\ref{eq:expsq}). It can be seen in eqns. (\ref{eq:exptl}) and (\ref{eq:exptq}) that both $\widetilde{T}_{lepton}$ and $\widetilde{T}_{quark}$ are always positive. Also contribution to these quantities from any mirror lepton and mirror quark generation (respectively) increases with the mass splitting within the doublet of the mirror generation. These behaviors are expected in \ewnur model so that the total $\widetilde{S}$ and $\widetilde{T}$ satisfy the experimental constraints given in section \ref{unconstrained}.
%
%
\FloatBarrier
%
%
%
%

\end{document}